\pdfoutput=1
\documentclass[11pt,twoside,a4paper,cmspaper,final,collab]{cms-tdr}
\def\svnVersion{7aaf069}\def\svnDate{2022/03/20}\def\cmsCernNoTag{CERN-EP-2021-177}\def\cmsCernDate{\today}\def\cmsMessage{Published in the Journal of Instrumentation as \href{http://dx.doi.org/10.1088/1748-0221/17/03/P03014}{\doi{10.1088/1748-0221/17/03/P03014}.}}
\begin{document}\cmsNoteHeader{BTV-20-001}

\cmsNoteHeader{BTV-20-001}

\title{A new calibration method for charm jet identification validated with proton-proton collision events at \texorpdfstring{$\sqrt{s}=13\TeV$}}

\date{\today}

\abstract{
   Many measurements at the LHC require efficient identification of heavy-flavour jets, \ie jets originating from bottom ({\PQb}) or charm ({\PQc}) quarks.
   An overview of the algorithms used to identify {\PQc} jets is described and a novel method to calibrate them is presented.
   This new method adjusts the entire distributions of the outputs obtained when the algorithms are applied to jets of different flavours.
   It is based on an iterative approach exploiting three distinct control regions that are enriched with
   either {\PQb} jets, {\PQc} jets, or light-flavour and gluon jets. Results are presented in the form of correction factors evaluated using proton-proton collision data with an integrated luminosity of 41.5\fbinv at $\sqrt{s}=13\TeV$, collected by the CMS experiment in 2017.
   The closure of the method is tested by applying the measured correction factors on simulated data sets and checking the agreement between the adjusted simulation and collision data. Furthermore, a validation is performed by testing the method on pseudodata, which emulate various mismodelling conditions.
   The calibrated results enable the use of the full distributions of heavy-flavour identification algorithm outputs, \eg as inputs to machine-learning models. Thus, they are expected to increase the sensitivity of future physics analyses.
}

\hypersetup{
pdfauthor={CMS Collaboration},
pdftitle={A new calibration method for charm jet identification validated with proton-proton collision events at sqrt(s) = 13 TeV},
pdfsubject={CMS},
pdfkeywords={CMS, jets, jet flavour tagging, charm}}

\maketitle

\section{Introduction \label{sec:intro}}
Quarks and gluons are abundantly produced in proton-proton ($\Pp\Pp$) collisions at the CERN LHC and an energetic interaction typically yields collimated streams of particles in the detector, referred to as a jet. The particles in a jet arise from the showering, fragmentation, and hadronisation of an initially coloured particle and constitute a colour-neutral final state. Several measurements of standard model processes, as well as searches for beyond the standard model physics, rely on the identification of the flavour of the coloured particle that initiates a jet. The jets originating from bottom ({\PQb}) and charm ({\PQc}) quarks are identified using heavy-flavour tagging, which distinguishes them from jets initiated by light flavour quarks (up, down, strange) or gluons~\cite{Chatrchyan:2012jua,Sirunyan:2017ezt}. These algorithms exploit the hard fragmentation, long lifetimes, and relatively high masses of {\PQb} and {\PQc} hadrons to identify heavy-flavour jets. Although {\PQb} jet identification algorithms have been deployed for several decades, the more challenging task of identifying {\PQc} jets at CMS was addressed for the first time during the preparation for the data taking with the first 13\TeV centre-of-mass energy $\Pp\Pp$ collisions~\cite{Sirunyan:2017ezt,CMS-PAS-BTV-16-001}.

The heavy-flavour tagging identification process calculates the probability (discriminator) that a jet is initiated by a specific quark flavour. Such tagging typically uses machine-learning classification algorithms to exploit the vast information available for a jet, such as the properties of the individual constituents as well as the jet as a whole. Currently the CMS Collaboration uses two such algorithms, called DeepCSV~\cite{Sirunyan:2017ezt} and DeepJet~\cite{bols2020jet,CMS-DP-2018-058}, which are based on the use of deep neural networks (DNNs)~\cite{Guest_2016}; these algorithms differ in the amount of input information, as well as the internal structure of the network, as explained in Section~\ref{sec:ctagging}. Simulated $\Pp\Pp$ collisions are used to train these algorithms, and care must be taken to avoid spurious effects resulting from mismodelling along the simulation chain. Discrepancies between simulation and data can arise from a variety of sources ranging from the imperfect simulation of the detector to the matrix element calculation, which has limitations in the modelling of the parton shower, hadronisation and fragmentation processes. When such algorithms are applied in physics analyses, a calibration of the heavy-flavour tagging output probabilities in simulated events is needed to match those in actual collision data.

Traditional approaches of flavour tagging usually involve labelling each jet in an event as "tagged" or "untagged" depending on whether the discriminator output for the jet is higher or lower than a fixed threshold (working point), thereby making it possible to count the number of {\PQb}- or {\PQc}-tagged jets in each event. To ensure agreement of this number between simulation and data, on average, the efficiency of selecting (rejecting) each tagged (untagged) jet in a simulated event is adjusted by a scale factor (SF), that quantifies the difference in selection (rejection) efficiencies of various flavours of jets between simulation and data for the working point being used. Such SFs are measured for each flavour of jet for various working points using different selections of jets in data. SFs for {\PQb} jets are derived from top quark pair (\ttbar) production and/or from quantum chromodynamics (QCD) multijet events where a jet contains a low-energy muon. A selection of events in which a W boson is produced in association with a charm jet ({\PW}+{\PQc}) is used to derive SFs for {\PQc} jets. Finally, light-flavour and gluon jet SFs are derived using QCD multijet events~\cite{Sirunyan:2017ezt}.

However, additional information can also be gained from the full output distribution (discriminator distribution shapes) of the heavy-flavour tagging discriminators, \eg by using the discriminator output for each jet as an input to a machine-learning classifier or by performing a fit of the discriminator distributions to data. This gives rise to the need for a second type of calibration that maps the full simulated distribution to the one observed in data. Such a shape calibration technique has already been developed for the identification of {\PQb} jets, and has been successfully applied in the observation of the Higgs boson ${\PH}\to{\PQb}\bar{{\PQb}}$ decay mode~\cite{Sirunyan:2018kst}.

This paper presents for the first time a calibration method to correct the differential {\PQc} tagging discriminator distribution shapes. This novel technique is based on an iterative fit procedure that exploits three control regions that are enriched in {\PQc}, {\PQb}, and light-flavour or gluon jets. The first successful attempt to perform such a charm tagging shape calibration was reported in a study of \ttbar production with additional {\PQc} jets~\cite{Moortgat:2676133}. The present paper describes a more advanced strategy that uses higher-purity control regions and an optimised granularity in the discriminator binning to derive shape calibration SFs. A similar technique has been demonstrated to work in a search for associated production of a Higgs boson with a vector boson, where the Higgs boson decays into a pair of charm quarks~\cite{Sirunyan:2019qia}.

In Sections \ref{sec:CMS} to \ref{sec:objectreco}, the CMS detector, details of the data and simulated events, and the object reconstruction are discussed. The {\PQc} tagging algorithms are introduced in Section~\ref{sec:ctagging}, followed by a description of the event selection used to derive three control regions in Section~\ref{sec:Event-Selection}. The iterative fit strategy is outlined in Section~\ref{sec:Method}, and the relevant sources of systematic uncertainties are discussed in Section~\ref{sec:Systematics}. The results are discussed in Section~\ref{sec:Results}, the validation tests are documented in Section~\ref{sec:Validation}, and the paper is summarised in Section~\ref{sec:conclusion}.

\section{The CMS detector \label{sec:CMS}}

The central feature of the CMS apparatus is a superconducting solenoid with an internal dia\-meter of 6\unit{m}, providing a magnetic field of 3.8\unit{T}. Within the solenoid volume are a silicon pixel and strip tracker, a lead tungstate crystal electromagnetic calorimeter (ECAL), and a brass and scintillator hadron calorimeter (HCAL), each composed of a barrel and two endcap sections. The silicon tracker consists of 1440 silicon pixel and 15\,148 silicon strip detector modules and its new inner part was installed in 2017 as a part of the Phase-1 pixel detector upgrade. The upgraded high-efficiency and low-mass detector with four barrel layers provides four-hit pixel coverage to efficiently detect charged particles within the pseudorapidity range $\abs{\eta} < 2.5$~\cite{CMS-TDR-011}. Transverse impact parameter resolution of each track ranges from 20 to 75\mum depending on the transverse momentum ($\pt$) and $\eta$ of the track~\cite{CMS-DP-2020-049}. Forward calorimeters, made of steel and quartz fibres, extend the $\eta$ coverage provided by the barrel and endcap detectors. Muons are detected in gas-ionisation chambers embedded in the steel flux-return yoke outside the solenoid. A more detailed description of the CMS detector, together with a definition of the coordinate system used and the relevant kinematic variables, can be found in Ref.~\cite{Chatrchyan:2008zzk}.

Events of interest are selected using a two-tiered trigger system~\cite{Khachatryan:2016bia}. The first level~\cite{Sirunyan:2020zal}, composed of custom hardware processors, uses information from the calorimeters and muon detectors to select events at a rate of around 100\unit{kHz} within a fixed latency of less than 4\mus. The second level, known as the high-level trigger (HLT), consists of a farm of processors running a version of the full event reconstruction software optimised for fast processing, and reduces the event rate to around 1\unit{kHz} before data storage.

\section{Data and simulated samples \label{sec:simulation}}

The data used to derive the calibration of the {\PQc} tagger discriminator distributions were collected from $\Pp\Pp$ collisions recorded at a centre-of-mass energy of 13\TeV with the CMS detector in 2017, and correspond to an integrated luminosity of 41.5\fbinv~\cite{CMS:2018elu}. The 2017 data set is the first data collected after the Phase-1 upgrade of the CMS pixel tracker \cite{CMS-TDR-011} and is expected to represent the current heavy-flavour tagging performance of CMS. The collision events are selected using a set of single-lepton and dilepton triggers. The single-electron (single-muon) trigger requires the presence of at least one isolated electron (muon) with a \pt above 32 (27)\GeV. The dielectron (dimuon) trigger requires at least two isolated electrons (muons), one with $\pt> 23 \, (17)\GeV$ and another with  $\pt> 12 \, (8)\GeV$. Finally, the electron-muon trigger requires the presence of at least one electron and at least one muon, where the lepton with the largest \pt is required to have $\pt>23\GeV$ and the one with the second largest \pt to have $\pt>12 \, (8)$\GeV if it is an electron (muon). These trigger requirements are also imposed on the simulated collision events.

The simulated events are produced using Monte Carlo (MC) generators, which provide a fixed-order perturbative QCD calculation of up to four noncollinear high-\pt hard partons, supplemented with parton showering (PS) and typical underlying event particles. For all the simulated events, the PS is simulated using \PYTHIA v8.230~\cite{SJOSTRAND2015159}, using the CP5 underlying event tune~\cite{Skands:2014pea} with the NNPDF~3.1~\cite{Ball:2014uwa} parton distribution function set.

The matrix element (ME) generation of the \ttbar simulation is performed with \POWHEG (v2)~\cite{Nason:2004rx,Frixione:2007vw,Alioli:2010xd, Campbell:2014kua} at next-to-leading order (NLO) accuracy in QCD, and its cross section is scaled to a theoretical prediction at next-to-NLO (NNLO) in QCD, including resummation of next-to-next-to-leading logarithmic soft-gluon terms. The ME generation of the $\PW+\text{jets}$ and Drell--Yan (DY) processes is performed with \MGvATNLO 2.4.2~\cite{Alwall:2014hca} at leading order (LO) precision, with MLM jet matching~\cite{Alwall:2007fs} and the cross sections normalised to NNLO calculations~\cite{Li:2012wna}. 
The single top quark production in the $s$-channel was also simulated using \MGvATNLO in the four-flavour scheme, whereas the single top quark production in the $t$-channel was simulated using \POWHEG also in the four-flavour scheme. The tW channel was simulated using \POWHEG in the five-flavour scheme~\cite{Frixione:2008yi,Re:2010bp}, with its cross section normalised to the NLO calculations~\cite{Kidonakis:2012rm}. The diboson samples are simulated at LO using \PYTHIA for the ME generation. Their cross sections are normalised to those calculated at NLO for {\PW}{\PZ} and {\PZ}{\PZ}~\cite{Campbell:2010ff}, and NNLO for {\PW}{\PW}~\cite{Gehrmann:2014fva}. 

The interactions between particles and the material of the CMS detector are simulated using \GEANTfour~\cite{AGOSTINELLI2003250}. The effect of additional $\Pp\Pp$ interactions within the same or nearby bunch
crossings (pileup) on top of the hard scattering processes is modelled by additional minimum bias collisions generated with \PYTHIA.

\section{Object reconstruction \label{sec:objectreco}}

The particle-flow (PF) event algorithm~\cite{CMS-PRF-14-001} reconstructs and identifies each particle (physics-object) in an event with an optimised combination of all subdetector information. In this process, the identification of the particle type (photon, electron, muon, charged and neutral hadrons) plays an important role in the determination of the particle direction and energy. Photons (\eg coming from \Pgpz\ decays) are identified as ECAL energy clusters not linked to the extrapolation of any charged-particle trajectory to the ECAL.
Electrons are identified as a primary charged-particle track and potentially many ECAL energy clusters corresponding to this track and to possible bremsstrahlung photons. Muons are identified as tracks in the central tracker consistent with either a track or several hits in the muon system, and associated with calorimeter deposits compatible with the muon hypothesis. Charged hadrons are identified as charged-particle tracks identified neither as electrons, nor as muons. Finally, neutral hadrons are identified as HCAL energy clusters not linked to any charged-hadron trajectory, or as a combined ECAL and HCAL energy excess with respect to the expected charged-hadron energy deposit.

The energy of photons is obtained from the ECAL measurement. The energy of electrons is determined from a combination of the track momentum at the main interaction vertex, the corresponding ECAL cluster energy, and the energy sum of all bremsstrahlung photons attached to the track. The energy of muons is obtained from the corresponding track curvature. The energy of charged hadrons is determined from a combination of the track curvature and the corresponding ECAL and HCAL energies, corrected by the response function of the calorimeters to hadronic showers. Finally, the energy of neutral hadrons is obtained from the corresponding corrected ECAL and HCAL energies. 

The missing transverse momentum vector, \ptvecmiss, is defined as the negative vector \pt sum of all the PF candidates in an event, and its magnitude is denoted as \ptmiss~\cite{Sirunyan:2019kia}. The \ptvecmiss is modified to correct the energy scale of the reconstructed jets in the event. 

The candidate vertex with the largest value of summed physics-object $\pt^2$ is the primary vertex (PV) of the $\Pp\Pp$ interaction. These physics objects are the jets, the leptons, and the \ptvecmiss. The jets are reconstructed by the jet-finding algorithm~\cite{Cacciari:2008gp,Cacciari:2011ma} with the tracks assigned to candidate vertices as inputs.

To identify and reconstruct the prompt leptonic decay modes of vector bosons, electrons (muons) are used, and are required to be well isolated from jet activity within a cone of radius $\Delta R = \sqrt{(\Delta\eta)^2 + (\Delta\phi)^2} = 0.3$ (0.4), where $\phi$ is the azimuthal angle in radians. The relative isolation is defined as the scalar \pt sum of the PF candidates within the cone divided by the lepton \pt. Additionally, a set of quality requirements are imposed on these leptons based on the quality of the track reconstruction, hit multiplicities in the tracking and muon subdetector layers, and the displacement of these particles with respect to the PV. Electrons and muons that arise from (semi)leptonic decay modes of hadrons typically have a lower momentum and are surrounded by hadronic activity of the underlying jet in which these hadrons are created. The branching fraction of the (semi)leptonic decay modes of heavy hadrons gives rise to the presence of an electron or a muon inside 20 (10)\% of {\PQb} ({\PQc}) jets~\cite{Sirunyan:2017ezt}. A set of relaxed kinematic and inverted isolation criteria is imposed to identify such low-energy (soft) leptons inside jets.

For each event, jets are clustered from these reconstructed particles using the infrared- and collinear-safe anti-\kt algorithm~\cite{Cacciari:2008gp, Cacciari:2011ma} with a distance parameter of 0.4. Jet momentum is determined as the vector sum of all particle momenta in the jet, and is found from simulation to be, on average, within 5-10\% of the true momentum over the entire \pt spectrum and detector acceptance. To mitigate the effect of pileup contributing additional tracks and calorimetric energy depositions to the jet momentum, the charged particles identified as originating from pileup vertices are discarded, and an offset correction is applied to correct for remaining contributions. Jet energy corrections are derived from simulations to bring the measured average response of jets to that of generator-level jets. In situ measurements of the momentum balance in dijet, $\PGg+\text{jet}$, $\PZ+\text{jet}$, and multijet events are used to correct any residual differences in jet energy scale in data and simulation~\cite{Khachatryan:2016kdb}. The jet energy resolution amounts typically to 15--20\% at 30\GeV, 10\% at 100\GeV, and 5\% at 1\TeV~\cite{Khachatryan:2016kdb}. Additional selection criteria are applied to each jet to remove jets potentially dominated by anomalous contributions from various subdetector components or reconstruction failures. Jets overlapping with isolated electrons or muons within $\Delta R<0.4$ are disregarded to remove isolated charged leptons reconstructed as jets. Since heavy-flavour tagging algorithms strongly rely on the track reconstruction, only jets within the tracker acceptance ($\abs{\eta} < 2.5$) are used throughout this work. Additionally, a lower threshold on the jet \pt at 20\GeV is imposed, and jets are required to pass tight identification as well as tight pileup rejection criteria~\cite{Sirunyan:2020foa}.

For simulated jets, a generator-level definition of the jet flavour is needed to train the heavy-flavour tagging algorithms, as well as to calibrate them using data. This definition is based on a procedure referred to as ghost-matching~\cite{Cacciari:2007fd}, which involves reclustering the generator-level jet constituents after adding the intermediately decayed {\PQb} or {\PQc} hadrons to the list of particles used for the clustering, as already used in Ref.~\cite{Sirunyan:2017ezt}. In case there are multiple {\PQb} ({\PQc}) hadrons in a decay chain, only the last {\PQb} ({\PQc}) hadron in the chain, \ie, the {\PQb} ({\PQc}) hadron that does not further decay into another {\PQb} ({\PQc}) hadron, is reclustered. The moduli of the four-momenta of these generated hadrons are rescaled to a very small number (resulting in so-called ghost hadrons), to ensure that they do not affect the reconstructed jet momentum, and only their directional information is kept. If at least one ghost {\PQb} hadron has been clustered inside the jet, it is referred to as a bottom jet. If no {\PQb} hadron is found, but instead at least one {\PQc} ghost hadron is clustered inside the jet, it is referred to as a charm jet. In all other cases the jet is categorised as a light-flavour jet, and hence this category includes jets originating from light quarks ($\PQu\PQd\PQs$), as well as gluon-initiated (\Pg) jets. Particles from pileup interactions are not included in the reclustering.

Displaced secondary vertices (SV) from the decays of {\PQb} or {\PQc} hadrons are reconstructed using the inclusive secondary vertex finding algorithm~\cite{Khachatryan:2011wq}. This algorithm does not a priori assume any matching of the jets to a SV, but instead takes as input all reconstructed tracks in the event with $\pt>0.8\GeV$, and having a distance of closest approach from PV to track, projected onto the direction of the beam axis, below 0.3\unit{cm}. After the iterative procedure of track clustering, the final set of tracks is matched to the reconstructed jets if the distance between the direction of the SV displacement (pointing from PV to SV) and the jet axis satisfies $\Delta R<0.4$.

\section{Charm jet identification \label{sec:ctagging}}
\subsection{The c tagging algorithms \label{ctaggingAlgorithm}}

The identification of {\PQc} jets relies on the long lifetime and the mass of the {\PQc} hadron. The average lifetime of such charmed hadrons in the rest frame is typically a picosecond, resulting in a displacement of the decay vertex in the detector frame by a few millimetres to a centimetre from the position of the primary vertex. The reconstructed invariant mass, computed from the four-vectors of the particles that are assigned to such a secondary vertex, is strongly correlated with the {\PQc} hadron mass.
The existing heavy-flavour tagging algorithms exploit these properties by combining the features of tracks inside the jet (momentum, displacement, multiplicity) with the features of reconstructed SVs (mass and both the direction and the magnitude of the SV displacement from the PV).

These are the same properties that are used to identify {\PQb} jets. Since the average lifetime and mass of {\PQb} hadrons is larger than that of {\PQc} hadrons, the signatures of {\PQb} jets are more easily distinguishable from those of light-flavour jets than those of {\PQc} jets. The fact that the discriminating properties of {\PQc} jets are distributed midway between those of light-flavour and {\PQb} jets requires the definition of two discriminating variables; one to distinguish {\PQc} jets from light-flavour jets (CvsL) and one to distinguish {\PQc} jets from {\PQb} jets (CvsB). Historically, the first {\PQc} jet identification algorithm used in the CMS Collaboration was based on a combination of two boosted decision trees (BDTs), each trained for one of the above mentioned discrimination purposes~\cite{CMS-PAS-BTV-16-001}. The state-of-the-art heavy-flavour tagging algorithms are currently based on multi-class deep neural network architectures and predict separate probabilities for each jet to originate from a given quark flavour (or gluon). Through an appropriate combination of the probability values computed by these algorithms, these taggers can function either as a bottom jet or as a charm jet identification algorithm.

The DNN architecture of the DeepCSV algorithm is composed of five fully connected hidden layers with 100 nodes in each layer. It takes as input a set of 66 reconstructed observables related to the charged-particle tracks and secondary vertices that are assigned to a given jet, and outputs four probabilities, $P$({\PQb}), $P$($\PQb\PQb$), $P$({\PQc}) and $P$($\PQu\PQd\PQs\Pg$), that denote the probability of a jet to originate from one {\PQb} quark, two {\PQb} quarks merged into the same jet, one or more {\PQc} quarks or a light-flavour quark or gluon, respectively. The DeepJet algorithm uses an architecture composed of subsequent convolutional, recurrent and fully connected hidden layers. Its input is composed of a set of up to 613 observables related to charged and neutral PF candidates (without a priori selection criteria and without explicitly classifying charged PF candidates as charged hadron or leptons, and neutral PF candidates as photons or neutral hadrons) as well as the SVs that are assigned to the jet. Apart from the fact that DeepJet exhibits a higher-dimensional input space and a more complex architecture, it further subdivides the output classes into additional categories. In addition to the DeepCSV output categories, $P$(${\PQb}_{\text{lep}}$) is added to identify leptonic {\PQb} hadron decays and $P$($\PQu\PQd\PQs\Pg$) is split further into $P$($\PQu\PQd\PQs$) and $P$($\Pg$) with the goal of separately identifying jets originating from light quarks and gluons, respectively. More detailed information on the inputs, architecture, and training of these algorithms can be found in Refs.~\cite{Sirunyan:2017ezt,bols2020jet,CMS-DP-2018-058}. 

\begin{table}[!ht]
	\center
	\topcaption{\label{tab:taggerDefinitions}
		Summary of the heavy-flavour tagging definitions for both {\PQb} and {\PQc} tagging using the DeepCSV and DeepJet taggers. $P$(a) represents the probability of having an a-type jet (see text).}
	\begin{tabular}{c c c c}
		\hline 
		Tagger &  BvsC/L  &  CvsB & CvsL  \\
		\hline
		DeepCSV & $P$({\PQb})+$P$($\PQb\PQb$) & $\frac{P({\PQc})}{P({\PQc})+P({\PQb})+P(\PQb\PQb)}$ & $\frac{P(\text{{\PQc}})}{P(\text{{\PQc}})+P({\PQu}{\PQd}{\PQs}{\Pg})}$ \\
		DeepJet & $P$({\PQb})+$P$($\PQb\PQb$)+$P$($\PQb_{\text{lep}}$) & $\frac{P(\PQc)}{P(\PQc)+P(\PQb)+P({\PQb}{\PQb})+P(\PQb_{\text{lep}})}$  & $\frac{P(\PQc)}{P(\PQc)+P(\PQu\PQd\PQs)+P(\Pg)}$ \\
		\hline
	\end{tabular}
\end{table}

\begin{figure}[htb!]
	\centering
		\includegraphics[width=0.48\textwidth]{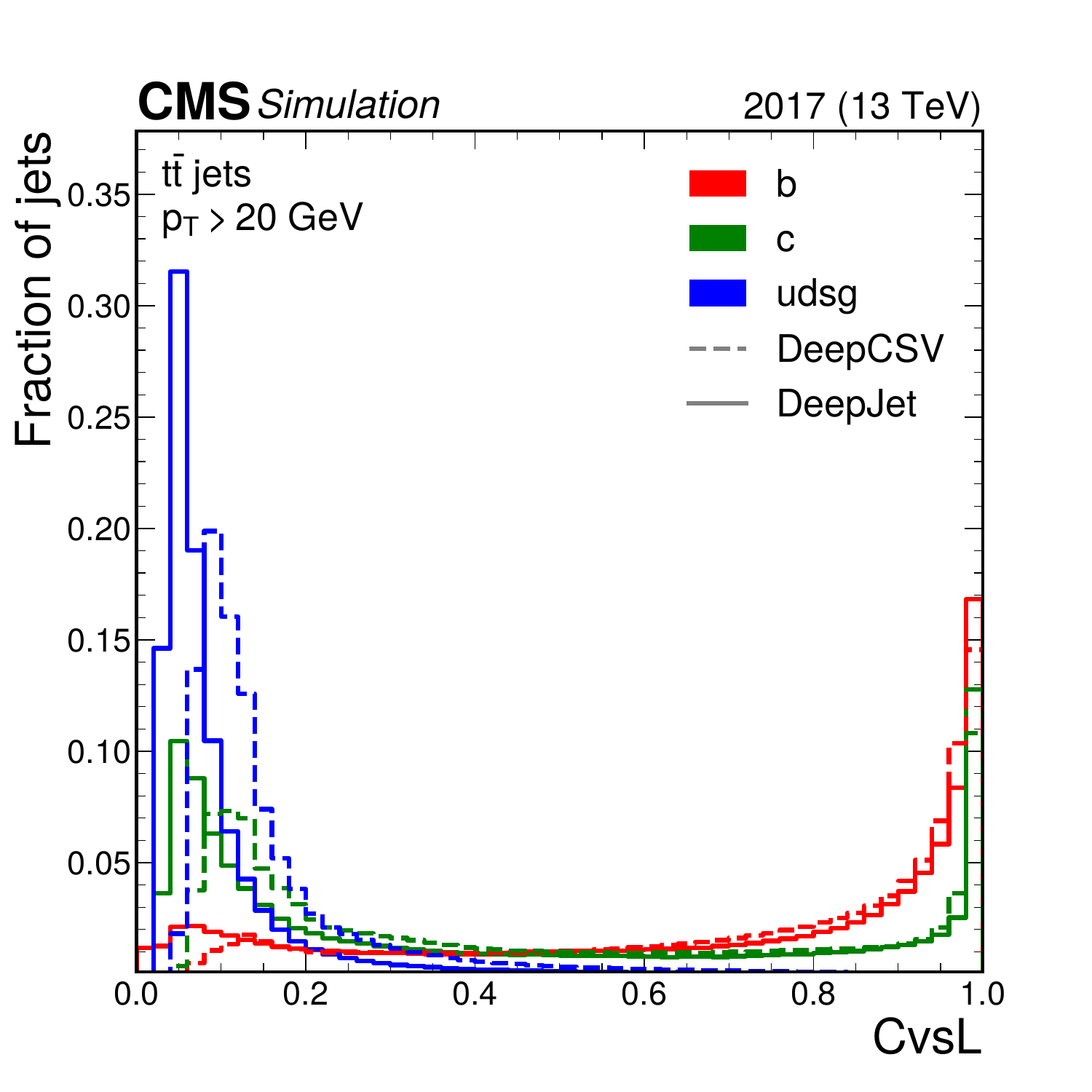}
		\includegraphics[width=0.48\textwidth]{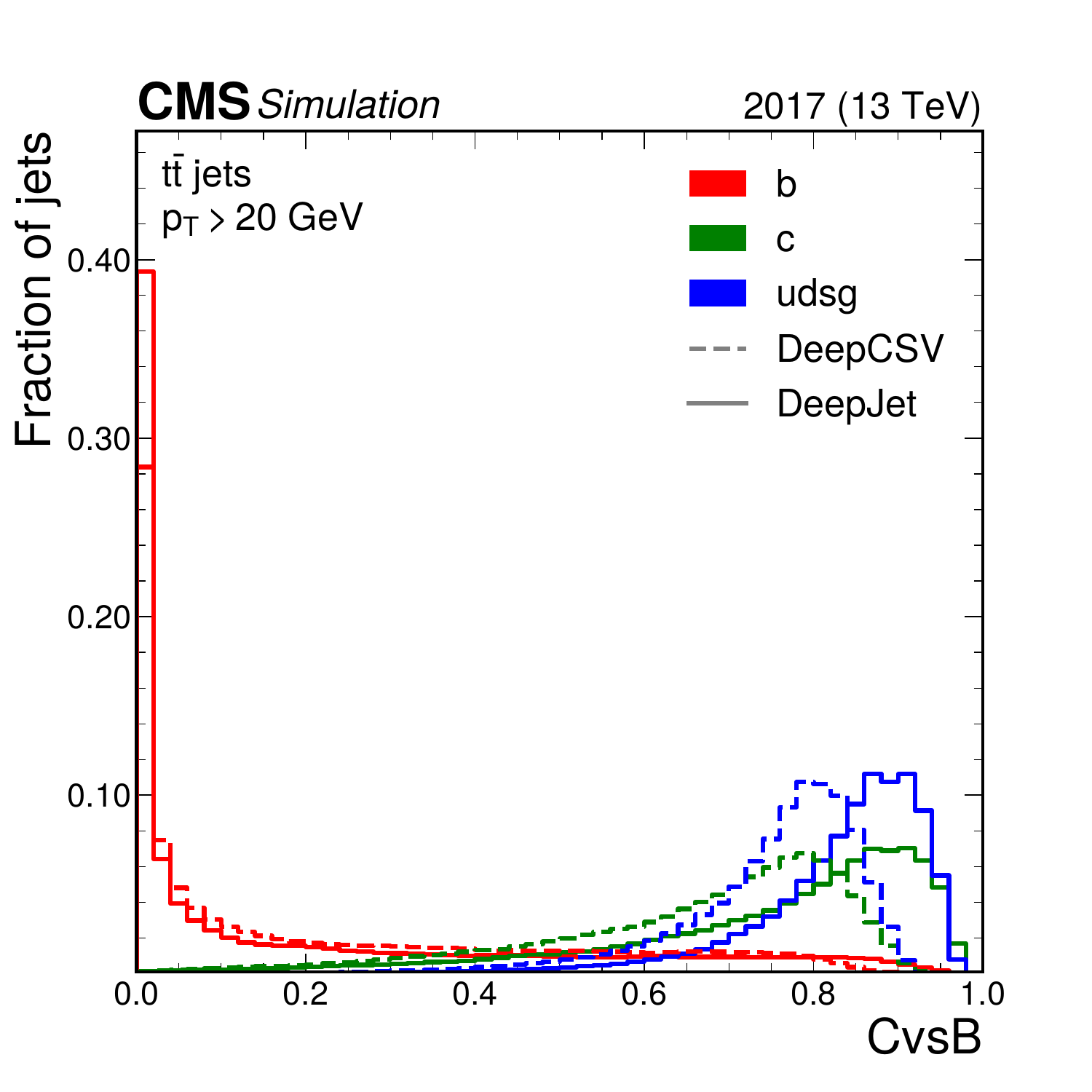}
			\caption{\label{fig:ctagdiscriminants} Unit-normalised distributions of the CvsL (left) and CvsB (right) discriminators for the DeepCSV (dashed) and DeepJet (solid) algorithms using jets from simulated hadronic \ttbar events with $\pt>20\GeV$ and $\abs{\eta} < 2.5$. The distributions are shown for {\PQb} (red), {\PQc} (green) and light-flavour jets (blue) separately.}
\end{figure}

\begin{figure}[htb!]
\centering
\includegraphics[width=0.475\textwidth]{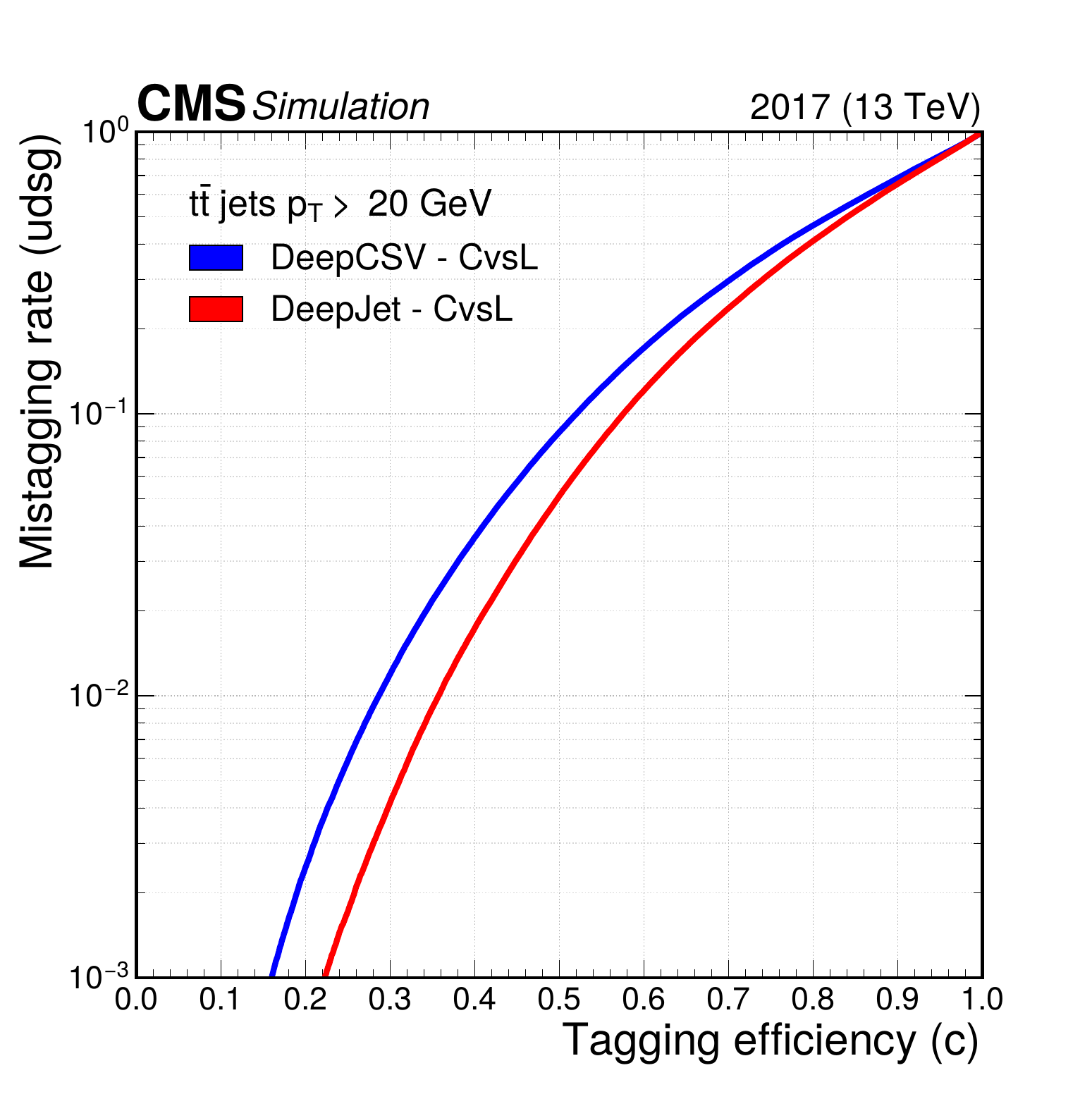}
\includegraphics[width=0.475\textwidth]{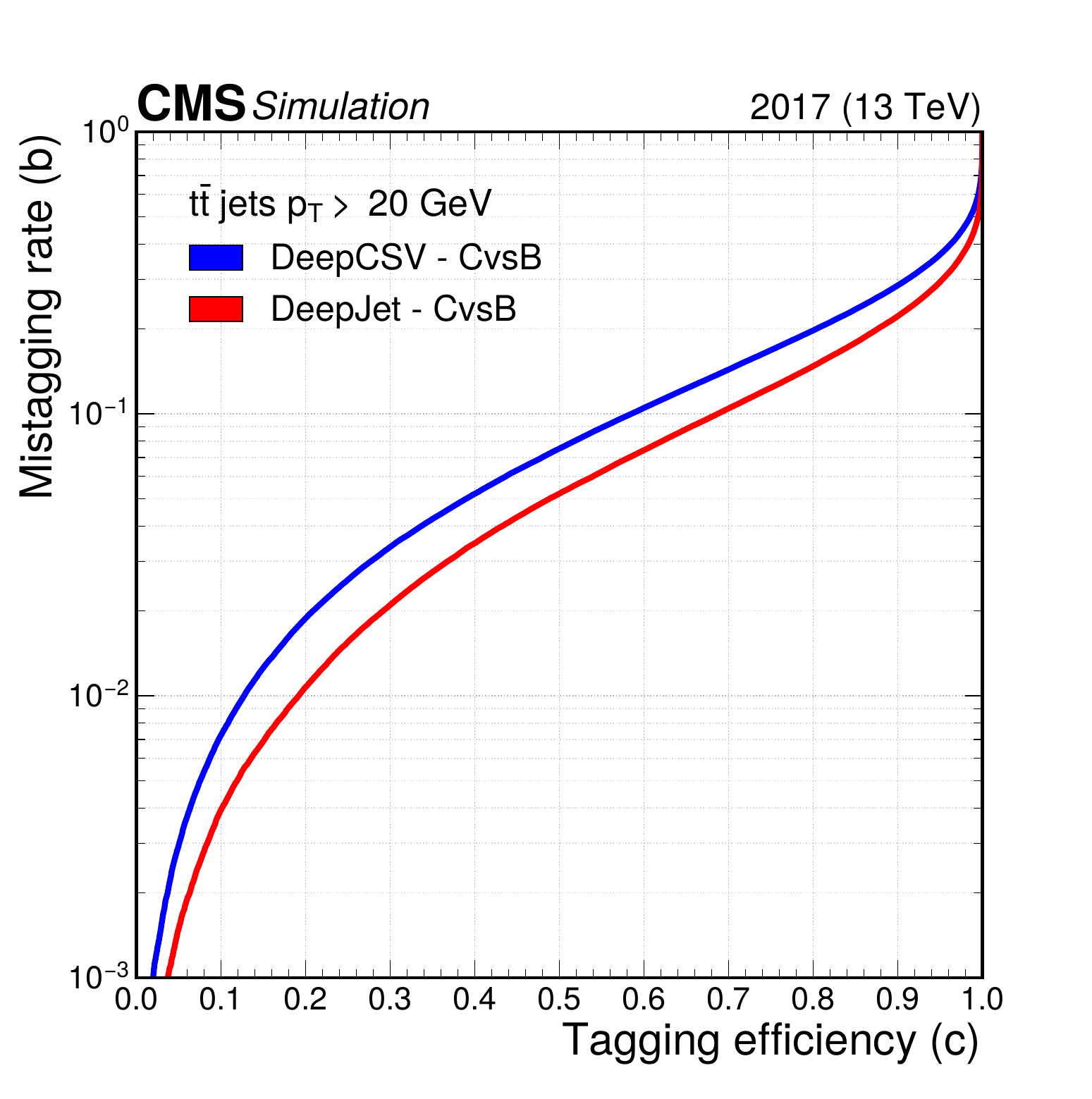}\\
\caption{\label{fig:ctagrocs}The ROC curves showing the individual performance of the CvsL (left) and CvsB (right) discriminators for the DeepCSV (blue) and DeepJet (red) algorithms using jets from simulated hadronic \ttbar events with $\pt>20\GeV$ and $\abs{\eta} < 2.5$.}
\end{figure}

\begin{figure}[htb!]
	\centering
		\includegraphics[width=0.6\textwidth]{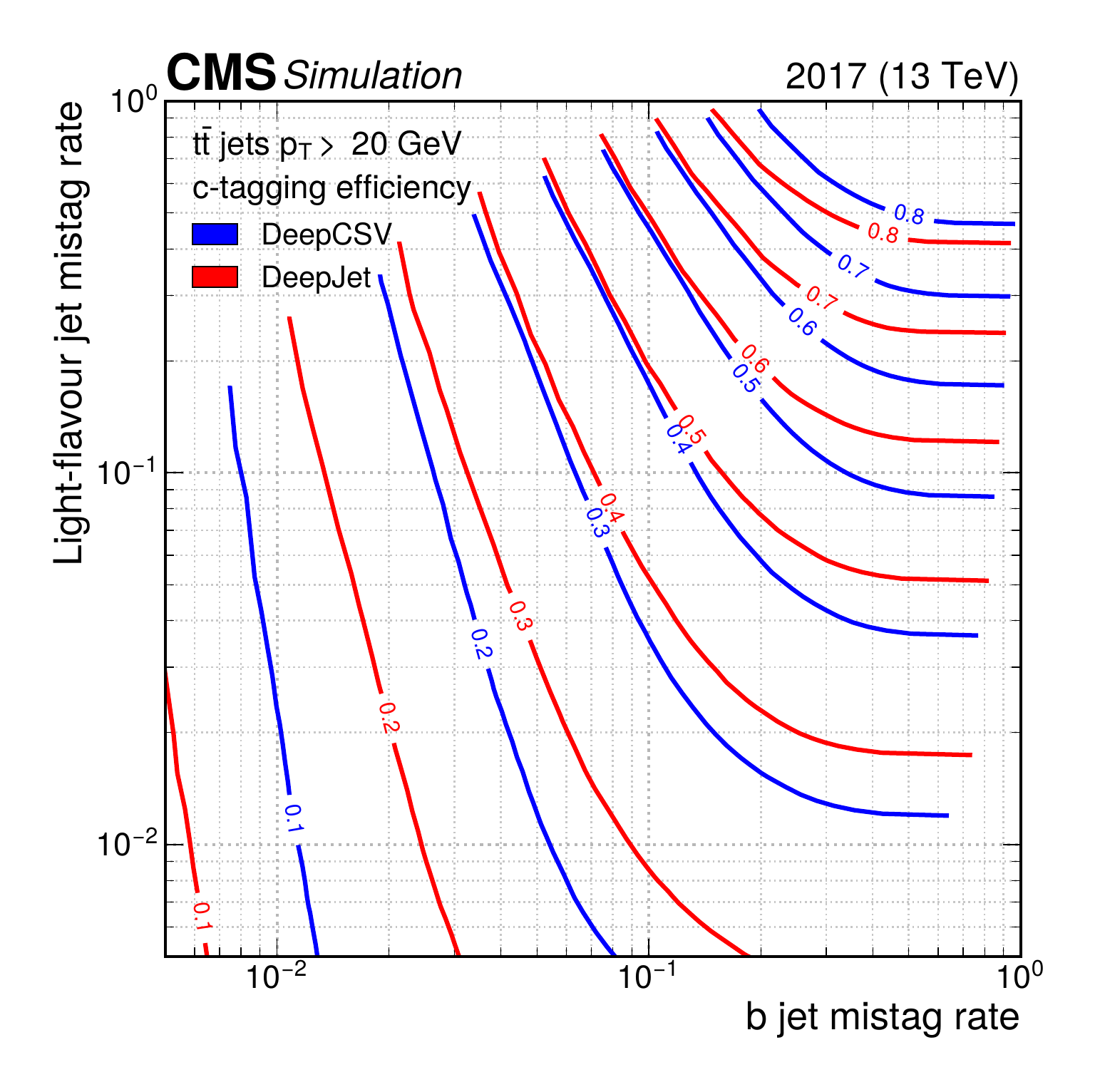}
	\caption{\label{fig:ctagroc2D} Two-dimensional ROC contours showing the {\PQc} tagging efficiency as simultaneous functions of {\PQb} jet and light-flavour jet mistagging rates for DeepCSV (blue lines) and DeepJet (red lines) algorithms using jets with $\pt>20\GeV$ and $\abs{\eta} < 2.5$,  from simulated hadronically decaying \ttbar events. Each line represents points in the plane that correspond to a fixed value of the {\PQc} tagging efficiency, which is shown as a number at the centre of each line.}
\end{figure}

These output probabilities can be appropriately combined to define a set of {\PQb} and {\PQc} tagging discriminators as summarised in Table~\ref{tab:taggerDefinitions}. For {\PQb} jet identification, a discriminant is defined to distinguish {\PQb} jets from either {\PQc} or light-flavour jets using one single discriminator (BvsC/L). For {\PQc} jet identification, two distinct discriminators are defined as the ratios in the second and third columns in Table~\ref{tab:taggerDefinitions}. The normalised distributions of these discriminators for both algorithms are shown in Fig.~\ref{fig:ctagdiscriminants} for jets with $\pt>20\GeV$ and $\abs{\eta} < 2.5$ from simulated hadronically decaying \ttbar events. The performance of these algorithms can be assessed by evaluating the selection efficiency for {\PQc} jets as a function of the misidentification rate for either {\PQb} or light-flavour jets for different selection thresholds on the discriminator values. These result in a so-called receiver operating characteristic (ROC) curve which is shown in Fig.~\ref{fig:ctagrocs}. It can be seen that the DeepJet algorithm has a lower mistagging efficiency than the DeepCSV algorithm in both CvsL and CvsB discrimination. Furthermore, the selection efficiency for {\PQc} jets can be evaluated as a simultaneous function of {\PQb} and light-flavour jet misidentification rates, which produces a two-dimensional (2D) ROC contour plot as shown in Fig.~\ref{fig:ctagroc2D}. This plot shows that the DeepJet algorithm outperforms the DeepCSV algorithm in simultaneous CvsL and CvsB discrimination over the entire 2D phase space, as well. The DeepCSV algorithm itself has already shown a significantly improved performance over the original {\PQc} tagging algorithm, which was based on a combination of two BDTs~\cite{Sirunyan:2017ezt}, demonstrating the significant advancements that have been made in heavy-flavour identification over the last five years.

\begin{figure}[p!]
\centering
\includegraphics[width=0.48\textwidth]{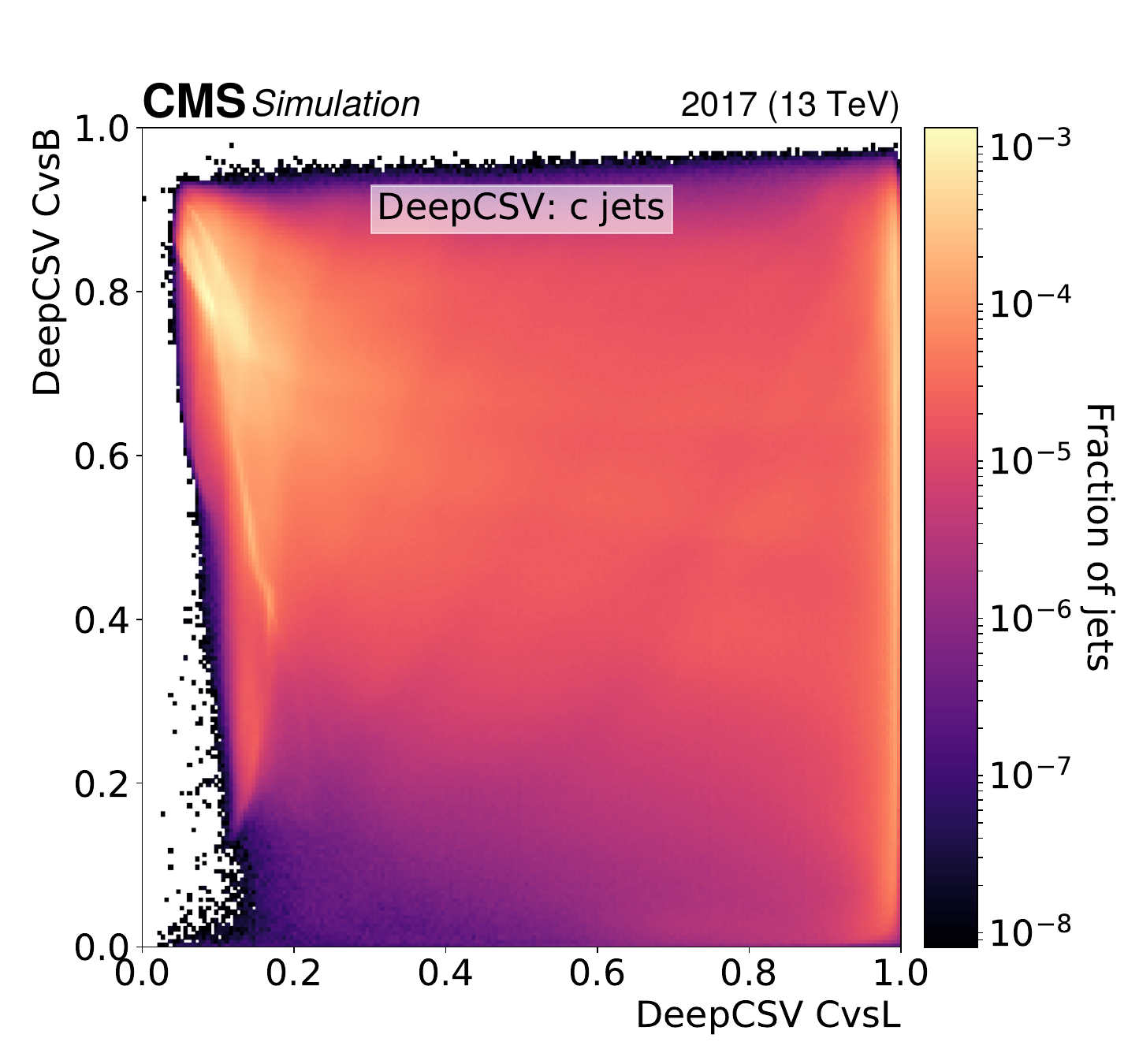}
\includegraphics[width=0.48\textwidth]{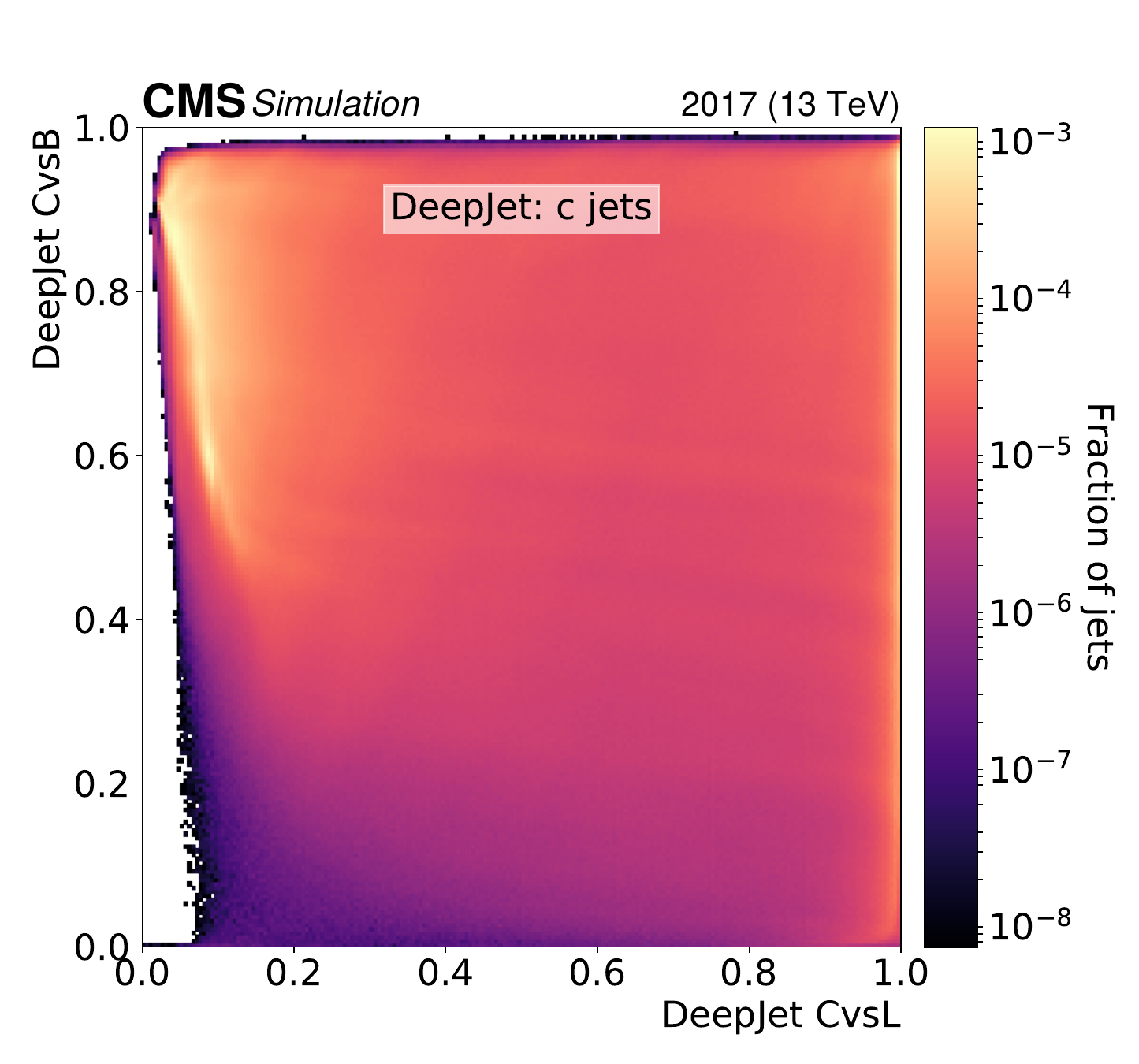}
\includegraphics[width=0.48\textwidth]{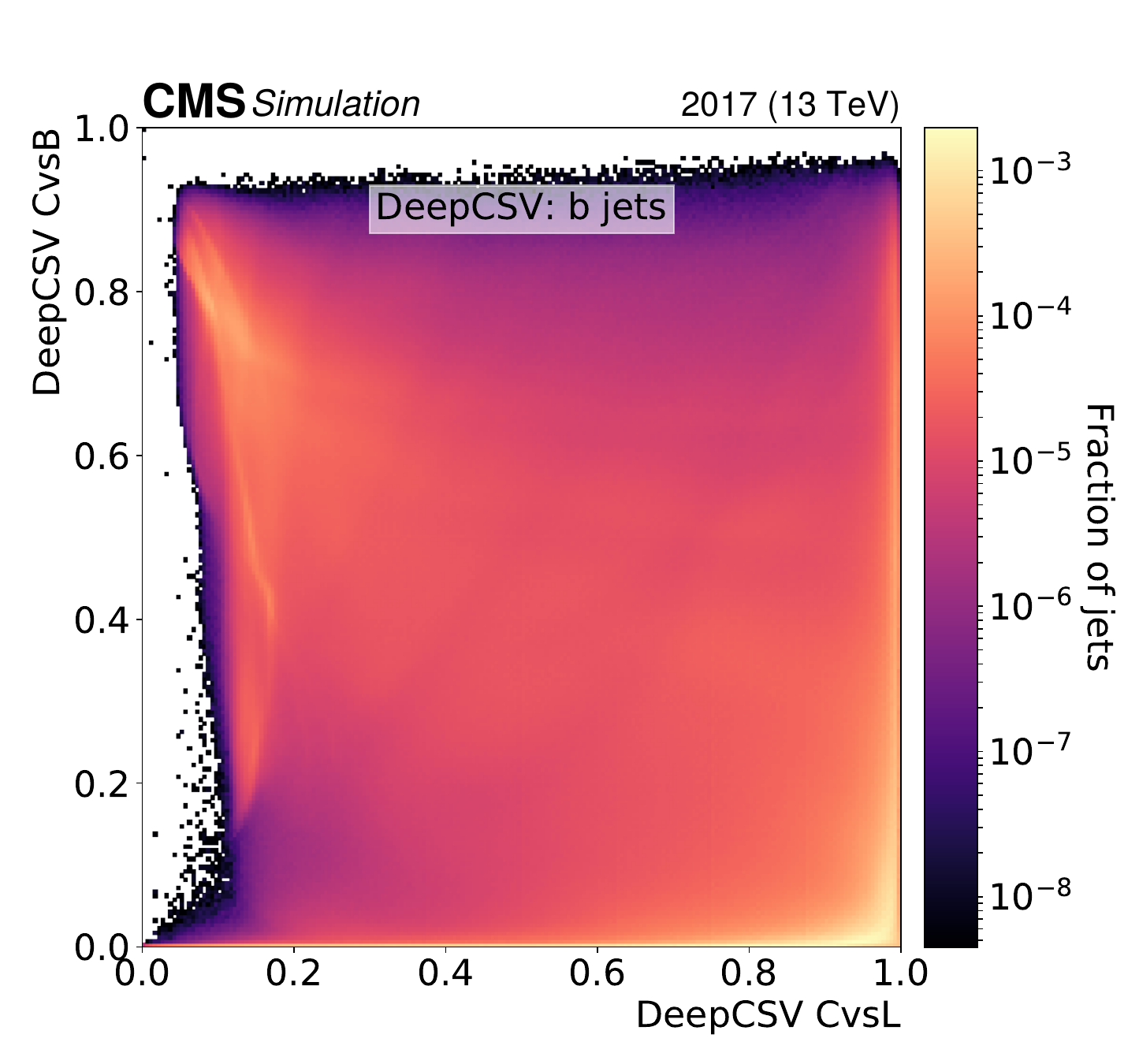}
\includegraphics[width=0.48\textwidth]{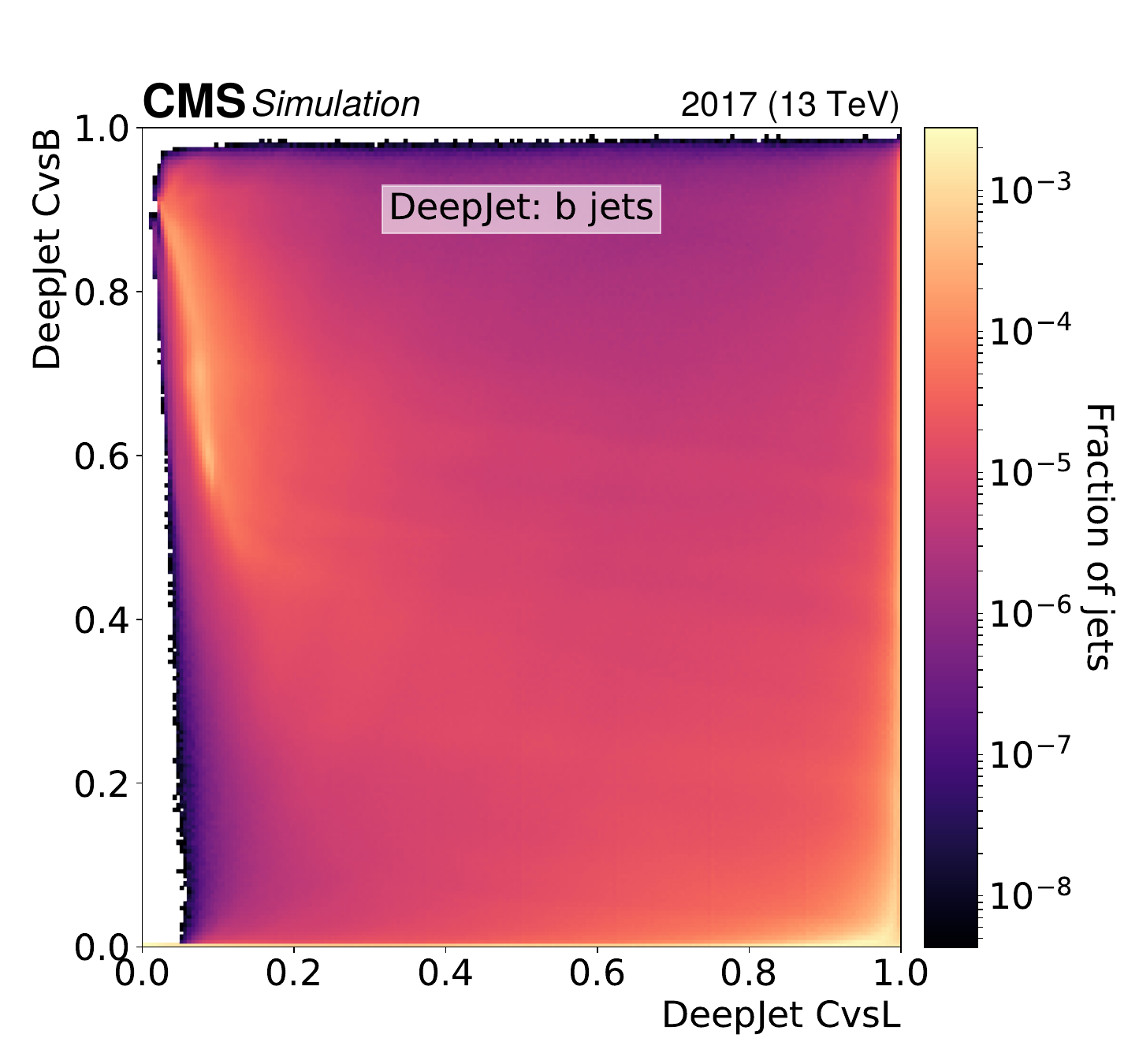}
\includegraphics[width=0.48\textwidth]{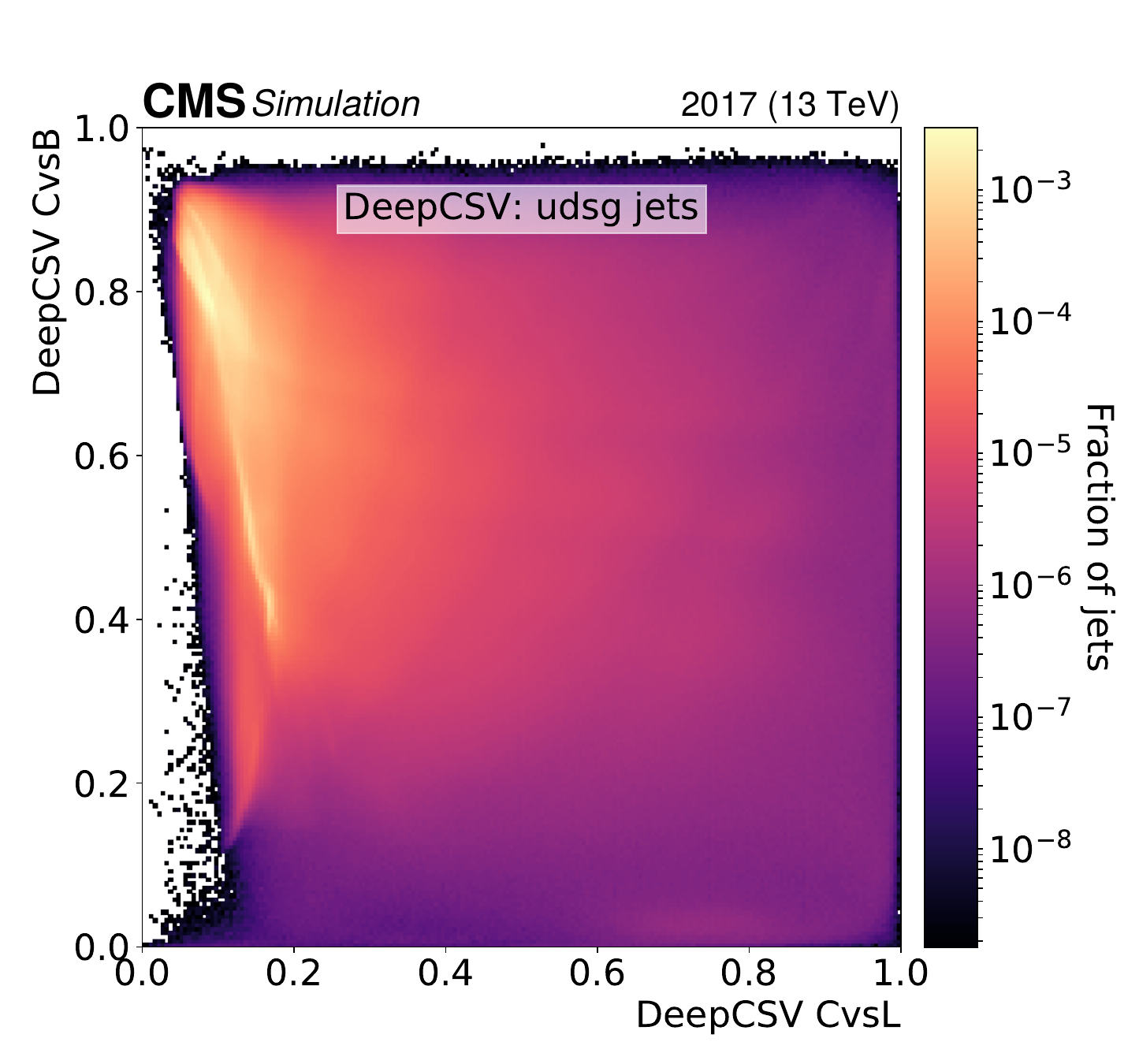}
\includegraphics[width=0.48\textwidth]{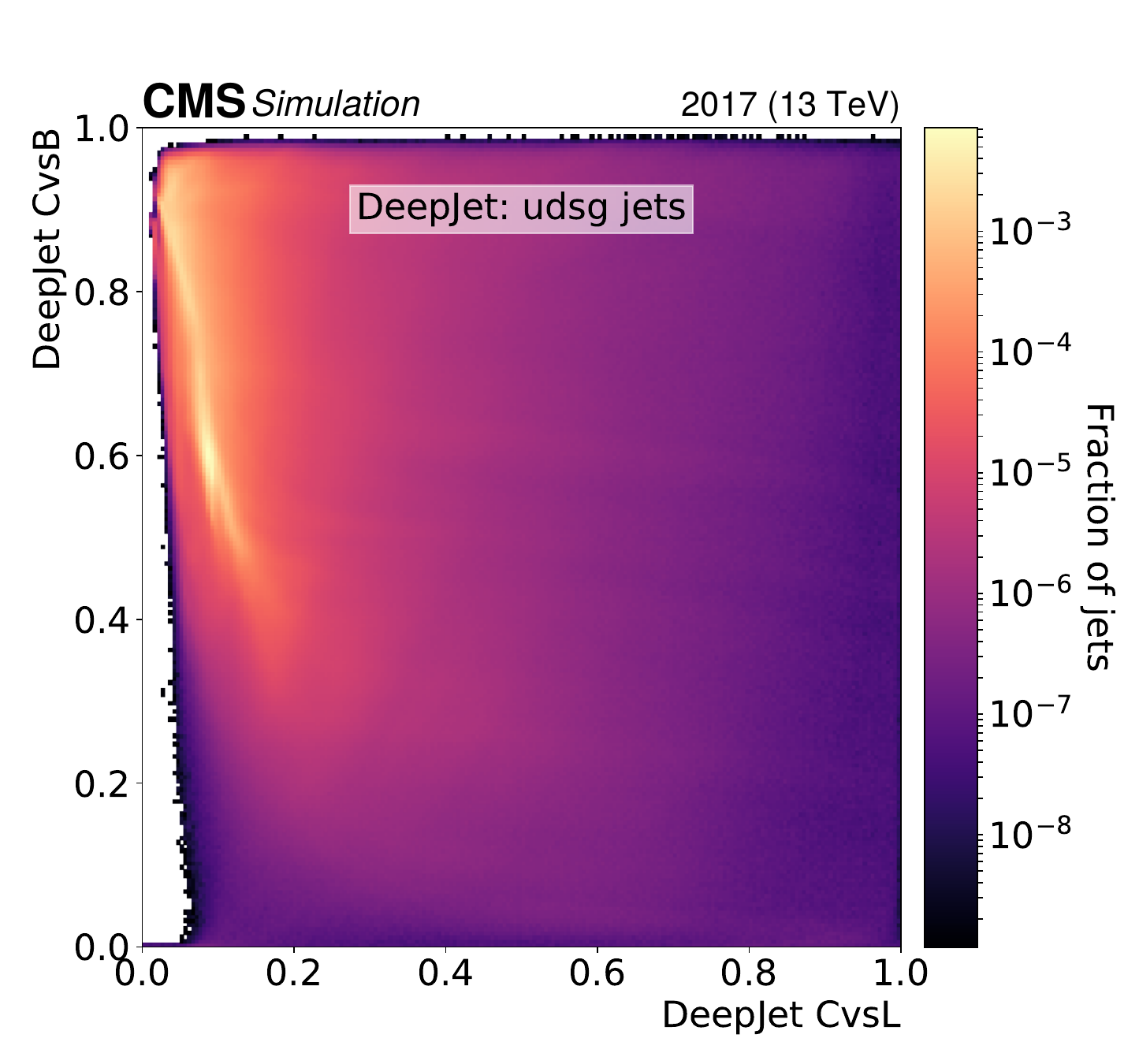}
\caption{\label{fig:2Dctaggingdistributions} Normalised 2D distributions showing the CvsL and CvsB discriminators on the $x$ and $y$ axes, respectively. Distributions are shown using {\PQc} (upper), {\PQb} (middle) and light-flavour (lower) jets with $\pt>20\GeV$ and $\abs{\eta} < 2.5$ from simulated hadronically decaying \ttbar events . The left-hand column shows the discriminators of the DeepCSV algorithm, whereas the right-hand column shows those of the DeepJet algorithm.}
\end{figure}

The 2D (normalised) distributions of the CvsL and CvsB discriminators of both algorithms are shown in 
Fig.~\ref{fig:2Dctaggingdistributions} for different jet flavours. In this 2D phase space spanned by the CvsL and CvsB discriminators, light-flavour jets are situated almost exclusively in the upper left corner, whereas {\PQc} jets have a significant fraction along the right edge, and {\PQb} jets are distributed largely towards the lower right corner, as expected.

An a priori track selection is applied to define the collection of tracks considered as input to the DeepCSV algorithm. This selection is optimised to identify tracks from {\PQb} and {\PQc} hadrons, while rejecting tracks from pileup interactions and poorly reconstructed tracks that do not match any genuine particle passing through the detector. It is therefore possible that a given jet has no tracks that pass this preselection and therefore no information is present to calculate a heavy-flavour tagging discriminant. These jets are assigned a default output value of $-1$. Given the known differences in track reconstruction efficiency and the poorly reconstructed track rate between the simulation and collision data, it is important to calibrate the rate of jets with such default discriminator values.

No such preselection criteria are used in the DeepJet algorithm, which instead takes as input the entire collection of PF candidates associated with a given jet. Therefore, no default values are expected to appear in the output probabilities of the DeepJet algorithm. Nevertheless, because of the definition of the CvsL and CvsB discriminators shown in Table~\ref{tab:taggerDefinitions}, an undefined discriminator value can appear if the denominator becomes zero. This situation is observed when the {\PQb} jet probability, $P$({\PQb})+$P$($\PQb\PQb$)+$P$(${\PQb}_{\text{lep}}$), is evaluated to be exactly 1 by the DeepJet algorithm, resulting by construction in an output value of 0 for all other output probabilities. Consequently, this appears almost exclusively for {\PQb} jets. Whenever such a situation appears for a given jet, its {\PQc} tagging discriminator values, CvsL and CvsB, are evaluated to be undefined and 0, respectively, and hence constitute a special category of jets separated from the continuous jet distribution in the 2D CvsL-CvsB plane. For simpler representation, both CvsL and CvsB values of these jets are defaulted to a value of $-1$ in this paper.

\subsection{Mismodelling of the simulated c tagging discriminators   \label{ctaggingCalibration}}

The need for calibrating the heavy-flavour tagging discriminants arises from possible mismodelling of the simulated inputs. Important discriminating properties such as the track displacement or the SV displacement are not very well modelled in simulation and are subject to changes in the detector alignment. As already evident from other heavy-flavour calibration methods~\cite{Sirunyan:2017ezt}, the performance of the heavy-flavour tagging algorithms is overestimated in the simulation, resulting in higher simulated {\PQb} and {\PQc} tagging efficiencies and lower misidentification probabilities compared with those observed in the data. Rather than correcting the efficiency of a selection in the discriminator, the full simulated differential shape of that discriminator can be calibrated to match the shape observed in data. Such a strategy has been developed in the past for the {\PQb} tagging discriminator distribution~\cite{Sirunyan:2017ezt}. For the first time this paper presents a method of calibrating the differential {\PQc} tagging discriminator shapes. Since {\PQc} jet identification relies on both the CvsL and CvsB discriminators simultaneously, the calibration of the {\PQc} tagger is performed as functions of both the CvsL and CvsB distributions. Jets with defaulted values of CvsL and CvsB are calibrated as a separate category. The rest of this paper is devoted to the discussion of this novel calibration method.

\section{Event selections for calibration\label{sec:Event-Selection}}

The strategy for calibrating the DeepCSV- and DeepJet-based {\PQc} taggers
involves identifying three samples of jets in the data that are enriched respectively
in {\PQc}, {\PQb}, and light-flavour jets. This has resulted in the
definition of three jet samples enriched either in {\PW}+{\PQc} events
({\PQc} enriched), \ttbar events ({\PQb} enriched), and $\text{DY}+\text{jet}$ events (udsg enriched).
The flavour tagging algorithms being calibrated are not used in the construction of these
selections to keep the resulting calibration free from potential biases.
However, the sampled {\PQc} and {\PQb} jets are required to contain a soft muon
inside the jet cone, to enrich the samples with jets containing semileptonically decaying
{\PQc} and {\PQb} hadrons, which increases signal purity. Since soft muons within jets are treated as any other charged
PF candidate in both DeepCSV and DeepJet trainings, as opposed to
training the network with explicit soft muon information, the bias in the discriminator responses arising from
using muon-containing {\PQc} and {\PQb} jets as a proxy for all {\PQc} and {\PQb} jets,
is expected to be minimal. In Section~\ref{subsec:MuBias} we further demonstrate the applicability
of the correction factors to inclusive samples despite the nonuniversality of the
{\PQc} and {\PQb} jets used for calibration.

The SM physics processes, namely, {\PW}+{\PQc}, \ttbar and $\text{DY}+\text{jet}$, chosen to yield the three jet samples,
were experimentally studied with great precision in previous publications \cite{CERN-EP-2018-282,CMS:2021vhb,CMS:2019zct,Sirunyan_2020}.
These papers did not observe any
significant deviation from the SM expectations. Therefore, effects of BSM physics, if any, on these samples are expected to be
insignificant when compared with the statistical and systematic uncertainties associated with the data and simulation.
Hence, the distributions of the {\PQc} tagger outputs for these jet samples in data are expected to agree
with the corresponding calibrated distributions in simulation within uncertainties.

In this section, the event selection criteria to obtain each of these three topologies
are outlined, followed by a discussion on the purity of the different jet flavours after each selection.

\subsection{The c jet enriched selection}

Charm jets in data are selected from events with a jet produced in association
with a {\PW} boson, following the strategy discussed in the CMS {\PW}+{\PQc} cross section analysis 
\cite{CERN-EP-2018-282} and the {\PQc} jet identification SF
derivation \cite{CMS-PAS-BTV-16-001}. The relevant events involve a
leptonically decaying {\PW} boson ($\PW \to l\nu$, $l=\mathrm{e},\mu$) and
a jet originating from the hadronisation of a charm quark. These charm
jets are identified using the semileptonic decay of the {\PQc} hadron, which
produces a soft muon within the jet in the final state.
Charm hadron decays into electrons are not considered, because of the lower
efficiency of electron reconstruction at low energies \cite{CMS-DP-2018-017}.
Thus, the targeted final state consists of an isolated charged lepton
(electron or muon) from the {\PW} boson decay, \ptmiss due to the
presence of a neutrino from the {\PW} boson decay, and at least one jet with
a soft, nonisolated muon inside it.

The selected events are divided into two categories:
\begin{itemize}
\item \textit{Opposite-sign (OS)
events}, where the soft muon inside the jet and the isolated lepton
from the {\PW} boson have opposite charges;
\item \textit{Same-sign (SS) events},
where the soft muon inside the jet and the isolated lepton from the
{\PW} boson have similar charges.
\end{itemize}
If there are multiple soft muons inside the jet, the muon with the highest
\pt determines the sign of the charge.

\begin{figure}
\centering
\includegraphics[width=0.95\textwidth]{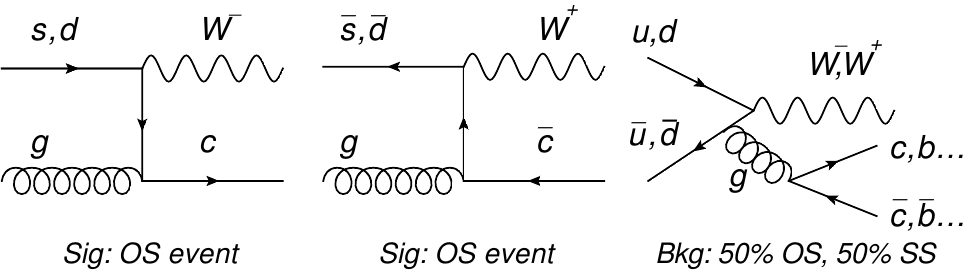}
\caption{\label{fig:Feynman-diagrams}Feynman diagrams showing production of
charm quarks in association with a {\PW} boson (left and middle) and the major background (right).}
\end{figure}

The major background for the {\PW}+{\PQc} selection, as described above,
is the production of a {\PW} boson associated with the radiation of a
virtual gluon that decays into a pair of {\PQb} or {\PQc} quarks, with one or
more of the resulting jets in the final state having a soft muon inside it.
These backgrounds have 50\%-50\% probabilities of being tagged as OS or SS. In contrast, the leading-order contributions to the
production of a {\PW} boson with one {\PQc} jet always result in an OS signature,
as illustrated in Fig.~\ref{fig:Feynman-diagrams}. Hence, an OS-SS
subtraction largely reduces the major background and yields a
distribution enriched in {\PW}+{\PQc} events. All distributions related to {\PW}+{\PQc} selection
in this paper are OS-SS subtracted. 

A few details of the selection are described as follows:
\begin{description}
\item [{Isolated~lepton:}] A charged lepton (electron or muon)
from the {\PW} boson decay that originates close to the PV and has a relative isolation
smaller than 0.05 is required. The events are
categorised into electron and muon channels, depending on the flavour
of each charged lepton. The electron channel requires exactly one isolated electron
with a transverse momentum of at least 34\GeV
and no isolated muon, and the muon channel requires exactly one isolated muon with
a \pt of at least 30\GeV and no
isolated electron. The events are triggered by a single-electron and/or single-muon HLT. The {\PW} bosons
decaying into $\tau$ leptons are not explicitly excluded, and may enter
the selection through their leptonic decay products.
\item [{{\PW} boson candidate:}] The {\PW} boson transverse mass is defined as
\begin{equation*}
	\mT \equiv \sqrt{{2 \pt^{\text{lep}} \ptvecmiss \bigl[1 - \cos(\phi_{\text{lep}} - \phi_{\ptvecmiss})\bigr]}},
\end{equation*}
where $\pt^{\text{lep}}$ is the \pt of the isolated lepton and $\phi_{\text{lep}}$ ($\phi_{\ptvecmiss}$) is the $\phi$ of the isolated lepton (missing momentum). An \mT of at least 50\GeV is required to ensure the presence of a neutrino produced from a {\PW} boson decay.
\item [{Jets:}] Events are required to have at least one and at most three
jets lying inside the tracker coverage ($\abs{\eta}<2.5$) and with $\pt>20\GeV$. At least one of these jets
is required to contain a low-energy muon with $\pt<25\GeV$ and a relative isolation of at least 0.2 (0.5) in the electron (muon) channel. If there are multiple such jets in an event, the jet with the highest \pt (referred to as the muon jet) is selected. Only jets separated from the isolated lepton (from {\PW} boson decay) by $\Delta R > 0.5$ are considered.
\item [{DY~suppression:}] To discard events where a prompt muon is misidentified
as the muon jet, an upper threshold of 0.4 (0.6) is applied to the muon energy
fraction of the muon jet in the muon (electron) channel. For only the muon channel, the sum of the muon and the neutral electromagnetic energy fractions of the muon jet is required to be smaller than 0.7. These two requirements heavily suppress
the DY background in the muon channel by rejecting events where
one of the prompt muons is misidentified as the muon jet, \eg when the prompt muon
undergoes final-state photon radiation. Furthermore, the invariant mass of the muon inside the jet
and the isolated muon in the muon channel cannot match
that of the {\PZ} boson (in between 80 and 100\GeV) or other dimuon resonances
(below 12\GeV).
\end{description}
\begin{figure}[htb!]
\centering
\includegraphics[width=0.48\textwidth]{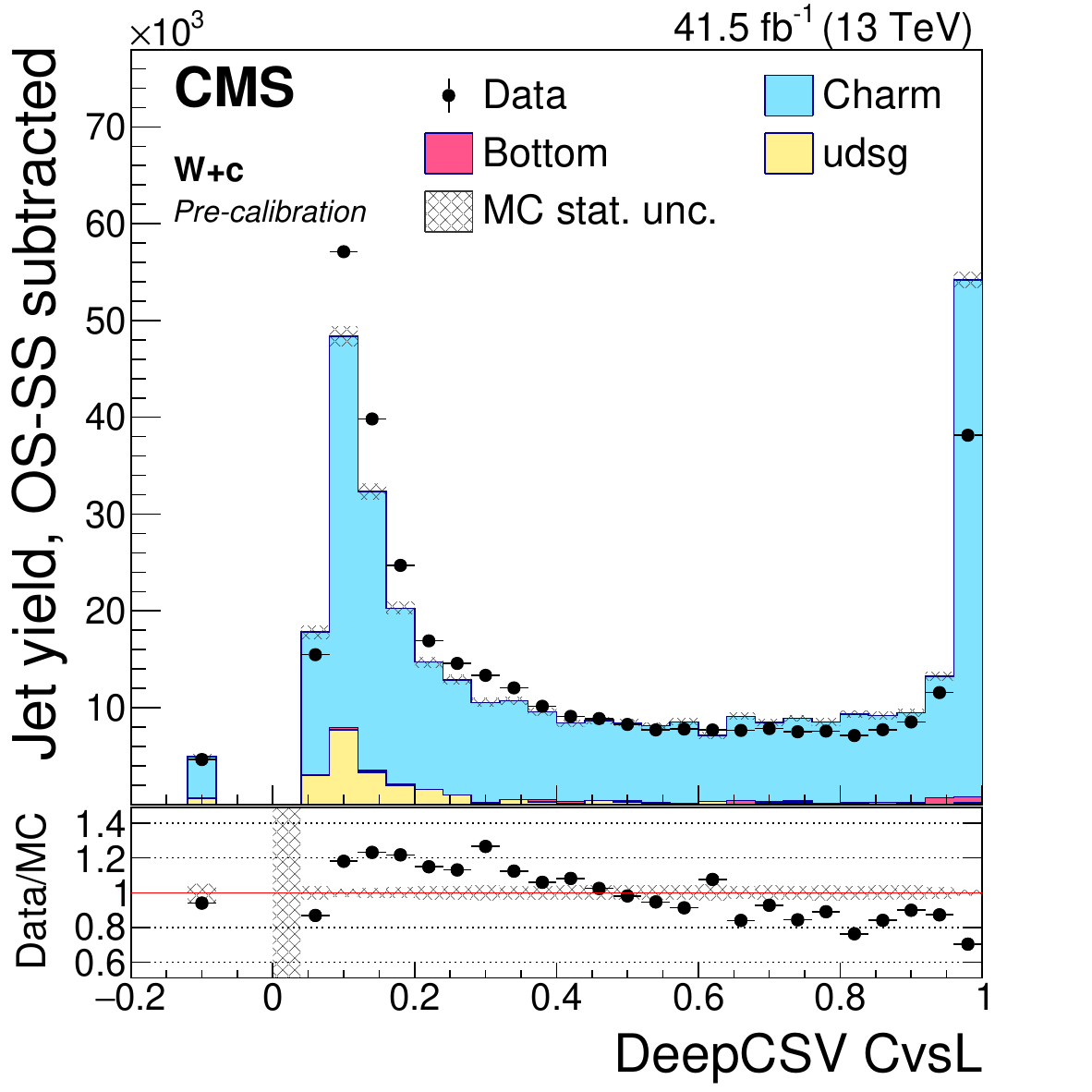}
\includegraphics[width=0.48\textwidth]{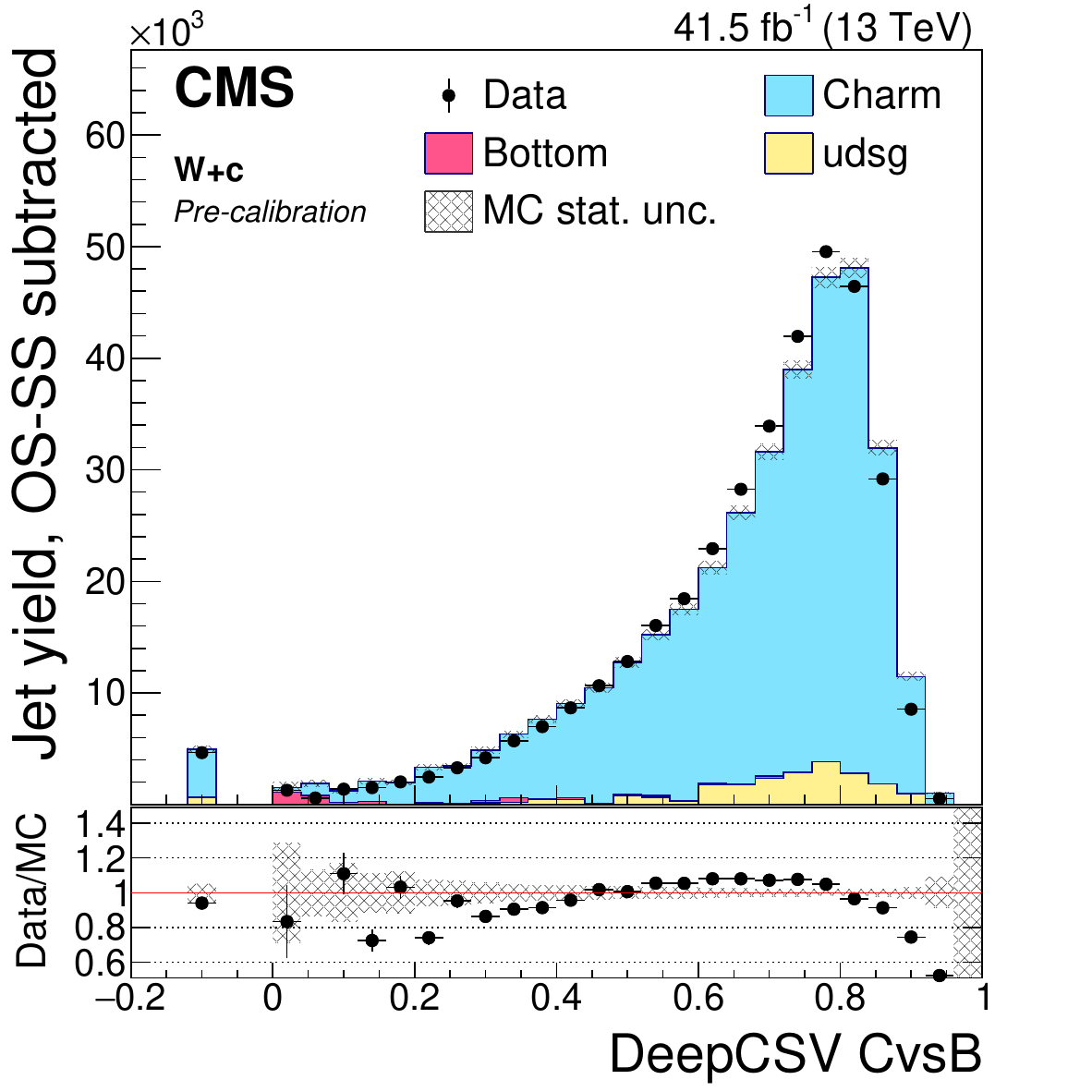}
\includegraphics[width=0.48\textwidth]{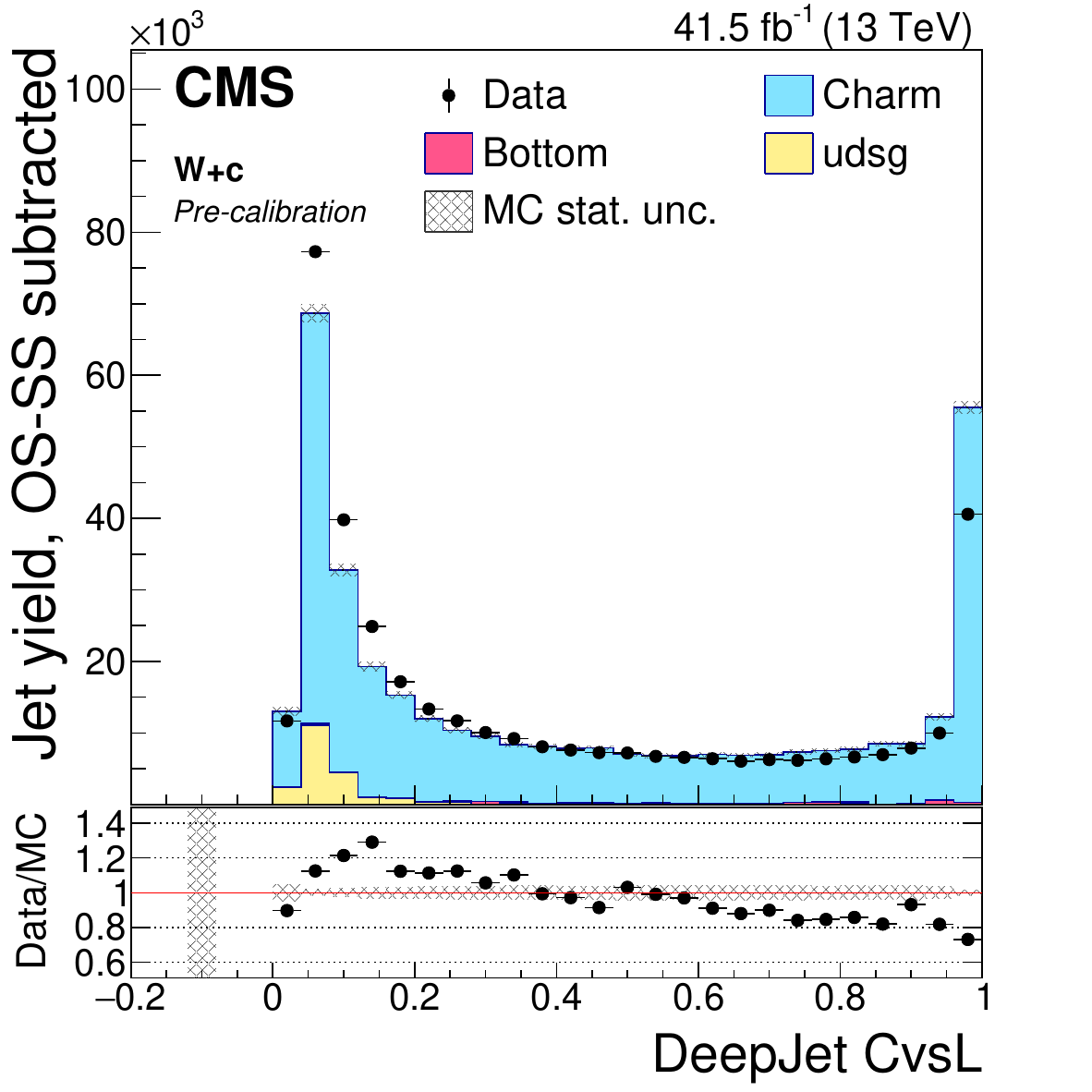}
\includegraphics[width=0.48\textwidth]{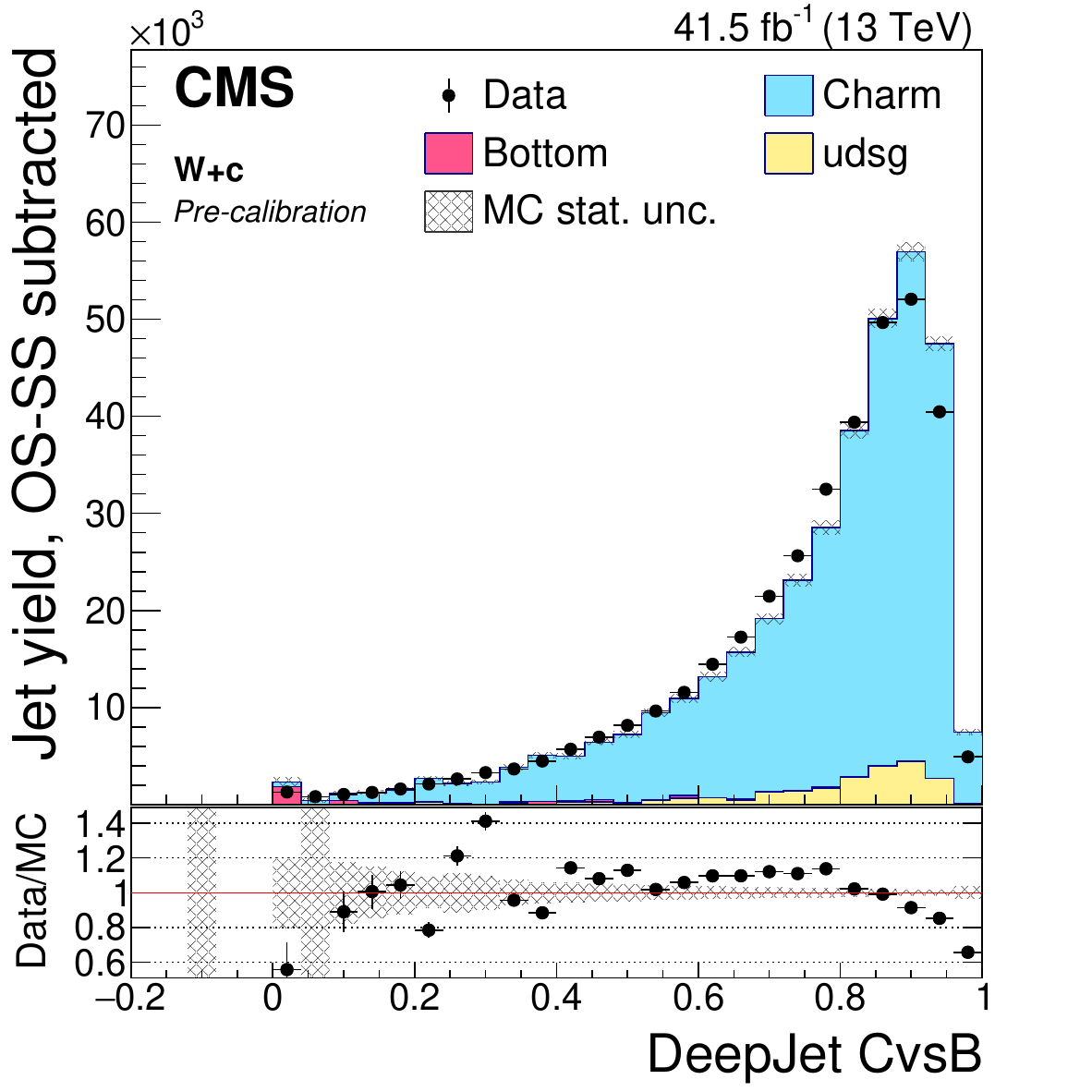}
\caption{\label{fig:WcDeepJet}Precalibration distributions of CvsL (left) and CvsB (right) obtained
from the DeepCSV (upper) and DeepJet (lower) taggers for jets selected in the {\PW}+{\PQc} (OS-SS) selection.
The bin corresponding to a tagger value of $-1$ is plotted at $-0.1$. Vertical error bars in data
represent statistical uncertainties in data. The simulations are shown as stacked histograms.}
\end{figure}

The muon jets selected in these events are the {\PQc} jet
candidates. The OS-SS distributions of the CvsL and CvsB variables of these candidates are
presented in Fig.~\ref{fig:WcDeepJet} for both DeepCSV and DeepJet taggers.
These plots, as well as all other jet distribution plots in this paper,
are plotted after scaling the simulated jets such that the total number of 
simulated jets matches the number of jets in data in every channel of every
selection. This ensures that any remaining discrepancies between data and simulation, in terms of
cross sections of the contributing processes, are largely reduced.

\subsection{The b jet enriched selection}

The {\PQb} jet enriched sample is obtained from \ttbar  events by
selecting one of the {\PQb} jets originating from the decay of the two
top quarks. The selection includes both the semileptonic and
dileptonic decay channels of the \ttbar process, each of which
is discussed separately below.

\subsubsection{Semileptonic \ttbar selection\label{SemittSel}}

The strategy for the semileptonic \ttbar  selection is to select
one of the {\PQb} jets produced in a semileptonic \ttbar  decay
by identifying a leptonically decaying {\PW} boson, a jet with a soft muon
from a bottom hadron decay chain inside it, and several additional jets
from the other {\PQb} quark and the hadronically decaying {\PW} boson.
The selection is similar to that of the {\PW}+{\PQc} discussed
in the previous subsection and is kept orthogonal to the latter by
requiring at least four jets in the event.

\subsubsection{Dileptonic \ttbar selection}

The dileptonic \ttbar selection chooses one of the two
{\PQb} jets in dileptonic \ttbar events. The target final state
consists of two isolated, oppositely charged leptons from
the two {\PW} boson decays, and a minimum of two jets, at least one of which
has a soft muon inside it. The selection can be subdivided into three
channels depending on the flavour of the isolated leptons, namely dielectron,
dimuon and electron-muon channels. The events in those channels
are triggered using the dielectron, dimuon and electron-muon HLTs, respectively.

The dielectron (dimuon) channel requires exactly two electrons (muons) having \pt
of at least 27 and 15 (20 and 12)\GeV,
and no muons (electrons). The electron-muon channel requires either one muon with
$\pt^{\mu} > 25\GeV$ and one electron with $\pt^{\mathrm{e}} > 15\GeV$, or
one electron with $\pt^{\mathrm{e}} > 27\GeV$ and one muon with $\pt^{\mu} > 14\GeV$.
In all cases, the charged leptons are required to have a relative isolation less than 0.15.

Although the muon-electron channel thus selected is quite pure in {\PQb} jets,
the two other channels are contaminated by the DY background. Therefore,
in the dielectron and dimuon channels, the event is required to have
a \ptmiss of at least 40\GeV, and the invariant mass of the dilepton
candidate is required to be incompatible with that of the {\PZ} boson (i.e. outside the range 75--105\GeV)
or other dimuon resonances (below 12\GeV), to suppress the DY contamination.

\begin{figure}[htb!]
\centering
\includegraphics[width=0.48\textwidth]{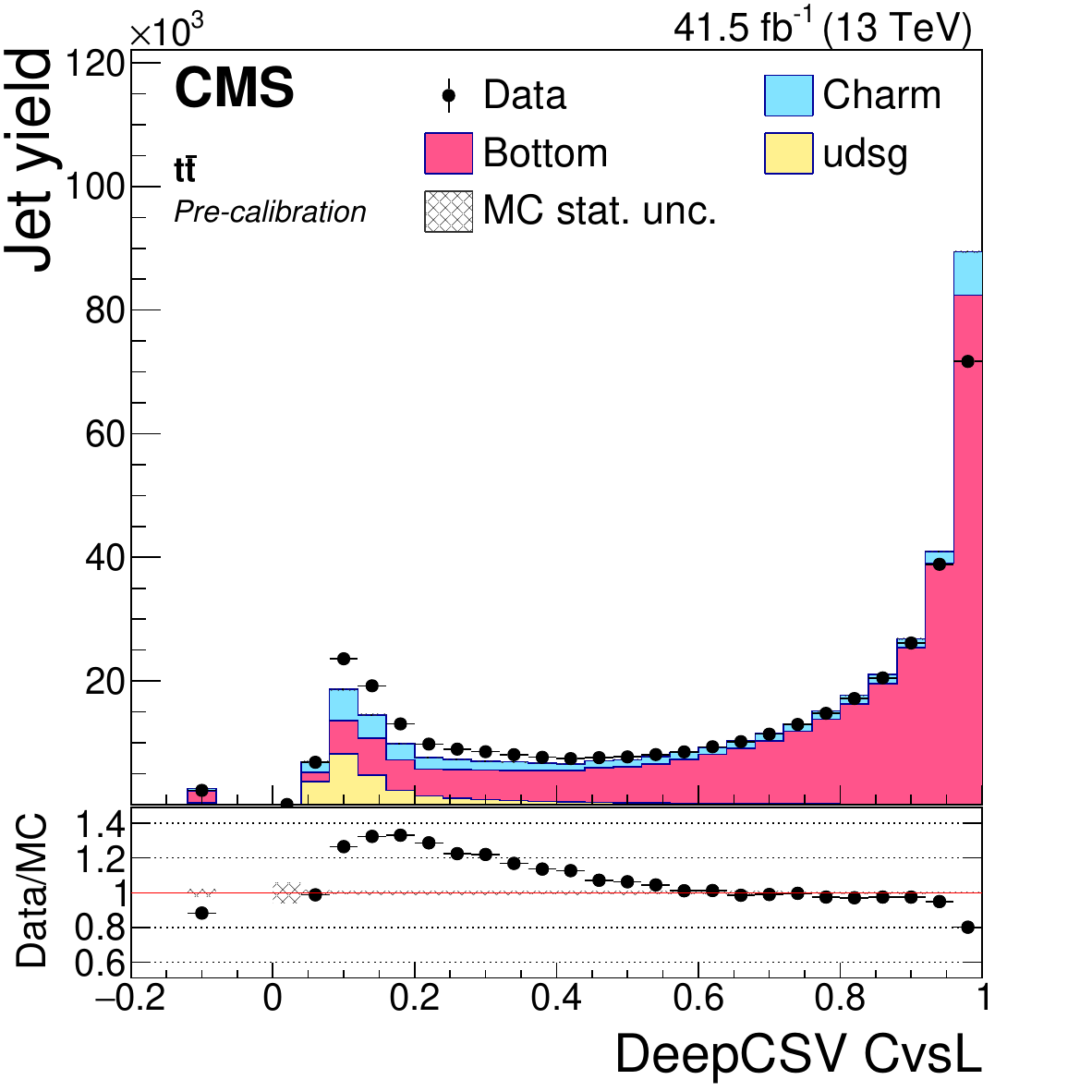}
\includegraphics[width=0.48\textwidth]{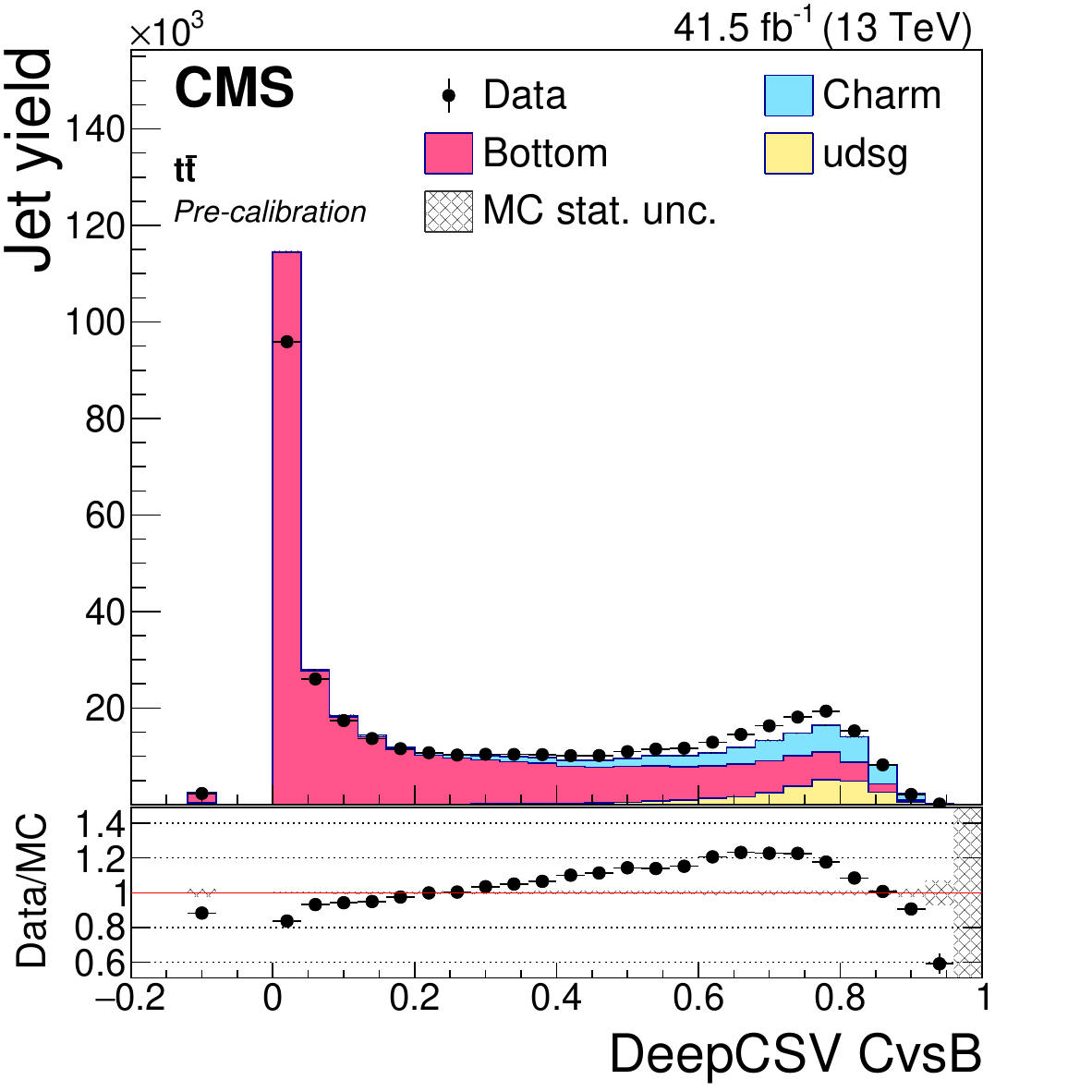}
\includegraphics[width=0.48\textwidth]{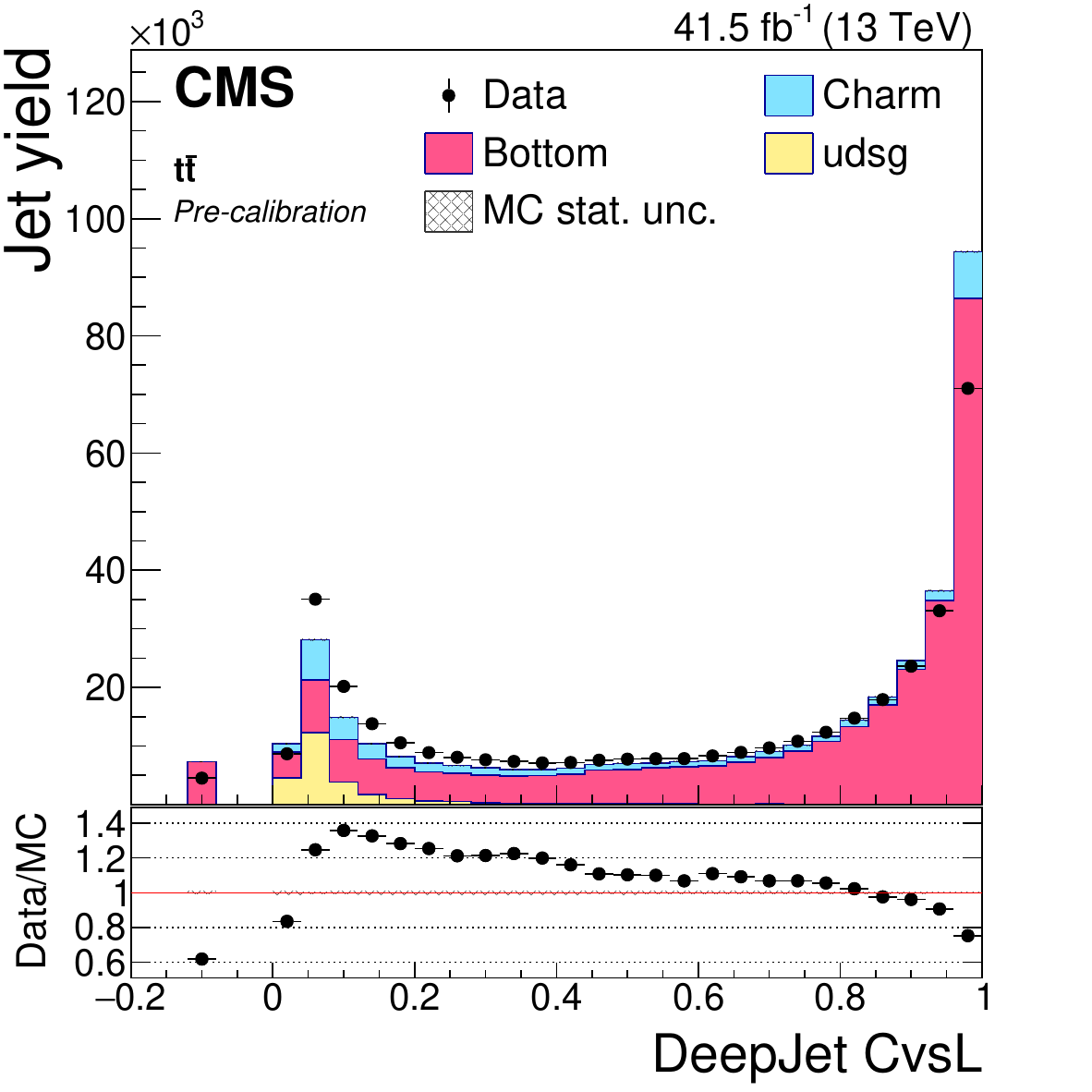}
\includegraphics[width=0.48\textwidth]{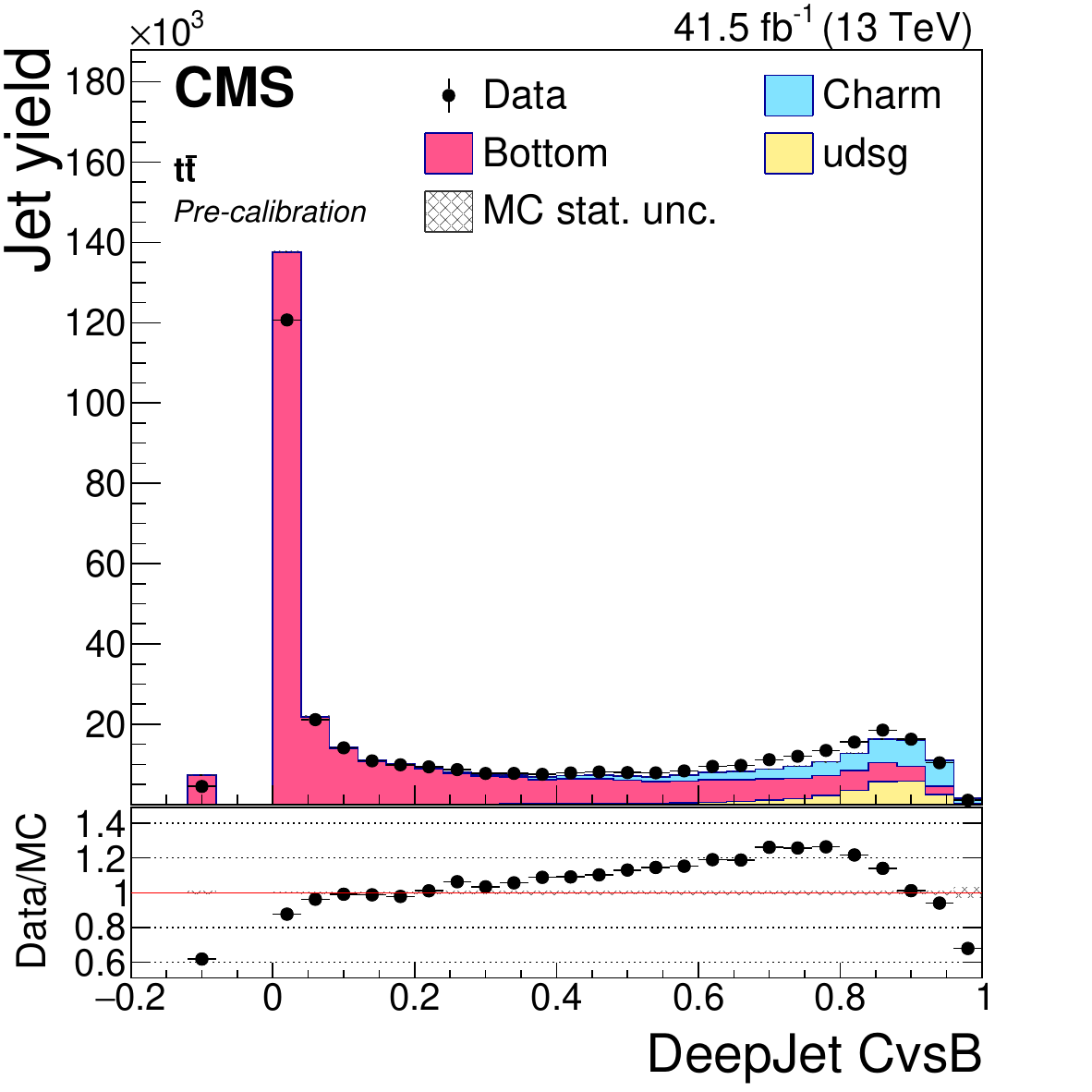}
\caption{\label{fig:TTDeepJet}Precalibration distributions of CvsL (left) and CvsB (right) obtained
from the DeepCSV (upper) and DeepJet (lower) taggers for jets in the \ttbar
 selection. The bin corresponding to a tagger value of $-1$ is plotted at $-0.1$. Vertical error bars in data
 represent statistical uncertainties in data. The simulations are shown as stacked histograms.} 
\end{figure}

The muon jet selected in each semileptonic or dileptonic \ttbar event
is the {\PQb} jet candidate. In case an event has multiple muon jets,
the jet with the highest \pt is selected. The CvsL and CvsB distributions
of these candidates combined from both semileptonic and dileptonic decays are presented in
Fig.~\ref{fig:TTDeepJet} for both DeepCSV and DeepJet taggers.
We have verified that the {\PQc} tagger discriminator shapes of individual flavours agree
quite well between semileptonic and dileptonic decay channels for both of the taggers.

\subsection{Light-flavour jet enriched selection}

Jets arising from light quarks ($\PQu\PQd\PQs$) and gluons ($\Pg$) are selected in
events containing one or more jets produced in association
with a leptonically decaying {\PZ} boson ($\text{DY}+\text{jet}$ events). The selection is split
into two channels (electron and muon), depending on the flavour of the charged leptons
from the {\PZ} boson. The {\PZ} bosons decaying into $\tau$ leptons are not
explicitly considered, but may enter the selection through the leptonic
$\tau$ decays.

\begin{figure}[htb!]
\centering
\includegraphics[width=0.48\textwidth]{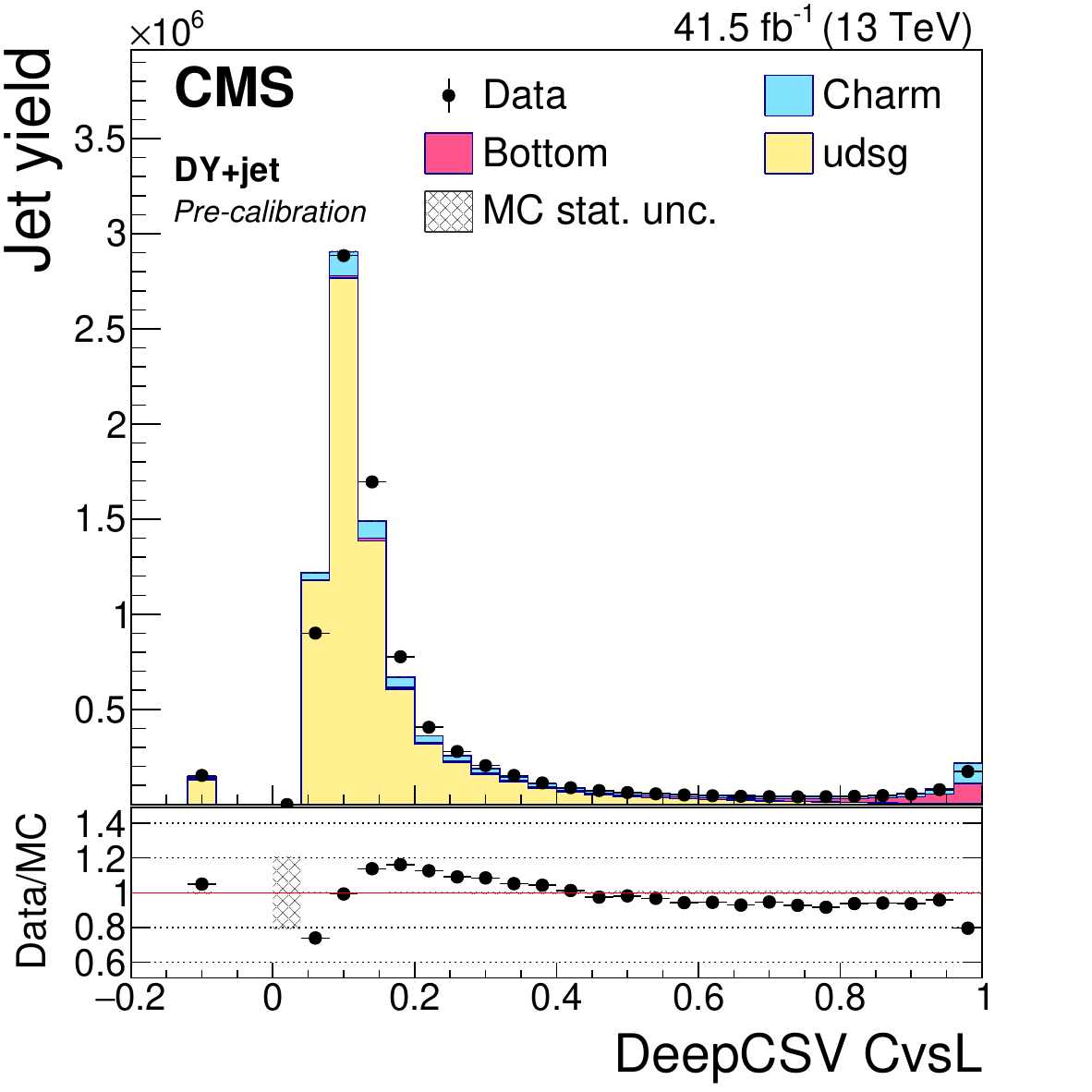}
\includegraphics[width=0.48\textwidth]{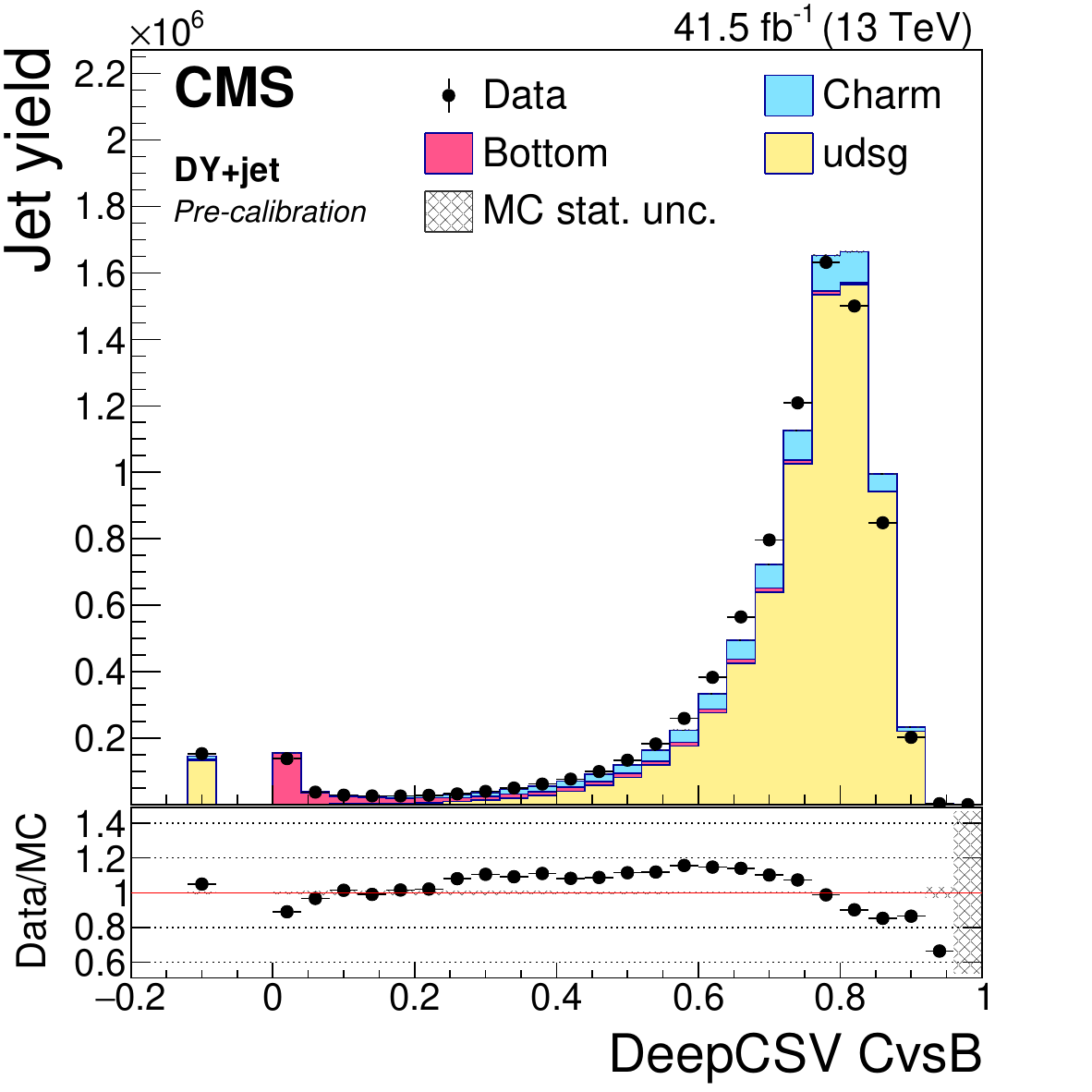}
\includegraphics[width=0.48\textwidth]{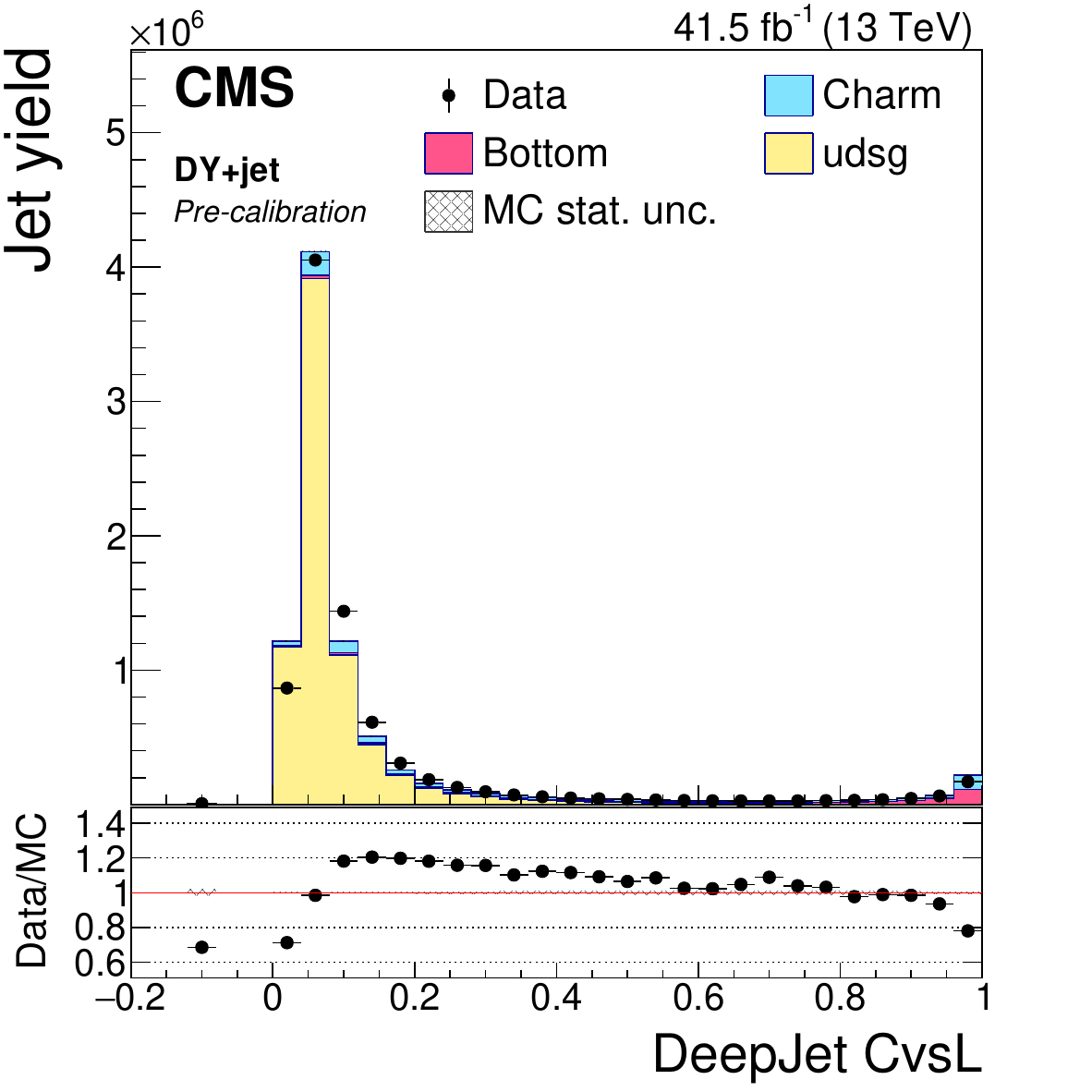}
\includegraphics[width=0.48\textwidth]{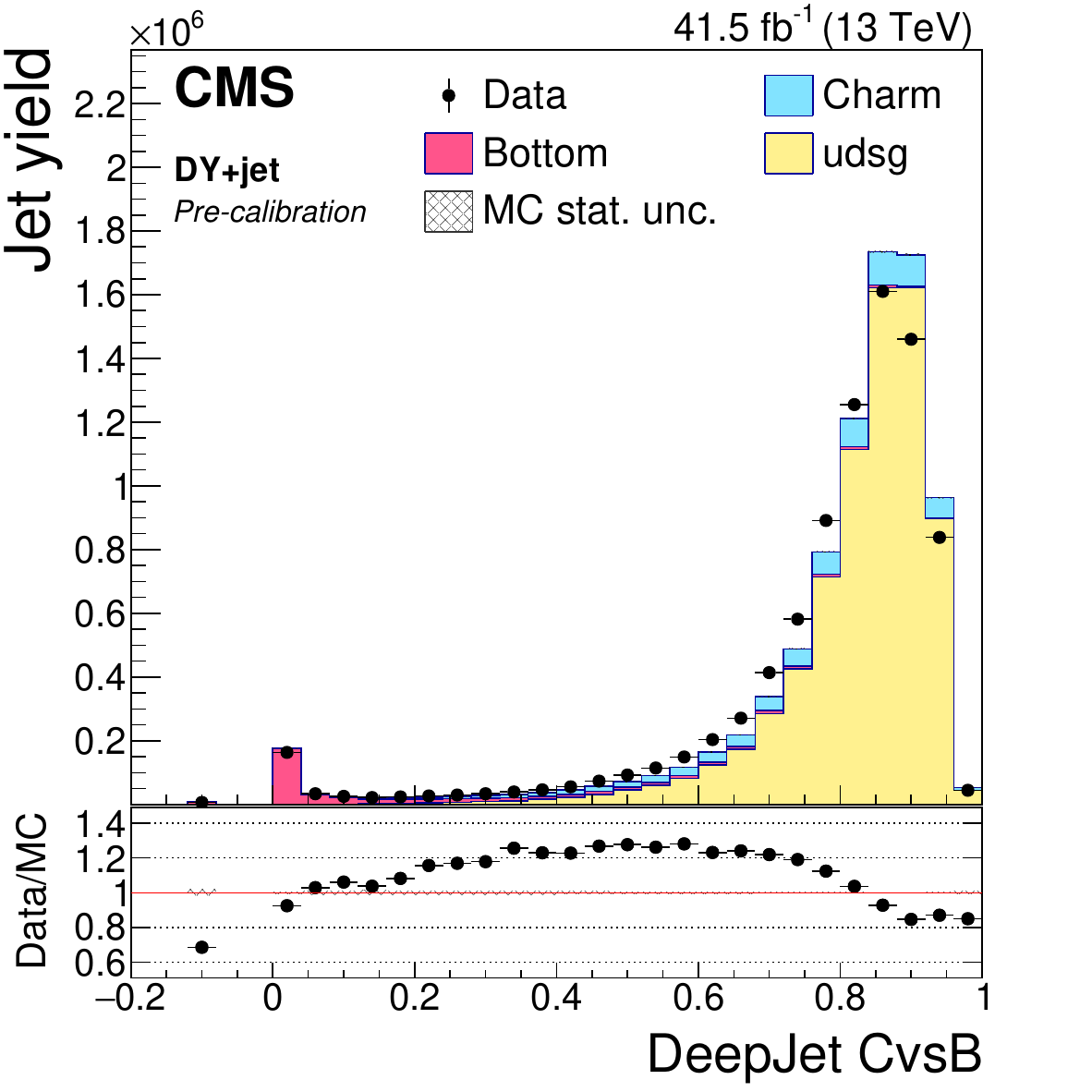}
\caption{\label{fig:DYDeepJet}Precalibration distributions of CvsL (left) and CvsB (right) obtained
from the DeepCSV (upper) and DeepJet (lower) tagger for jets in the $\text{DY}+\text{jet}$ selection.
The bin corresponding to a tagger value of $-1$ is plotted at $-0.1$. Vertical error bars in data
represent statistical uncertainties in data. The simulations are shown as stacked histograms.}
\end{figure}

The final state in the electron (muon) channel consists
of exactly two isolated, oppositely charged electrons (muons) having \pt
of at least 27 and 15 (20 and 12)\GeV
and at least one jet. Each of the two leptons is also required to have a relative
isolation less than 0.15 and their invariant mass is required to be within 10\GeV
of the {\PZ} boson mass. The events are triggered with the dielectron
and dimuon HLTs, respectively. The requirement of a soft muon inside the jet is not 
imposed for light-flavour jets. In case there is more than one
jet in the event, the jet with the highest \pt is
considered as the light-flavour jet candidate. The CvsL and CvsB distributions
of these candidates are presented in Fig.~\ref{fig:DYDeepJet}
for both the DeepCSV and DeepJet taggers.

\subsection{Summary of selections}

\begin{table}[!ht]
\centering
\topcaption{\label{tab:Selectionsummary}The combined jet yield 
	and contribution of each jet flavour to
	each selection is shown. The jet yield is reported from data,
	whereas the per-flavour contribution is determined from simulation.
	The ``purity'' of each selection (row) is highlighted in bold text.}
\begin{tabular}{ccccc}
\hline
Selection  & Jet yield  & {\PQc} \%  & {\PQb} \%  & $\PQu\PQd\PQs\Pg$ \%\\
\hline
{\PW}+{\PQc}  & 362\,002  & \textbf{92.9}  & 0.957  & 6.14\\
\ttbar  & 380\,366  & 12.1  & \textbf{81.0}  & 6.91\\
$\text{DY}+\text{jet}$  & 8\,509\,206  & 8.87  & 5.05  & \textbf{86.1}\\
\hline
\end{tabular}
\end{table}

To summarise, {\PW}+{\PQc} events (OS-SS subtracted), \ttbar events,
and $\text{DY}+\text{jet}$ events are used to probe the {\PQc} tagger distributions
of jet samples in data enriched in {\PQc} jets, {\PQb} jets
and light-flavour jets, respectively. The jet yield and percentage
of jets of each flavour in each of these samples are summarised
in Table~\ref{tab:Selectionsummary}. These selections yield a {\PQc} jet
purity of 92.9\%, {\PQb} jet purity of 81.0\% and light-flavour jet purity
of 86.1\% in their respective samples.

\section{Iterative calibration of the c tagging discriminators\label{sec:Method}}

A calibration of the full discriminator shape requires
deriving SFs to correct for the data-simulation discrepancy
at every value of the discriminator. Since {\PQc} tagging requires
using both CvsL and CvsB values of a jet simultaneously, this translates
into deriving data-to-simulation SFs as functions of both these parameters
to simultaneously adjust the discriminator shape in the 2D
plane of CvsL and CvsB. In addition, mismodelling in simulation is
expected to be different for light-flavour, {\PQc}, and {\PQb} jets.
This entails deriving different sets of SFs for different jet flavours.

The general idea of deriving SFs employed in this paper
is to perform a fit iteratively on three flavour-enriched jet samples
allowing the three flavour components (light, {\PQc} and {\PQb}) in
each to float freely until the difference between the distribution
of jets in data and simulation in all the samples is minimised.
The simultaneous usage of three samples with different flavour
compositions ensures the existence of a solution, while an iterative
fitting approach helps in converging to a physical value. After motivating
two distinct choices of binning in the 2D phase space of the CvsL
and CvsB discriminators below, the algorithm employed in the iterative
fit method is discussed in detail. Finally, an interpolation scheme
is discussed that combines the SF results obtained in two
different binning schemes into one set of SFs with finer binning.

\subsection{Adaptive binning\label{subsec:Adaptive-binning}}

Ideally, a perfect calibration would require deriving the flavour-wise correction
factors in infinitesimally small areas in the CvsL-CvsB plane. However,
the smaller the bin area, the lower the yield of jets in that bin,
and hence the larger the statistical uncertainty in the SF. A
large bin area, on the other hand, results in largely discontinuous SFs,
which could be unrepresentative and could result in an incorrect calibration.
To find a trade-off, an adaptive binning approach is adopted
with the aim of having smaller bin areas for SFs of a
given flavour in the part of the 2D plane where the yield of jets of that
flavour is large.
This results in a more continuous correction in high jet count regions,
and ensures optimised statistical uncertainties over the entire phase space.

In practice, the binning is kept constant along one axis and made
adaptive along the other in the first run, and then  the choice is
reversed in the second. The 2D plane of SFs
for all three flavours is first divided into ten equal-width
bin-slices along one axis (the ``fixed axis''). The algorithm treats
each of these slices independently and proceeds to determine the binning
along the other axis (the ``variable'' axis) on a per-bin-slice
basis and independently for each flavour. Thus, for each bin slice
of the SF map of a given flavour, the binning along the
variable axis is sequentially increased starting from a minimum threshold,
$b_{\text{min}}$, in steps of $b_{\text{min}}$, until at least one of the following two conditions is
fulfilled for each bin: 
\begin{itemize}
\item the resulting SF for that flavour in that bin has a relative
statistical uncertainty smaller than $\epsilon_{\text{max}}$, or, 
\item the bin width along the variable axis is equal to $b_{\text{max}}$. 
\end{itemize}
A dedicated scan of these three parameters ($b_{\text{min}}$, $b_{\text{max}}$,
and $\epsilon_{\text{max}}$) is performed and values of $b_{\text{min}}=0.02$, 
$b_{\text{max}}=0.10$, and $\epsilon_{\text{max}}=2\%$ are chosen as an optimal
compromise between the granularity of the binning and the statistical
uncertainties.

After the SF results are derived using such a binning scheme, the choice of
fixed and variable axes is reversed to yield another binning scheme.
For example, at first, the CvsB axis could be chosen as the fixed axis and CvsL
as the variable axis to yield a binning scheme and perform the
iterative fit. Next, the CvsL axis could be chosen as the fixed axis and the CvsB axis
as the variable one. This would yield a second binning scheme using which the
iterative fit can be performed.

\subsection{Iterative fit\label{subsec:Iterative-fit}}

The iterative fit method minimises a global $\chi^{2}$ value that
quantifies the data-to-simulation discrepancy in the projected CvsL distribution
over a given (fixed-width) CvsB bin slice, or vice versa. The method proceeds by iteratively evaluating
and applying SFs for all three jet flavours in the corresponding selected
topology in which they are enriched.

The algorithm begins by scaling the total number of simulated jets across all contributing processes in each channel
(i.e. muon, electron, or electron-muon, as applicable) of each
sample (i.e. {\PQc}, {\PQb} or light-flavour jet enriched) to match the number of jets selected in data in that channel.
This removes any remaining data-simulation discrepancies in terms of the
total jet yield, without altering the discriminator shape, as explained
in Section~\ref{sec:Event-Selection}, and simultaneously ensures that the
SFs thus derived correct only the discriminator shape without 
significantly altering the jet yield.

Then for each fixed bin slice:
\begin{itemize}
\item The three samples are ranked by the purity of the samples for the
given bin slice. The purity of a sample is defined as the percentage
contribution of the jet flavour that the sample is meant to enrich,
e.g. fraction of {\PQc} jets in the {\PW}+{\PQc} selection (see Table~\ref{tab:Selectionsummary}).
\item Starting with the purest sample, ``signal'' is defined as the CvsL (CvsB)
distribution of simulated jets of the flavour $f_{1}$ that
this sample aims to enrich, when CvsB (CvsL) is chosen as the fixed axis. Similarly, "background'' is defined as the sum
of the templates of the other two flavour components of
the same sample in simulation.
``Signal in data'' is defined as the difference between the simulated background
template and the data distribution. The ratio of the ``signal in
data'' to ``signal'', evaluated as a function of the quantity on the variable axis,
gives the SF for $f_{1}$ in this bin slice in this
iteration. The ratio (an array of SFs) is calculated in discrete
intervals following the binning scheme determined as explained
in Section~\ref{subsec:Adaptive-binning}. If the yield 
of simulated jets of flavour $f_{1}$ in the enriched sample
in any of these bins is less than one, or if the purity of the sample
in any bin is less than a threshold (10\%), the resulting ratio in this bin
may have an unphysical central value with a large statistical uncertainty.
In all such bins, a default value of 1 with a 100\% statistical
uncertainty is assigned as the SF.

In practice,
the array of SFs is initialised with the value 1 for all bins and
all flavours with the goal of updating them to the current best
estimate of the SF values after every ratio evaluation.
When the difference between the current and previous best
estimates after a ratio evaluation in a given bin is larger
than 0.002 (0.005), a change of only 0.002 (0.005)
is allowed in that bin for DeepCSV
(DeepJet). This ensures a smooth convergence of the $\chi^{2}$
parameter and prevents convergence to a local minimum.

\item Moving to the second purest sample for this bin slice, the
SFs for $f_{1}$ are applied to the simulated template of $f_{1}$
in this sample with the same discrete binning used to determine
the SFs. Again, ``signal'' is defined as the template of the
flavour $f_{2}$ that this distribution enriches and thus
the array of SFs for $f_{2}$ is evaluated in the same way.
The binning of these SFs can, in general, be different from
that of $f_{1}$, and will depend on the distribution of jets of flavour
$f_{2}$ in this bin slice. 
\item Proceeding to the third sample, the SFs for $f_{1}$ and
$f_{2}$ are applied to the respective simulated templates following their respective
binnings, and the SFs for $f_{3}$ corresponding to
this sample are evaluated. This completes one full iteration for this bin slice. 
\item The next iteration begins by applying all three SF arrays
to the respective simulated templates of the first sample. The SFs for $f_{1}$ 
are then updated. Proceeding to the next samples,
$f_{2}$ and $f_{3}$ are updated likewise.
\item The iterations are continued until a global $\chi^{2}$ is minimised and 
no longer improves with further iterations. The
$\chi^{2}$ for a given sample, ``$\text{sel}$'' ($\text{sel} = \text{Wc}$, \ttbar, DY), in this bin slice
is defined as 
\begin{equation*}
\chi_{\text{sel}}^{2}=\sum_{i=1}^{n}\frac{\left(N_{\text{sim},i}-N_{\text{data},i}\right)^{2}}{\sigma_{\text{sim},i}^{2}+\sigma_{\text{data},i}^{2}},
\end{equation*}
where $n$ is the number of bins of width $b_{\text{min}}$ in the distribution along the variable axis,
$N_{\text{sim},i}=\sum_{f=\text{{\PQc},{\PQb},{\PQu}{\PQd}{\PQs}{\Pg}}}\text{SF}_{f,i} N_{f,i}$
is the SF-adjusted simulated jet count in
the $i$-th bin along the variable axis, $N_{\text{data},i}$ is the jet count in data
in the $i$-th bin, with $\sigma_{\text{sim},i}^{2}$ and $\sigma_{\text{data},i}^{2}$
as the corresponding statistical uncertainties in simulation and data, respectively. 
This quantity is always well defined since the
width of any given bin is a multiple of $b_{\text{min}}$ by construction, as explained in
Section~\ref{subsec:Adaptive-binning}.
The global $\chi^{2}$ of a bin slice is defined as the sum of the three $\chi_{\text{sel}}^{2}$ values
corresponding to the three samples in this bin slice, and is updated
after every iteration. 
\end{itemize}
The same procedure is used for every bin slice and for both DeepCSV
and DeepJet taggers. Thus, three 2D maps of SFs corresponding
to three different flavours are obtained for each tagger. Next,
the entire procedure is repeated with the fixed and variable
axes interchanged. This gives an additional set of SFs for the
same phase space, but with a different binning. The method of combining
the two sets of results is presented in Section~\ref{subsec:Interpolation}.

\subsubsection{Treatment of jets with default discriminator values}
Jets that are assigned a default value of $-1$, because they either do not have sufficient information to be processed by the
tagging algorithm (DeepCSV) or have an undefined value for at least one of the {\PQc} tagging discriminators (DeepJet), need to be treated separately in the calibration.

\paragraph*{DeepCSV}
Jets that have no tracks passing the DeepCSV preselection criteria, and hence
have been assigned default discriminator values $\text{CvsL} = \text{CvsB} = -1$, remain 
excluded from the iterative fit treatment described above. Therefore, the iterative fit
procedure is repeated in the same way for the jets with defaulted discriminator values
in the three samples. However, only the jets from the electron channel of the {\PW}+{\PQc} selection are used
as the {\PQc} jet enriched sample in this particular fit, because such jet candidates in the muon channel of {\PW}+{\PQc} are heavily
contaminated with prompt muons from DY events misidentified as jets and hence are
unsuitable for use in a three-flavour fit. The $\chi^{2}$ for a sample in this case is defined as:
\begin{equation*}
\chi_{\text{sel}}^{2}=\frac{\left(N_{\text{sim}}-N_{\text{data}}\right)^{2}}{\sigma_{\text{sim}}^{2}+\sigma_{\text{data}}^{2}},
\end{equation*}
and the global $\chi^{2}$, defined as the sum of the $\chi^{2}_{\text{sel}}$ values
of the three samples, is minimised iteratively. This yields three
correction factors corresponding to the three flavours. These SFs 
for the bin containing jets with default DeepCSV values,
when combined with the 2D maps of SFs
for the CvsL-CvsB plane, give the full set of SFs
required for the calibration of the DeepCSV {\PQc} taggers.

\paragraph*{DeepJet}
Since the CvsL $=$ CvsB $=-1$ bin of DeepJet consists almost entirely of {\PQb} jets,
a lack of data for light-flavour and {\PQc} jets in this bin makes
it both impossible and unnecessary to perform a three-flavour iterative
fit to find DeepJet SFs for udsg and {\PQc}. Therefore,
in this bin, {\PQc} and udsg SFs are both set to $1\pm1$,
and the {\PQb} SF is evaluated as:
\begin{equation*}
\mathrm{SF}_{\PQb}=\frac{N_{\text{data}}-N_{\text{sim},{\PQc}}-N_{\text{sim},{\PQu}{\PQd}{\PQs}{\Pg}}}{N_{\text{sim},{\PQb}}}
\end{equation*}
where $N$ represents the jet counts of the respective flavours in
the bin containing jets with default discriminator value in the \ttbar sample. 

\subsection{Interpolation\label{subsec:Interpolation}}

For each algorithm, an interpolation of the SF values in the 2D plane is performed on a per-flavour basis to find more representative corrections on a finer grid. Such an interpolation is feasible because the SFs have a fairly smooth distribution, as evident to a first degree of approximation from the data-to-simulation ratios in Figs.~\ref{fig:WcDeepJet}--\ref{fig:DYDeepJet}. Moreover, combining the two sets of SFs, which are derived using two different binning schemes with such an interpolation approach, is expected to improve the SF values, given the fact that the two sets of SFs potentially contain complementary information about rapidly varying SF values at certain parts of the CvsL--CvsB plane. For example, the SF set derived with an adaptive binning scheme along the CvsB axis could potentially reflect sharp SF variations along the CvsB axis for a certain flavour of jets in parts of the plane enriched in jets of that flavour.

At first, a nodal point for the interpolation (a point where the functional value is assumed to be known exactly) is defined for each bin in the 2D plane for each flavour in each SF set. The coordinates of this nodal point are defined as the means of the coordinates of all the jets of the corresponding flavour that are contained in the corresponding bin. Thus, the nodal point $(x,y)$ of a bin in an SF map for a given flavour containing $N$ jets of that flavour, each with a $(\text{CvsL},\text{CvsB})$ value of $(x_i,y_i)$, is given by:
\begin{equation*}
(x,y) = \left(\frac{\sum_{i=1}^{N}x_i}{N},\frac{\sum_{i=1}^{N}y_i}{N}\right).
\end{equation*}
The value of the SF corresponding to this bin is assigned to this nodal point and is assumed to be known exactly at this point for the purpose of the interpolation.

For each flavour, the nodal points obtained from the two SF sets are merged together to form an irregular grid of points in the 2D plane; an SF value is assigned to each from the respective SF set. When the same nodal point is found in both the SF sets, only one is used to form the grid. A bivariate interpolation is performed for each flavour using the SF values at the nodes of the corresponding irregular grid using the \textsc{interpolate} subpackage of \textsc{SciPy}~\cite{2020SciPy-NMeth}. The algorithm first triangulates the nodal points in the 2D plane using \textsc{Qhull}~\cite{Barber96thequickhull} and then constructs a piecewise cubic interpolating B\'ezier polynomial~\cite{bezier1970numerical} on each triangle using a Clough-Tocher scheme~\cite{clough1965finite}. Then the interpolant is constructed by choosing gradients that approximately minimize the curvature of the interpolating surface~\cite{10.2307/2007373,10.2307/44236796}.

However, such an interpolant is defined only inside the convex polygon (convex hull) containing all the nodal points. Furthermore, the interpolant is seen to have unphysical properties in some parts of the plane inside the convex hull but outside the concave hull of the nodal points. Therefore, a nearest-neighbour extrapolation (assigning the functional value of the nodal point nearest to a given point) is performed to assign values to all points in the 2D plane that lie outside the concave hull of the nodal points. The interpolant along with the extrapolant together define SF values at all points of the 2D plane (called interpolated SFs hereafter, for simplicity). Even though this allows for a continuous correction over all the plane, the interpolated SF values are computed at 50 discrete bins along each axis for convenience, which is a good approximation for a continuous correction.

We have verified that interpolating using nodal points from two sets of SFs derived with two different binning schemes provides a final set of SFs that exhibits a better closure of the method than SFs derived with interpolation using nodal points from any one set alone, in case of both DeepCSV and DeepJet algorithms. Furthermore, we also verified that the piecewise-cubic bivariate interpolant used here outperforms other methods of interpolation, such as piecewise-linear bivariate interpolation and smoothing bivariate spline approximation.

\section{Uncertainty estimation\label{sec:Systematics}}

The statistical uncertainty in the SF for a given flavour
in a given bin is evaluated from the statistical uncertainty in the
number of jets in both data and simulation in that bin in the selection enriched in that flavour.
In this regard, the per-bin
uncertainty in simulation, as well as data, of the {\PW}+{\PQc} OS-SS selection
accounts for the individual statistical uncertainties in the OS and SS selections,
and is hence propagated to the statistical uncertainties
in the {\PQc} jet SFs.

For systematic uncertainty estimation, all selections are performed
with different sources of systematic uncertainties shifted by $\pm1$ standard deviation
independently, and the whole iterative fit procedure is performed
for every systematic variation. This gives modified SF maps for every
source of uncertainty. The sources of systematic uncertainties considered
are:
\begin{itemize}
\item {\textit{Lepton identification:}}     Observed differences in electron and muon identification criteria~\cite{Sirunyan:2018fpa,Sirunyan:2019yvv,CMS-DP-2018-017} between data and simulation are included by $\pt$- and $\eta$-dependent SFs and the corresponding uncertainties are propagated to the calibration of the {\PQc} tagger discriminator, separately for electrons and muons.

\item {\textit{Pileup modelling:}}     A weighting is applied to the simulated pileup vertex multiplicity to match the profile observed in data. This weighting is based on the total inelastic pp cross section of 69.2\unit{mb}~\cite{Sirunyan:2018nqx}, whereas the corresponding uncertainty is computed by varying this inelastic cross section by $\pm4.6\%$.

\item {\textit{Jet energy resolution:}}     The jet energy resolution (JER)~\cite{Khachatryan:2016kdb,Sirunyan:2019kia} is observed to be worse in data than in simulation. To mimic more closely the resolution that is observed in data, an additional smearing is applied to the simulated jet energy. The uncertainties related to this smearing are a source of systematic uncertainties in the calibration of the {\PQc} tagger discriminator.

\item {\textit{Jet energy scale:}}     Similarly, differences in the observed jet energy scale (JES)~\cite{Khachatryan:2016kdb,Sirunyan:2019kia} between data and simulation result in corrections to the simulated jet four-momenta. The corresponding uncertainties related to these adjusted four-momenta are included in the calibration of the {\PQc} tagger discriminator.

\item {\textit{Factorisation and renormalisation scales:}}     The choice of the renormalisation and factorisation scales during the ME calculation may affect kinematical distributions of the final-state objects and therefore also the heavy-flavour tagging discriminants. Uncertainties in these scales are estimated by varying each of the two scales independently at the ME level by factors 0.5 and 2, simultaneously across all contributing processes.

\item {\textit{Initial- and final-state radiation in the parton shower:}}     During the parton shower, the uncertainty in the value of the strong coupling at a given energy scale is estimated by varying the renormalisation scale of QCD emissions separately in the initial-state radiation (ISR) and final-state radiation (FSR) up and down by factors of 2 and 0.5, respectively. The resulting uncertainty of these variations in the {\PQc} tagger calibration is included in the measurement.

\item {\textit{Cross sections of various physics processes:}} The cross sections of the different relevant processes are varied within the uncertainties of  the corresponding theoretical predictions~\cite{Li:2012wna,Frixione:2008yi,Re:2010bp,Kidonakis:2012rm,Campbell:2010ff,Gehrmann:2014fva}, up to the order in perturbation theory to which they are scaled in the simulation (see Section~\ref{sec:simulation}). Such variations are expected to alter the signal-to-background ratios in the selections and hence may affect the overall shapes of the heavy-flavour tagging discriminants in simulation.

\item {\textit{{\PQb} fragmentation:}} The uncertainties in the internal parameters of the Bowler--Lund model~\cite{Bowler:1981sb,Andersson:1983ia,Sjostrand:1984ic} that is used to parametrise the {\PQb} quark fragmentation in \PYTHIA, are obtained experimentally from the ALEPH~\cite{Heister:2001jg}, DELPHI~\cite{DELPHI:2011aa}, OPAL~\cite{Abbiendi:2002vt}, and SLD~\cite{Abe:2002iq} experiments. The effect of varying this parametrisation within the uncertainties in the kinematics of the {\PQb} jet is covered by a weighting of the ``transfer function'', $x_{\PQb} = \pt^{\text{{\PQb}~hadron}}/\pt^{\text{{\PQb}~jet}}$ at the generator level. The effect of this on the {\PQc} tagging discriminator shapes, in turn, can be parametrised as a 2D linear function of CvsL and CvsB values of {\PQb} jets, separately for DeepCSV and DeepJet. The up and down variations in this quantity are hence propagated to the {\PQb} jet template of each selection.

\item {\textit{Flavour composition of V+jet processes:}} Since the method can potentially be sensitive to the flavour composition of the background jets, additional uncertainties are assigned to the {\PQb} and {\PQc} jet yields in the DY+jet process and the {\PQc} jet yield in the {\PW}+{\PQc} process. These uncertainties are estimated to be 2.5 and 9\%, respectively, for {\PQb} and {\PQc} jets produced in association with a {\PZ} boson, and 8\% for {\PQc} jets produced in association with a {\PW} boson; these results are based on the uncertainties in the experimental measurement of {\PZ}+{\PQb}/{\PQc} processes \cite{Sirunyan_2020} and the experimental uncertainty in the measurement of the {\PW}+{\PQc} cross section \cite{CERN-EP-2018-282}.

\item {\textit{Interpolation:}} Although there is no straightforward method to estimate the bounds of the interpolation uncertainties at non-nodal points in the 2D plane in case of a piecewise-cubic bivariate interpolant, an approximate order-of-magnitude estimate is attempted. For each non-nodal point (probe point) in the concave hull of the SF nodal points for a given flavour, the interpolation uncertainty is quantified as the sum of 10 terms corresponding to the 10 control points of the cubic B\'ezier triangle that contains the point. Each of these terms is proportional to (a) the maximum absolute deviation of the directional derivative of the interpolant at each point between the control and probe points, from the average slope of the interpolant between the control and probe points, and (b) the linear distance between the control point and the probe point. This quantity is expected to be larger in parts of the plane where the interpolant has a high curvature, and at points farther away from the nodal points. In all cases, the quantity is computed approximately using numerical methods. This uncertainty is set to zero for points outside the concave hull of the nodal points.

\item {\textit{Extrapolation:}} The extrapolation uncertainty at a point outside the concave hull of the SF nodal points for a given flavour, is defined as the difference between the SF value obtained from a bicubic bivariate smoothing spline extrapolation~\cite{2020SciPy-NMeth} and the nominal SF value assigned (using nearest neighbour extrapolation). This uncertainty is set to zero for points inside the concave hull.

\end{itemize}

An estimation of the total uncertainty is made by adding
statistical and all systematic uncertainties in quadrature, assuming
no correlation. An estimation
of the relative contributions of each source of uncertainty is presented
in Section~\ref{sec:Results}.

\section{Results\label{sec:Results}}

\subsection{Scale factors}

\begin{figure}
\centering
\includegraphics[width=1\textwidth]{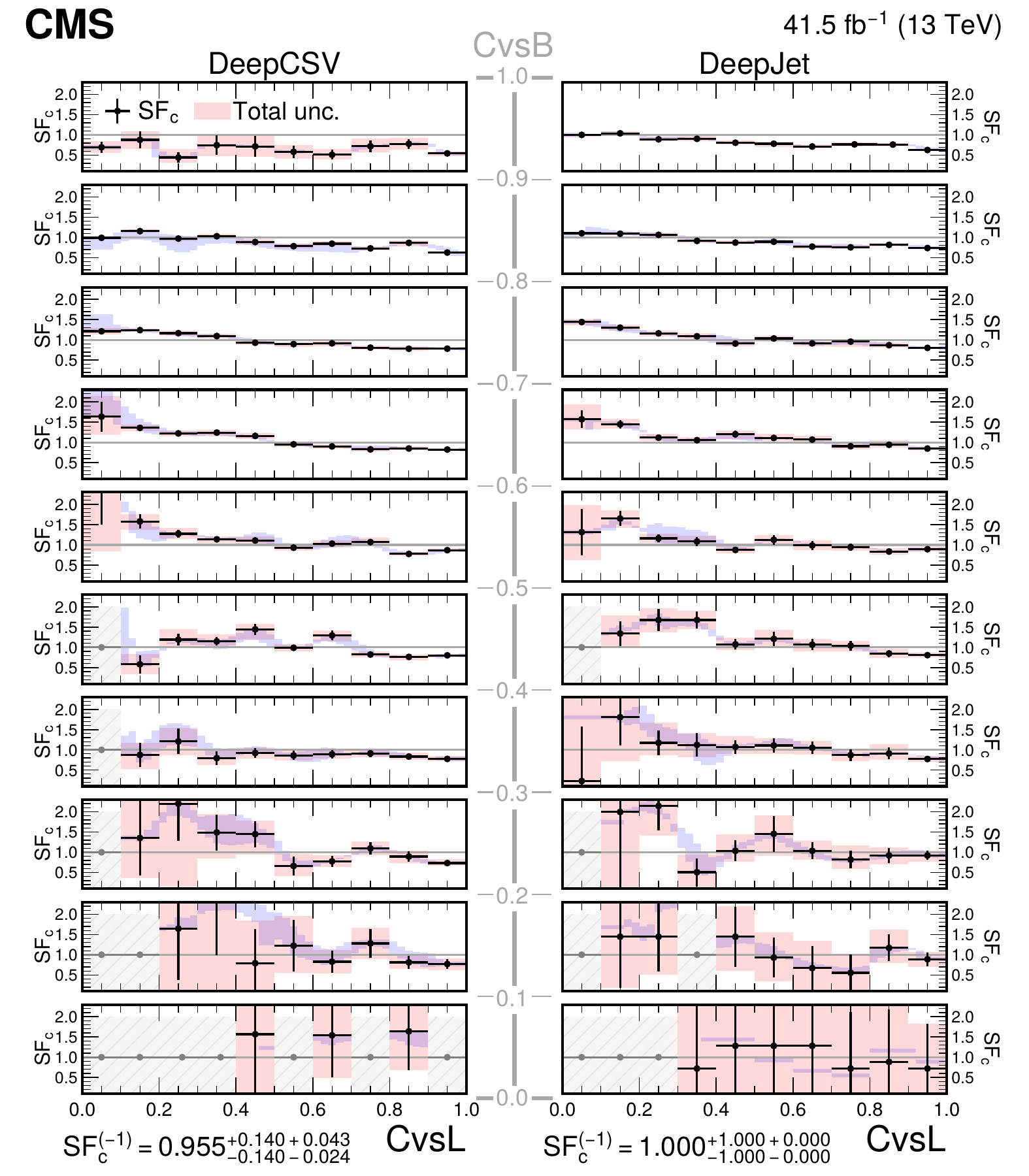}
\centering
\caption{\label{fig:SFc}The shape calibration SF values as a function of CvsL for DeepCSV- (left) and
DeepJet- (right) based {\PQc} taggers for {\PQc} jets in different ranges of CvsB are shown.
The black data points indicate the nominal SF values at the nodal points obtained with
a fixed bin width along CvsB and an adaptive binning scheme along CvsL. The total uncertainty
in the SFs at the nodal points is denoted by the red envelopes around the nominal values, whereas the statistical
uncertainties alone are denoted by the black vertical lines. Gray data points with the hatched
uncertainties denote bins with jet counts or signal purity insufficient for the SF evaluation.
The blue envelopes indicate the range of all nominal interpolated SF values in the corresponding CvsB range.
The quantity, SF$_\text{{\PQc}}^{(-1)}$, denotes the SF for {\PQc} jets with the
default discriminator value, along with the statistical (first term) and systematic (second term) uncertainties.}
\centering{} 
\end{figure}

\begin{figure}[p!]
\centering
\includegraphics[width=1\textwidth]{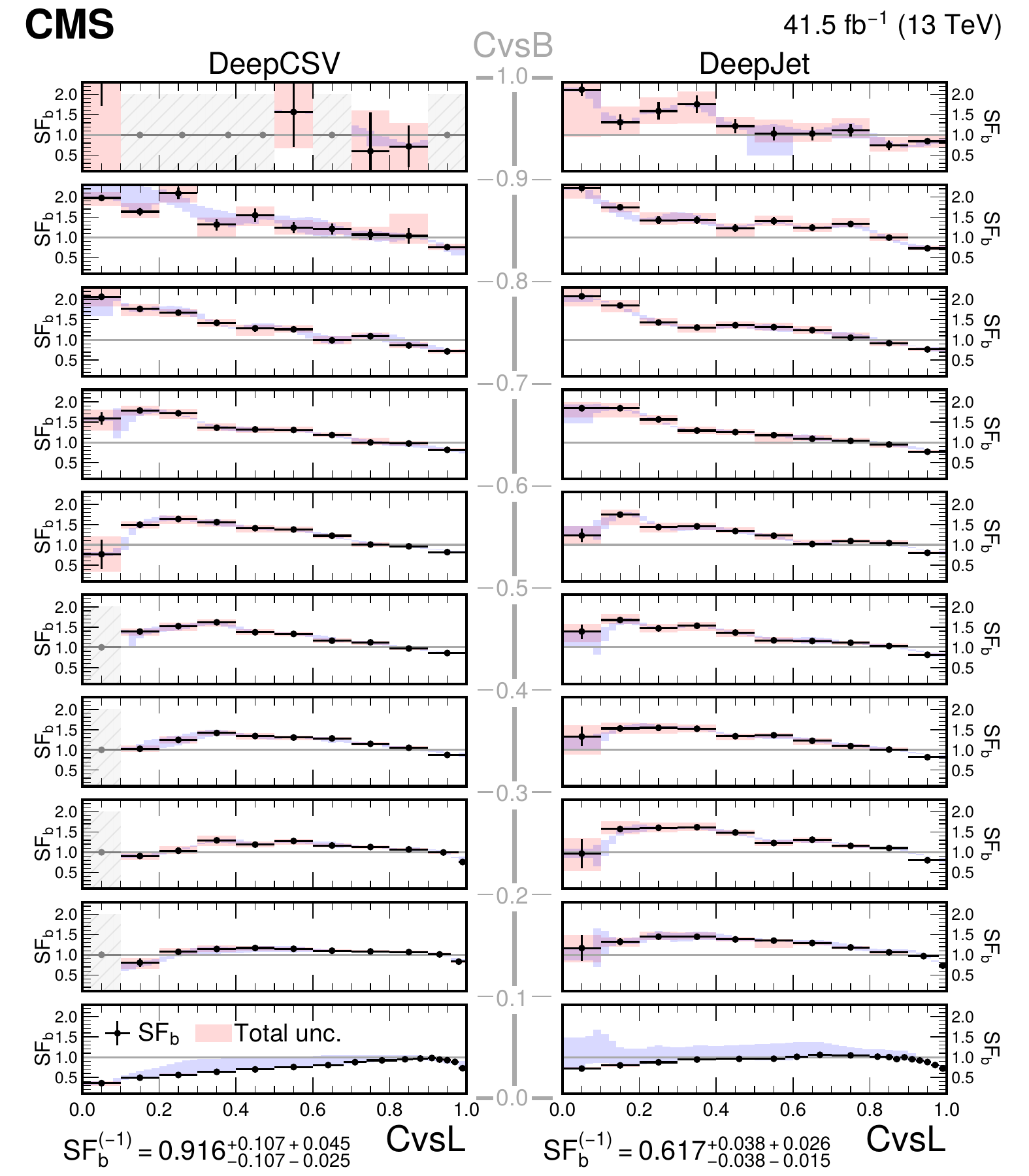} 
\centering
\caption{\label{fig:SFb}The shape calibration SF values as a function of CvsL for DeepCSV- (left) and
	DeepJet- (right) based {\PQc} taggers for {\PQb} jets in different ranges of CvsB are shown.
	The black data points indicate the nominal SF values at the nodal points obtained with
	a fixed bin width along CvsB and an adaptive binning scheme along CvsL. The total uncertainty
	in the SFs at the nodal points is denoted by the red envelopes around the nominal values, whereas the statistical
	uncertainties alone are denoted by the black vertical lines. Gray data points with the hatched
	uncertainties denote bins with jet counts or signal purity insufficient for the SF evaluation.
	The blue envelopes indicate the range of all nominal interpolated SF values in the corresponding CvsB range.
	The quantity, SF$_\text{{\PQb}}^{(-1)}$, denotes the SF for {\PQb} jets with the
	default discriminator value, along with the statistical (first term) and systematic (second term) uncertainties.}
\centering{}
\end{figure}

\begin{figure}[p!]
\centering
\includegraphics[width=1\textwidth]{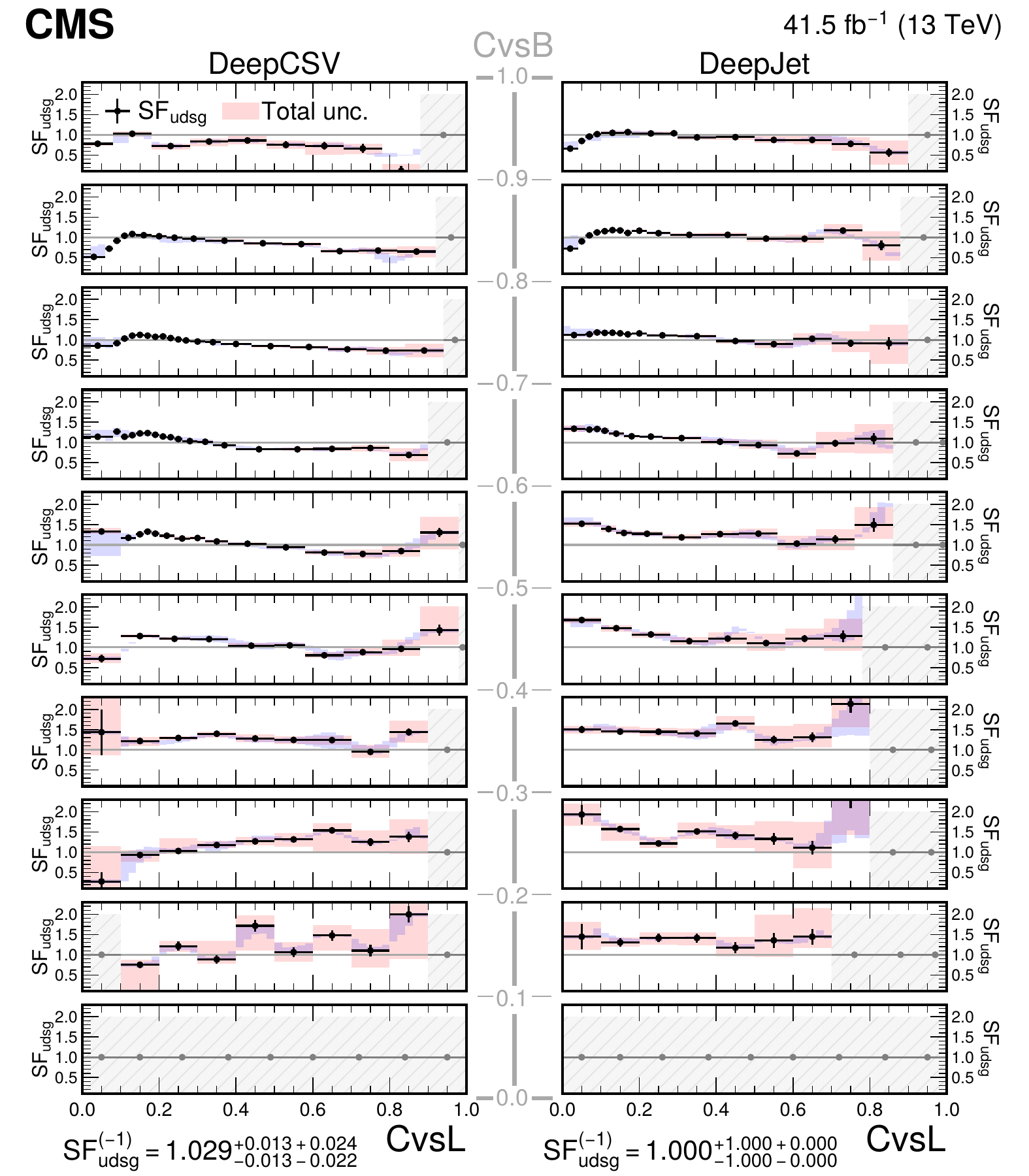} 
\centering
\caption{\label{fig:SFudsg}The shape calibration SF values as a function of CvsL for DeepCSV- (left) and
	DeepJet- (right) based {\PQc} taggers for light-flavour jets in different ranges of CvsB are shown.
	The black data points indicate the nominal SF values at the nodal points obtained with
	a fixed bin width along CvsB and an adaptive binning scheme along CvsL. The total uncertainty
	in the SFs at the nodal points is denoted by the red envelopes around the nominal values, whereas the statistical
	uncertainties alone are denoted by the black vertical lines. Gray data points with the hatched
	uncertainties denote bins with jet counts or signal purity insufficient for the SF evaluation.
	The blue envelopes indicate the range of all nominal interpolated SF values in the corresponding CvsB range.
	The quantity, SF$_\text{udsg}^{(-1)}$, denotes the SF for light-flavour jets with the
	default discriminator value, along with the statistical (first term) and systematic (second term) uncertainties.}
\centering{}
\end{figure}

The shape calibration SFs, obtained using the iterative fit method
for both DeepCSV- and DeepJet-based {\PQc} taggers, are shown in Figs.
\ref{fig:SFc}, \ref{fig:SFb}, and \ref{fig:SFudsg} for {\PQc}, {\PQb},
and light-flavour jets, respectively. These figures also show the
statistical and total systematic uncertainties associated with each SF value.
Tabulated results are provided in the HEPData record for this measurement~\cite{hepdata}.

\begin{figure}[p!]
	\centering
		\includegraphics[width=0.475\textwidth]{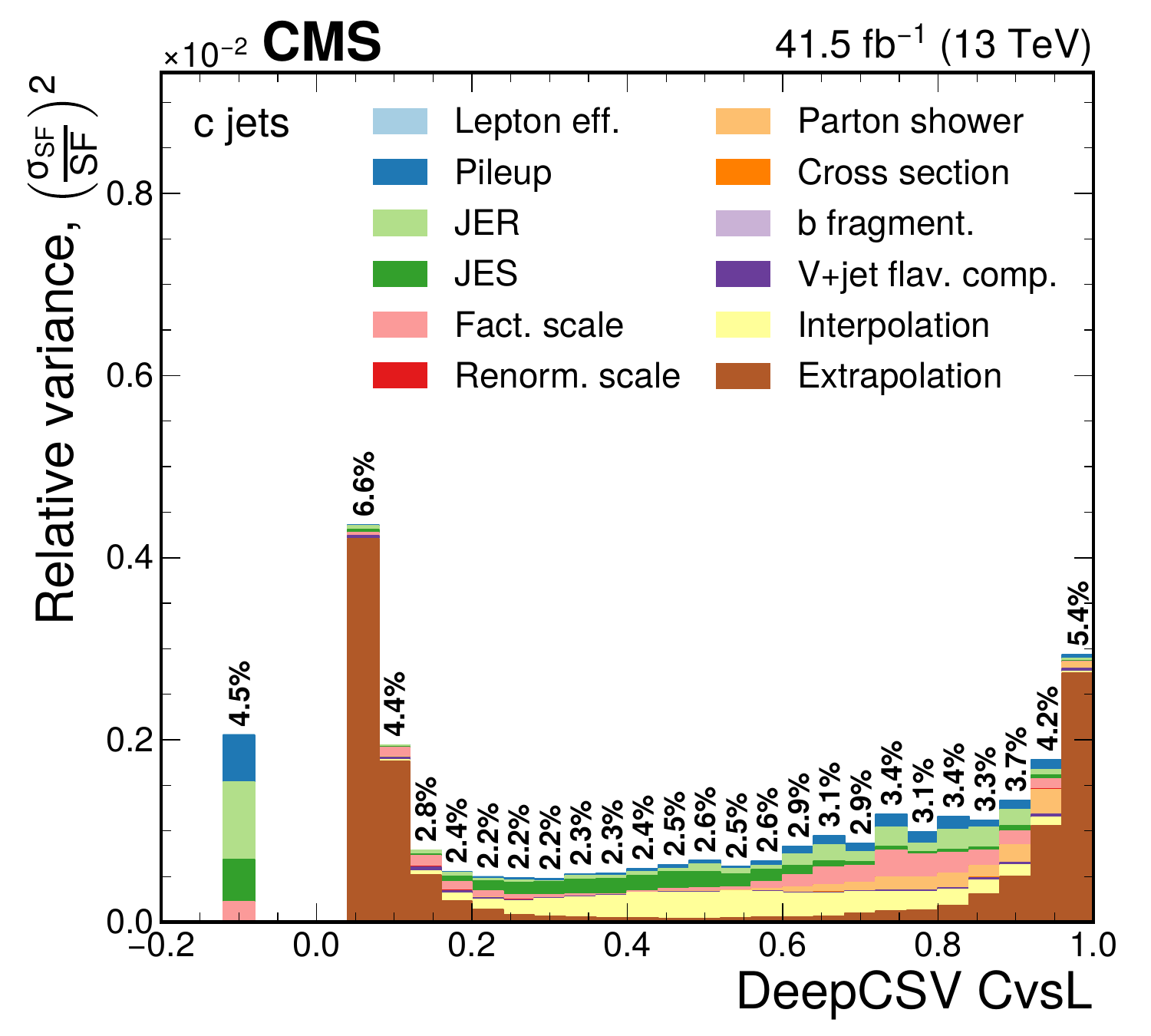}
		\includegraphics[width=0.475\textwidth]{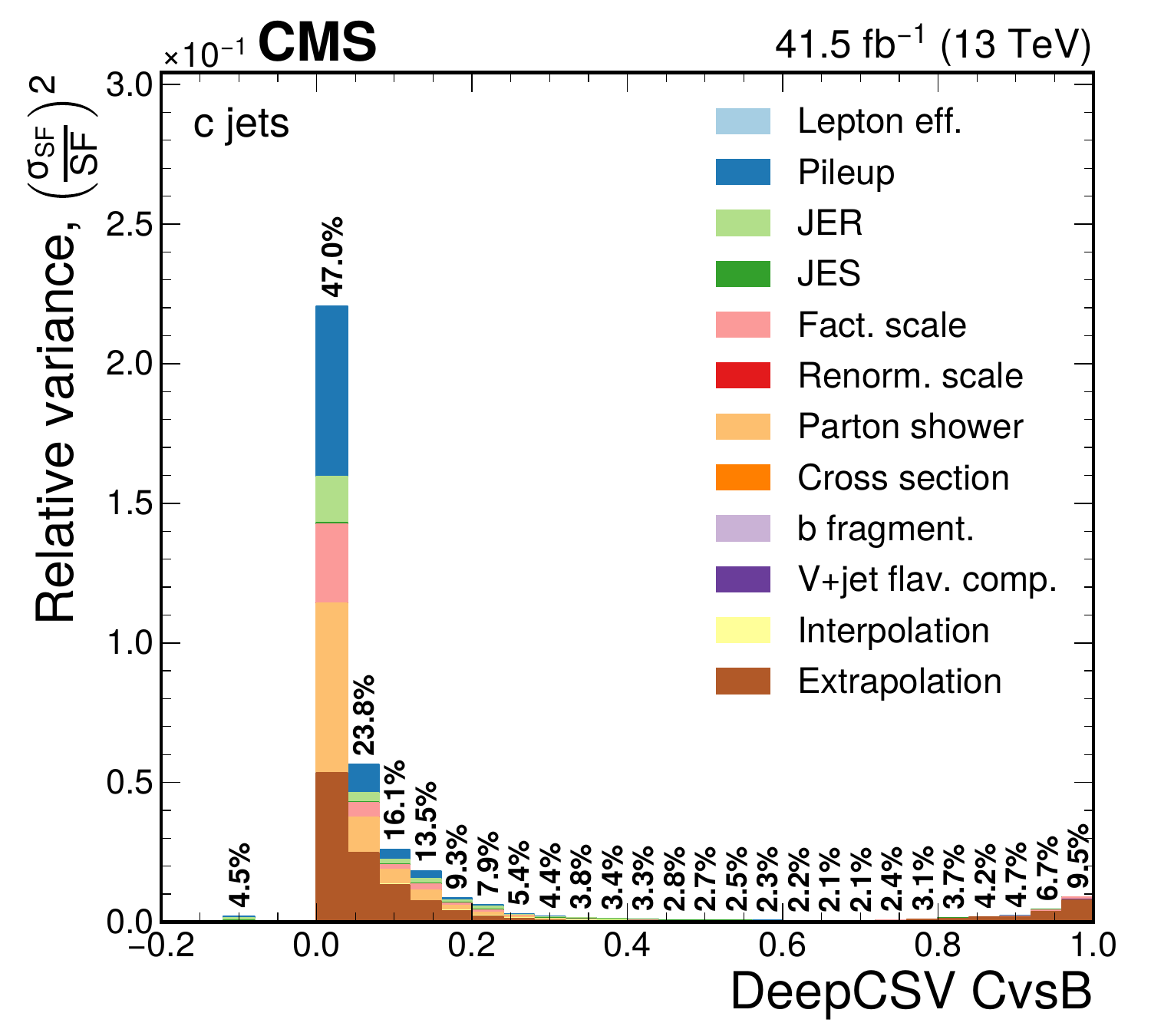}\\
		\includegraphics[width=0.475\textwidth]{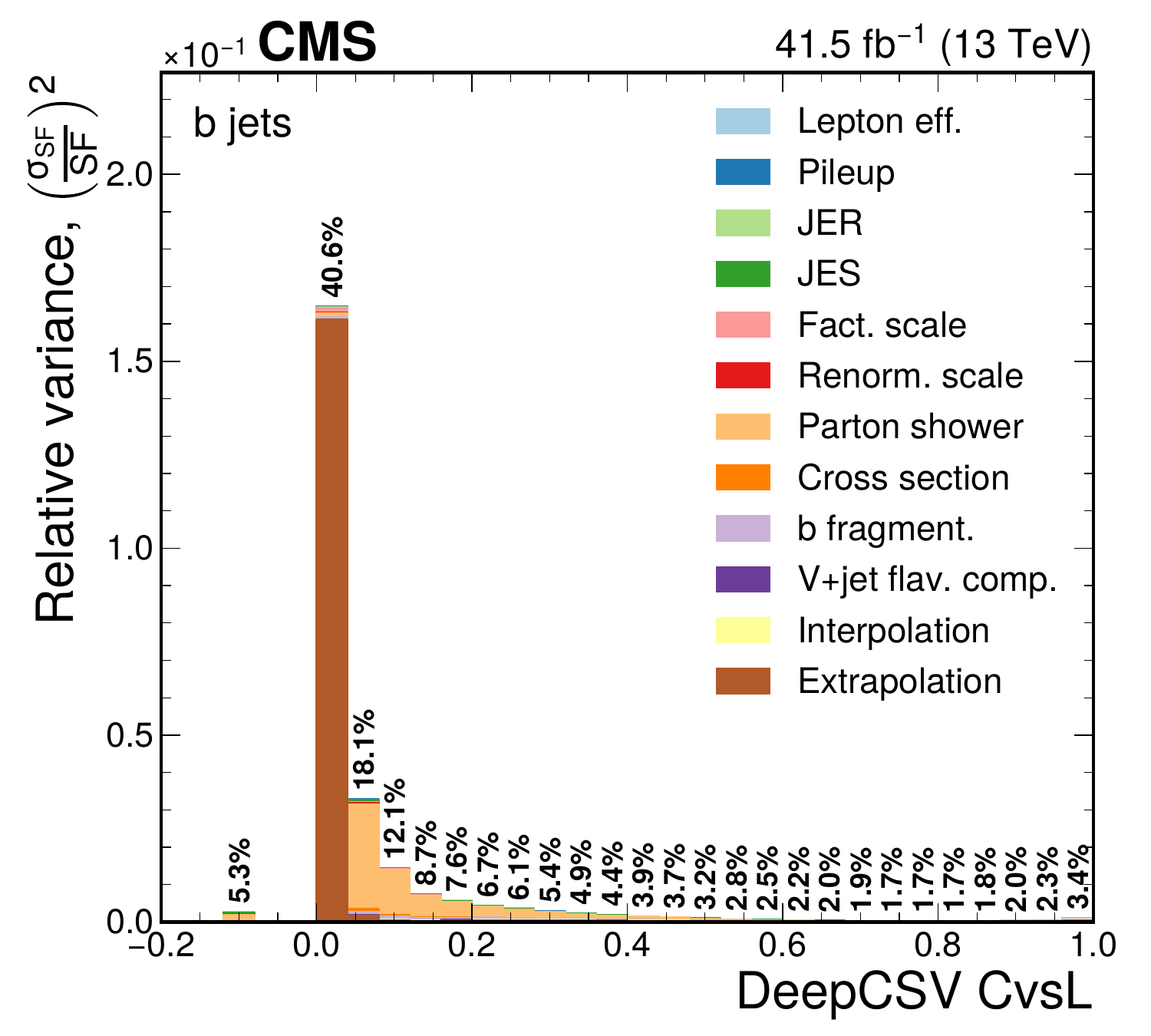}
		\includegraphics[width=0.475\textwidth]{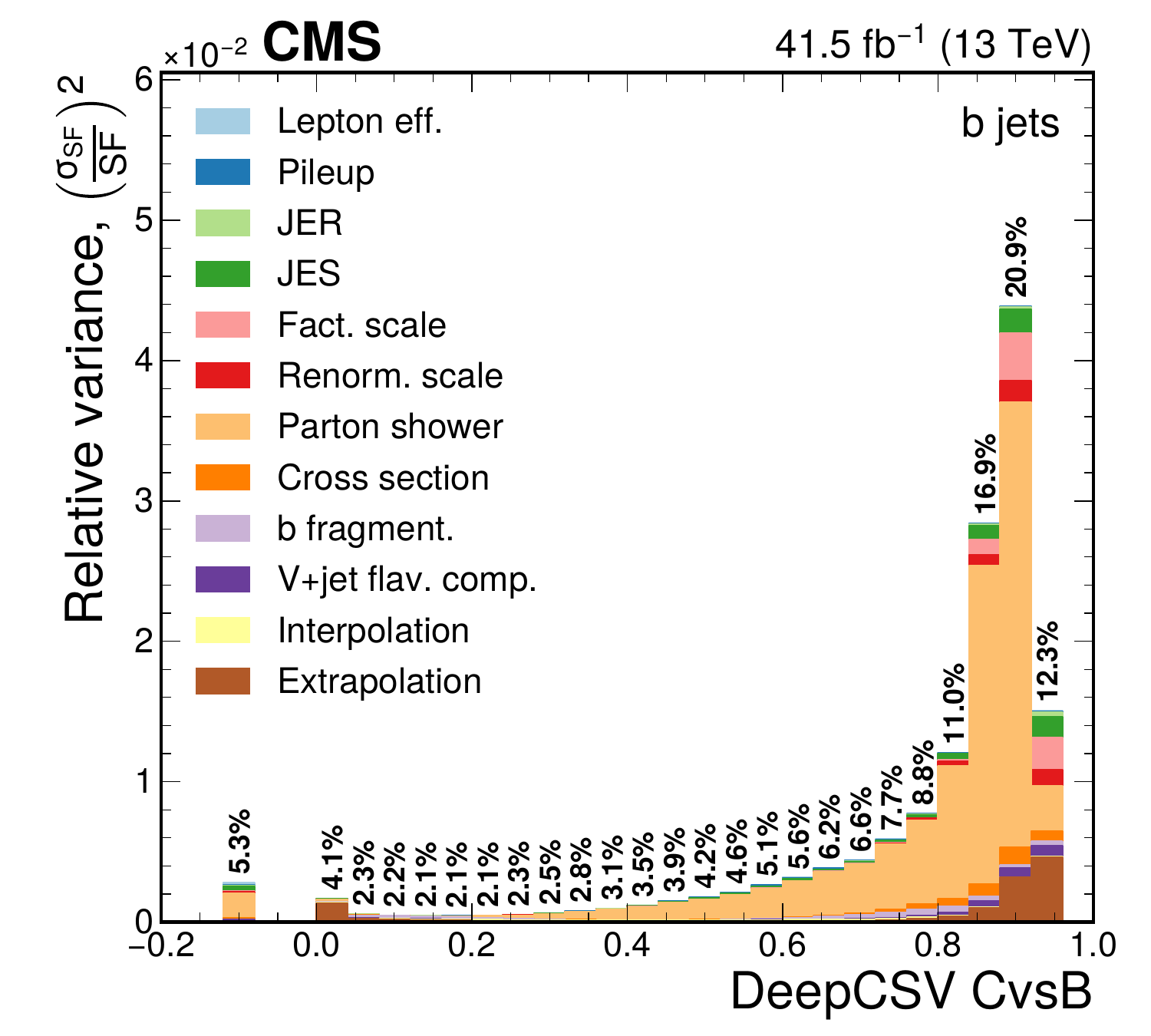}\\
		\includegraphics[width=0.475\textwidth]{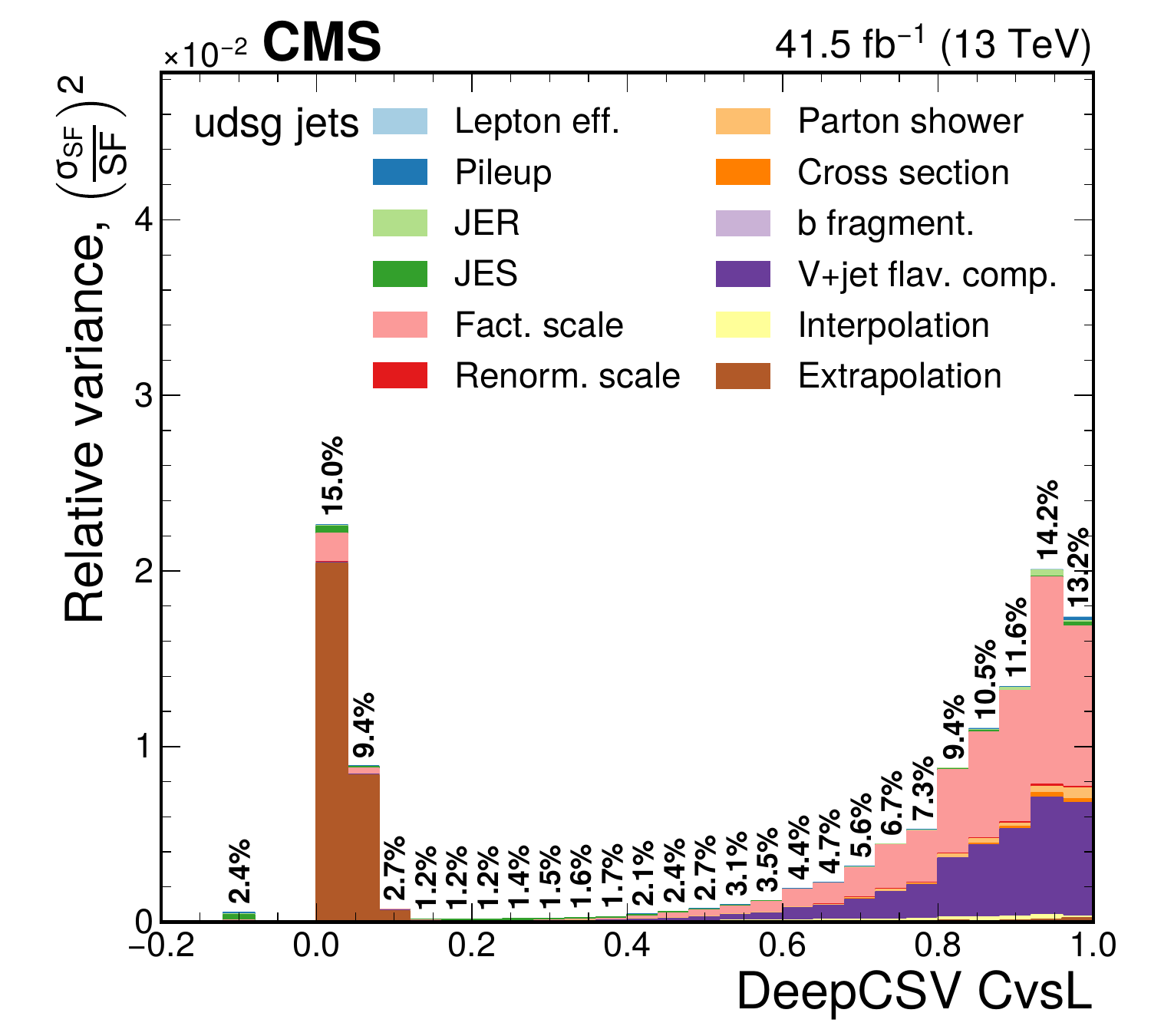} 
		\includegraphics[width=0.475\textwidth]{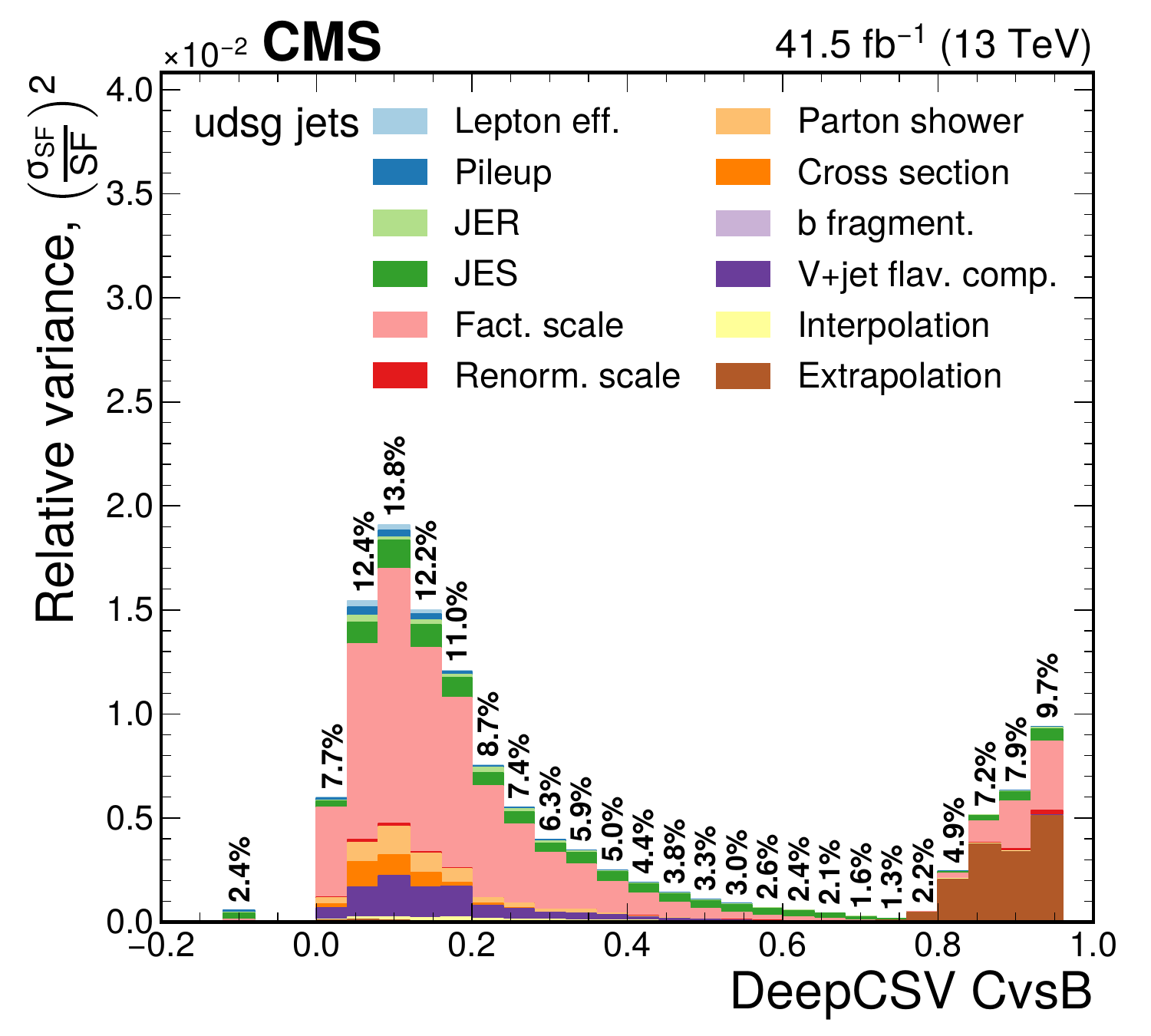}\\
	\caption{\label{fig:UncCont11}Contribution of each source of the SF uncertainty,
		calculated as the square of the relative uncertainty in the jet yield and expressed as the maximum
		of the up and down variations, at various values of the DeepCSV CvsL (left) and CvsB (right)
		discriminators for {\PQc} (upper), {\PQb} (middle), and light (lower) flavours. The
		effective total relative uncertainty values ($\sqrt{\sum\left(\frac{\sigma_{\text{SF}}}{\text{SF}}\right)^2}$) per bin are also shown in bold text, for reference. The
		bin corresponding to a tagger value of $-1$ is plotted at $-0.1$. Statistical
		uncertainties are not shown.}
\end{figure}

\begin{figure}[p!]
	\centering
		\includegraphics[width=0.48\textwidth]{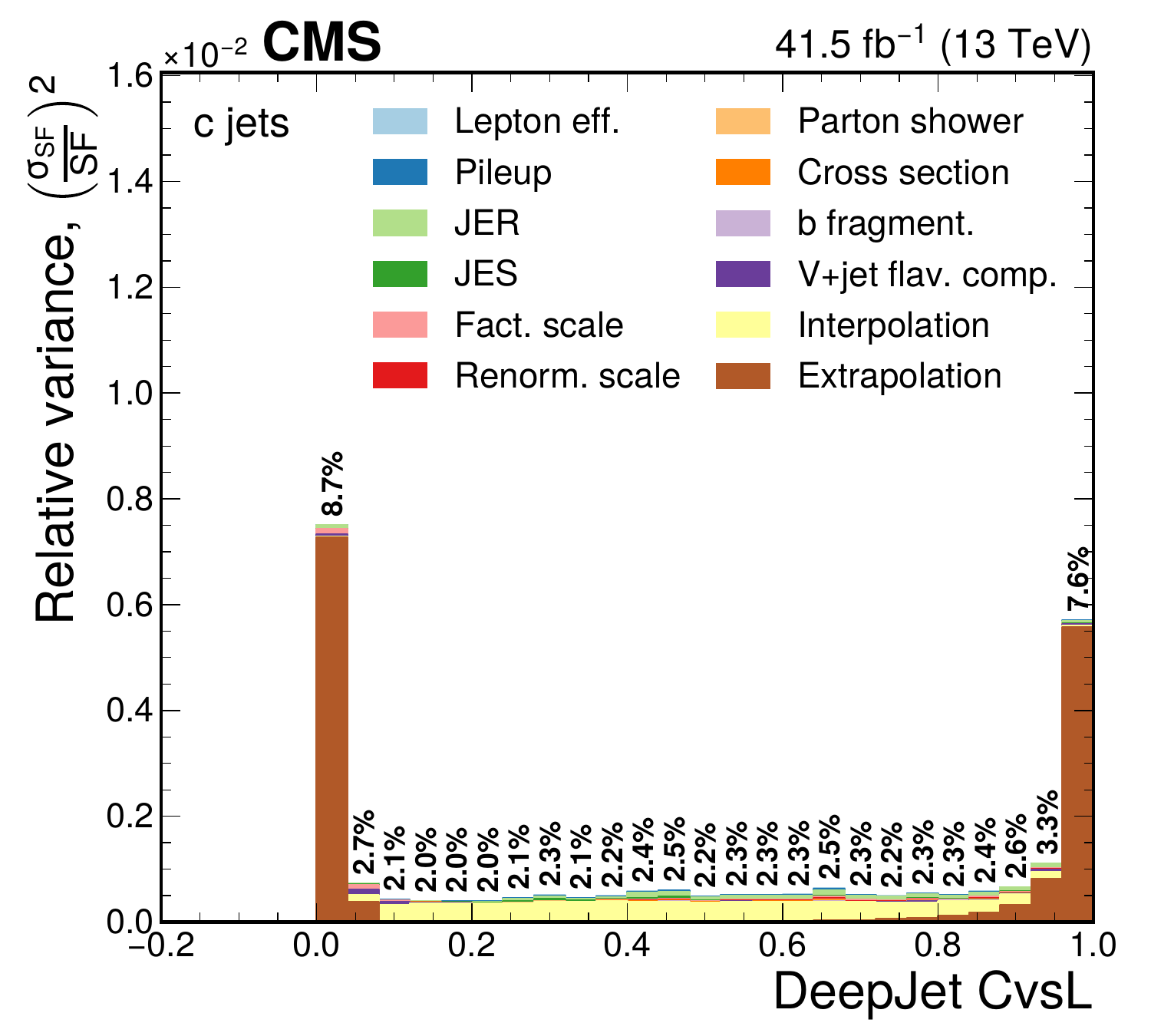}
		\includegraphics[width=0.48\textwidth]{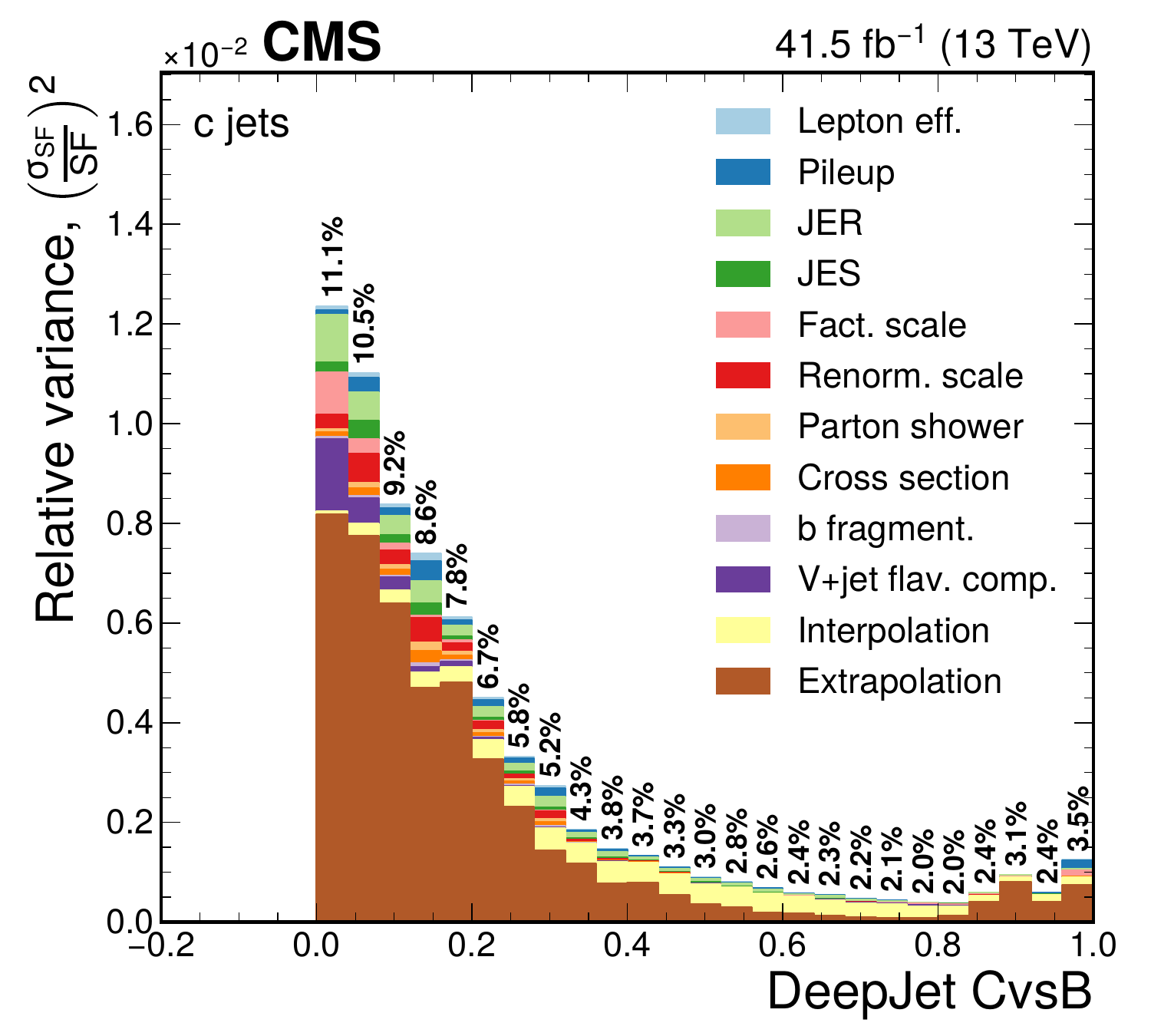}\\
		\includegraphics[width=0.48\textwidth]{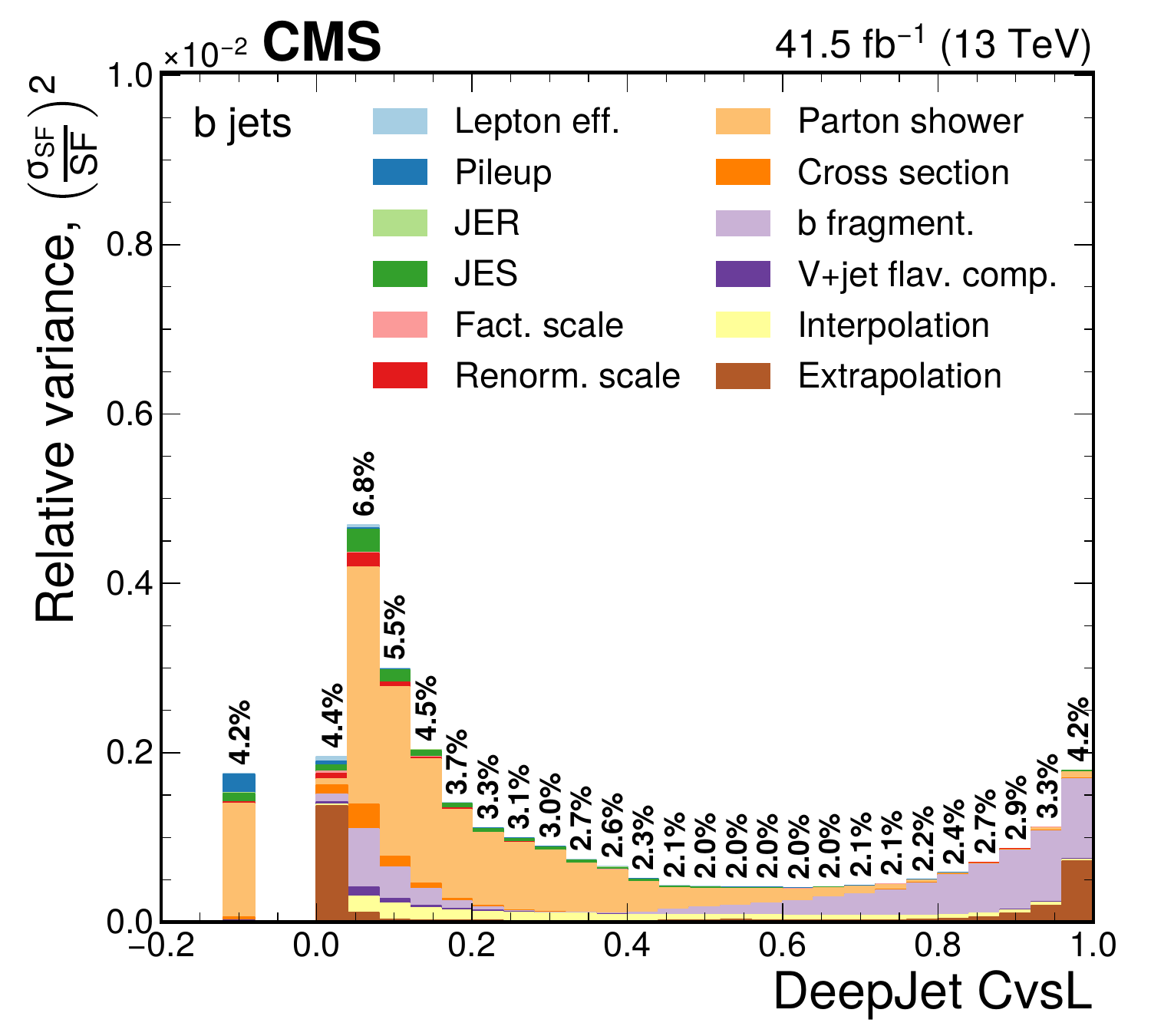}
		\includegraphics[width=0.48\textwidth]{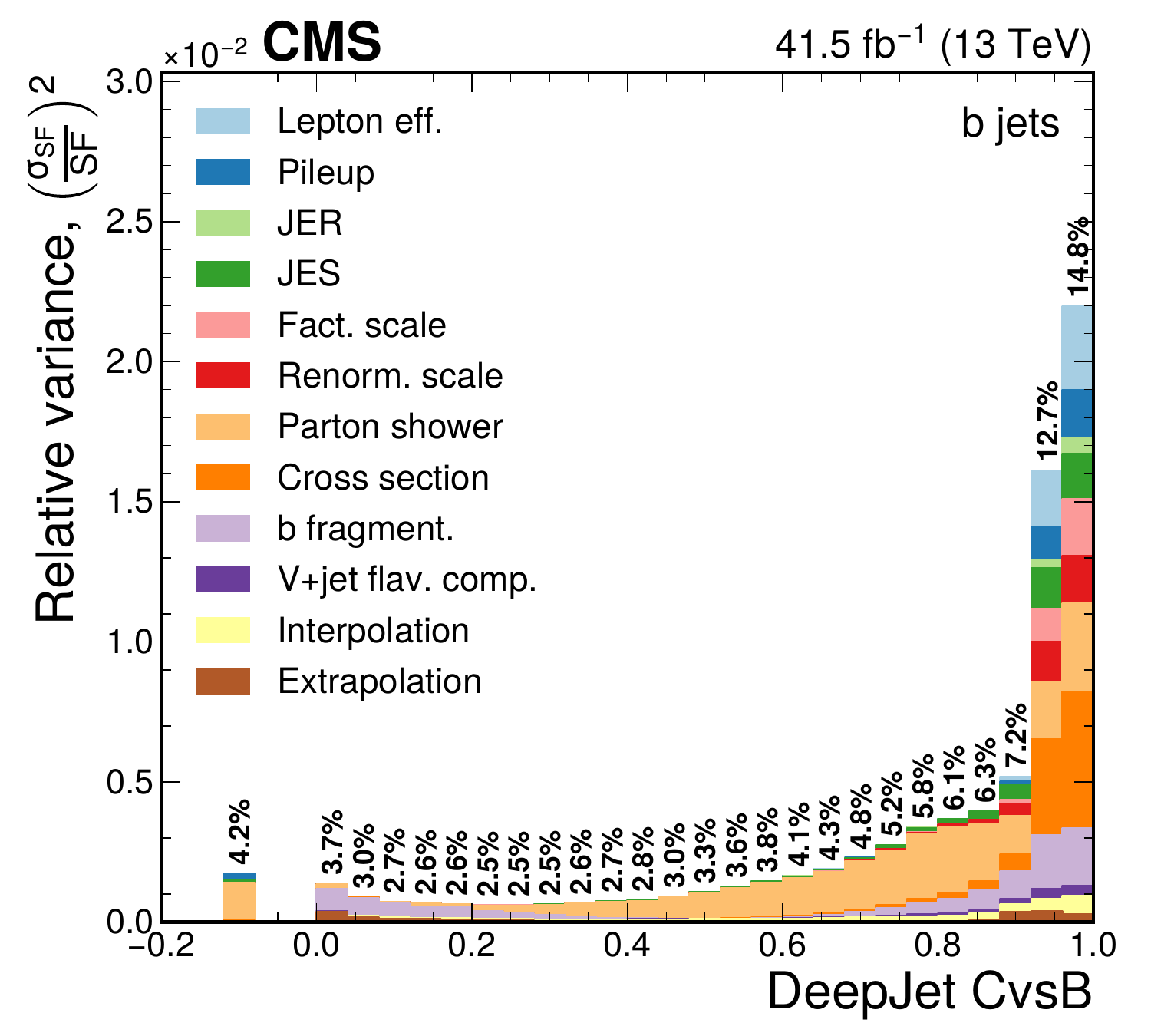}\\
		\includegraphics[width=0.48\textwidth]{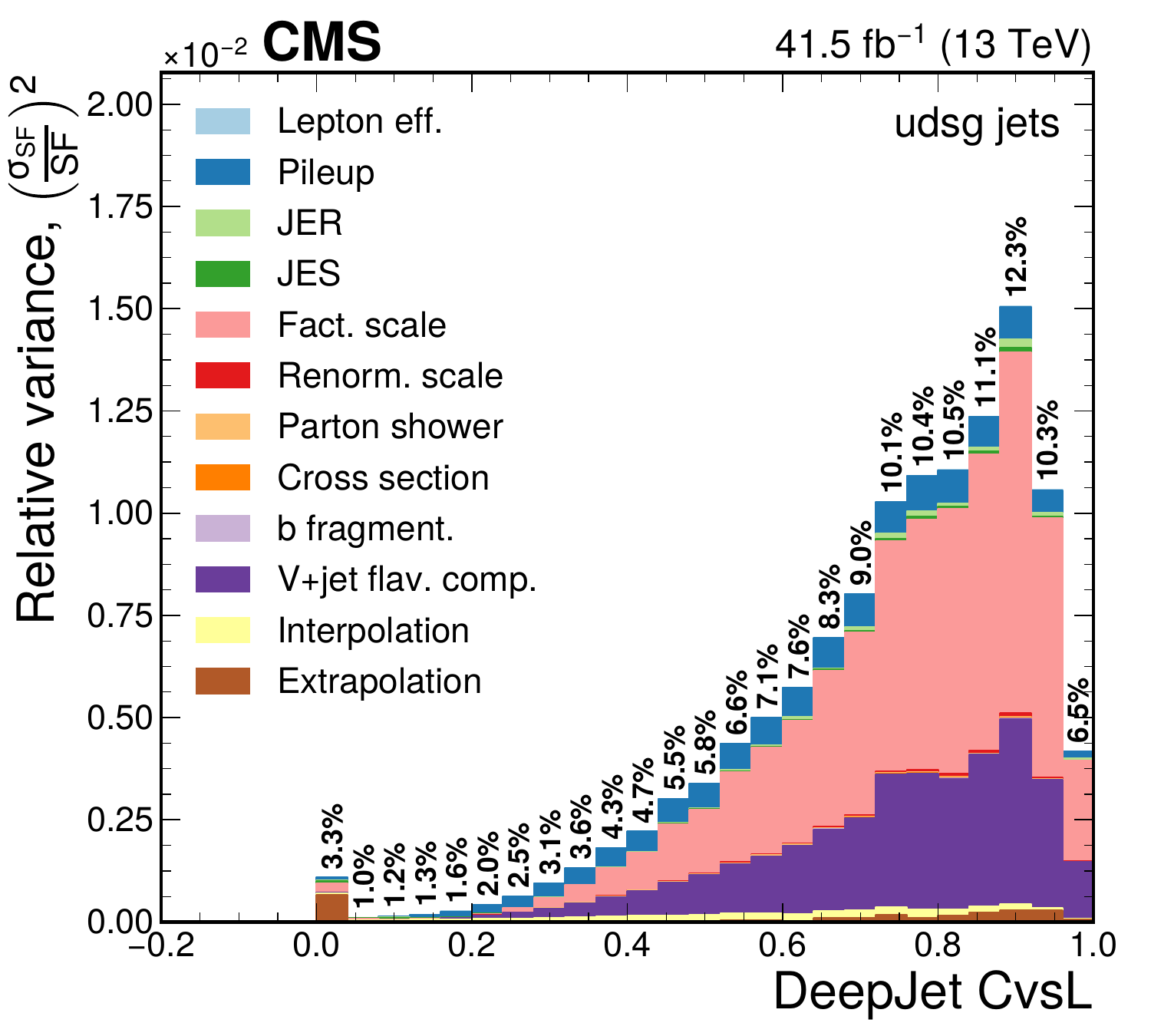}
		\includegraphics[width=0.48\textwidth]{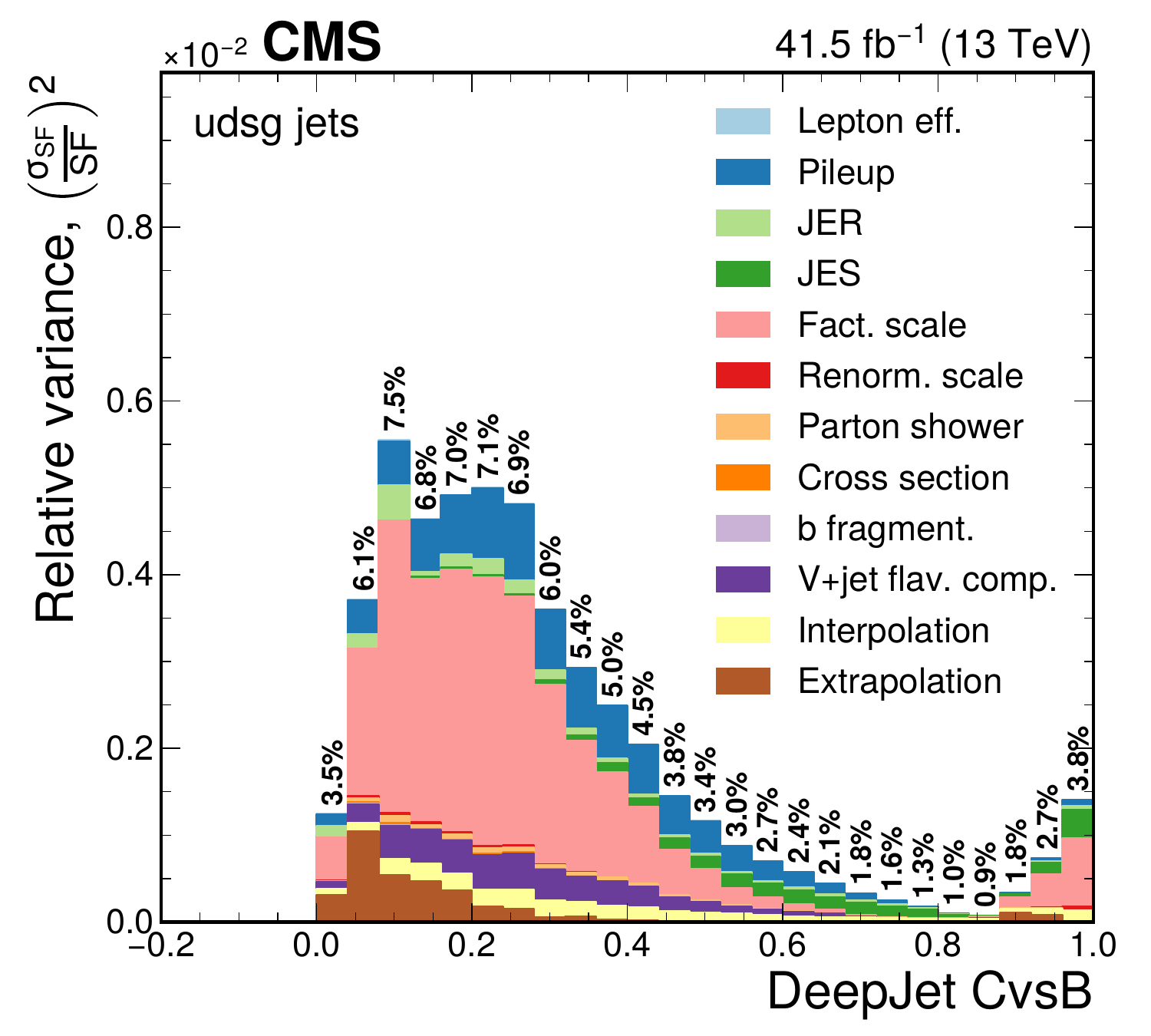}\\
	\caption{\label{fig:UncCont12}Contribution of each source of the SF uncertainty,
		calculated as the square of the relative uncertainty in the jet yield and expressed as the maximum
		of the up and down variations, at various values of the DeepJet CvsL (left) and CvsB (right)
		discriminators for {\PQc} (upper), {\PQb} (middle), and light (lower) flavours. The
		effective total relative uncertainty values ($\sqrt{\sum\left(\frac{\sigma_{\text{SF}}}{\text{SF}}\right)^2}$) per bin are also shown in bold text, for reference. The
		bin corresponding to a tagger value of $-1$ is plotted at $-0.1$. Statistical
		uncertainties are not shown.}
\end{figure}

Additionally, the effect of each source of systematic uncertainty in these SFs, when projected onto each of the
1D CvsL and CvsB distributions, is demonstrated by applying the nominal SFs
separately to {\PQc}, {\PQb}, and light-flavour jets taken from a simulated hadronically
decaying $\ttbar$ sample, and then measuring the relative changes in the jet yield
distributions by applying each systematic variation of the SFs.
The relative contributions of all systematic uncertainty sources thus projected for both the
taggers are shown in Figs.~\ref{fig:UncCont11} (DeepCSV) and \ref{fig:UncCont12} (DeepJet). 
Interpolation and extrapolation uncertainties have the highest contribution in the SFs
in case of {\PQc} jets. This is a result of the sparse binning of the nodal points in the {\PQc} jet
SFs all over the 2D plane, which, in turn, is a result of high statistical uncertainties
in the {\PW}+{\PQc} OS-SS selection that predominantly constrains the {\PQc} jet SFs.
Uncertainties in the factorisation scale
alone dominate among the light-flavour SF uncertainties, especially
in parts of the phase space having low 
yields for the corresponding flavour of jets. The uncertainties 
in the ISR and FSR in the PS have the highest
contribution to the SF uncertainties for {\PQb} jets, since these
variations are relatively larger in the $\ttbar$ selections from which
the SFs for {\PQb} jets are predominantly constrained.


\subsection{Adjusted c tagging performance in data}

\begin{figure}
\centering
\includegraphics[width=0.48\textwidth]{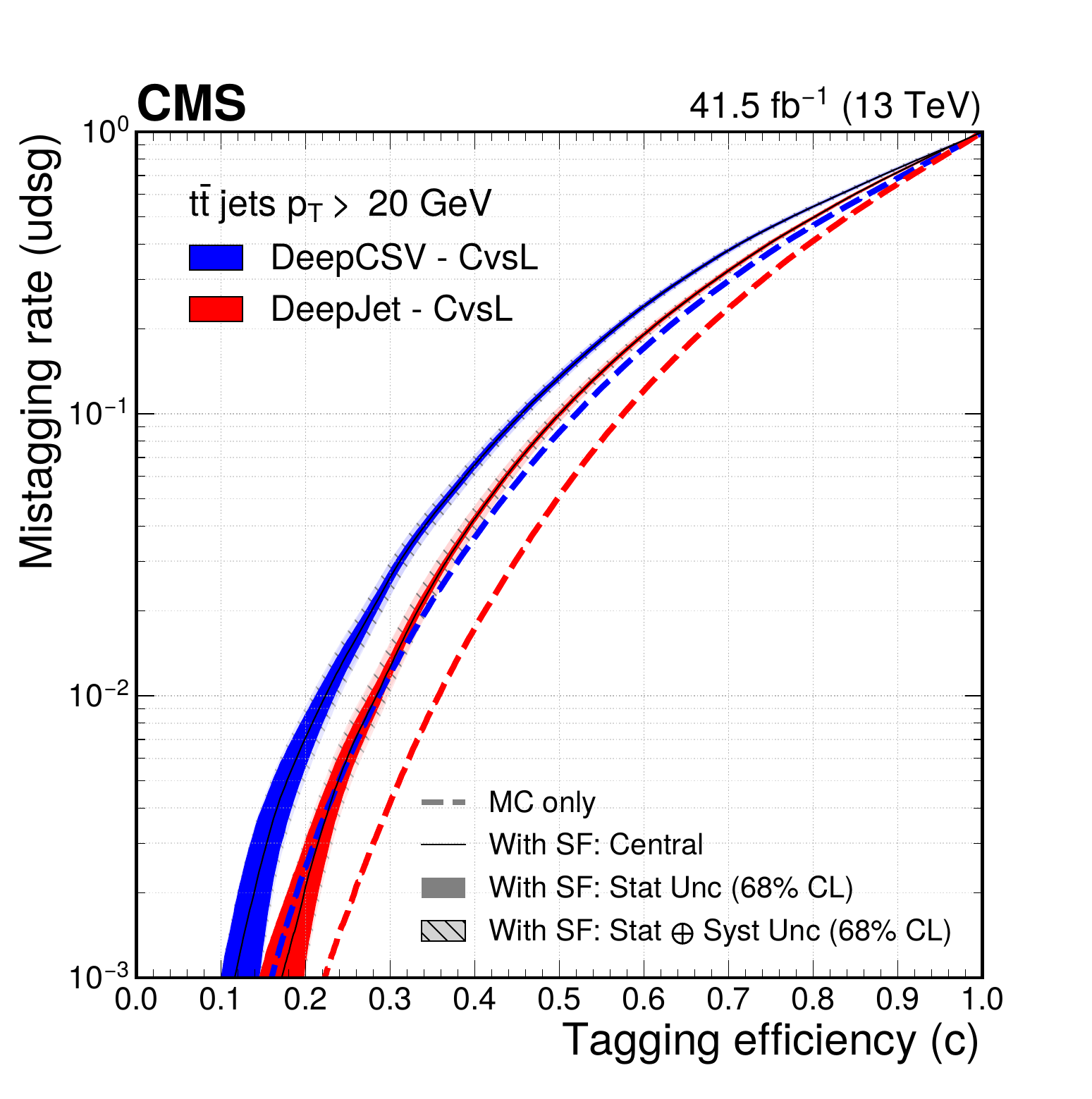} 
\includegraphics[width=0.48\textwidth]{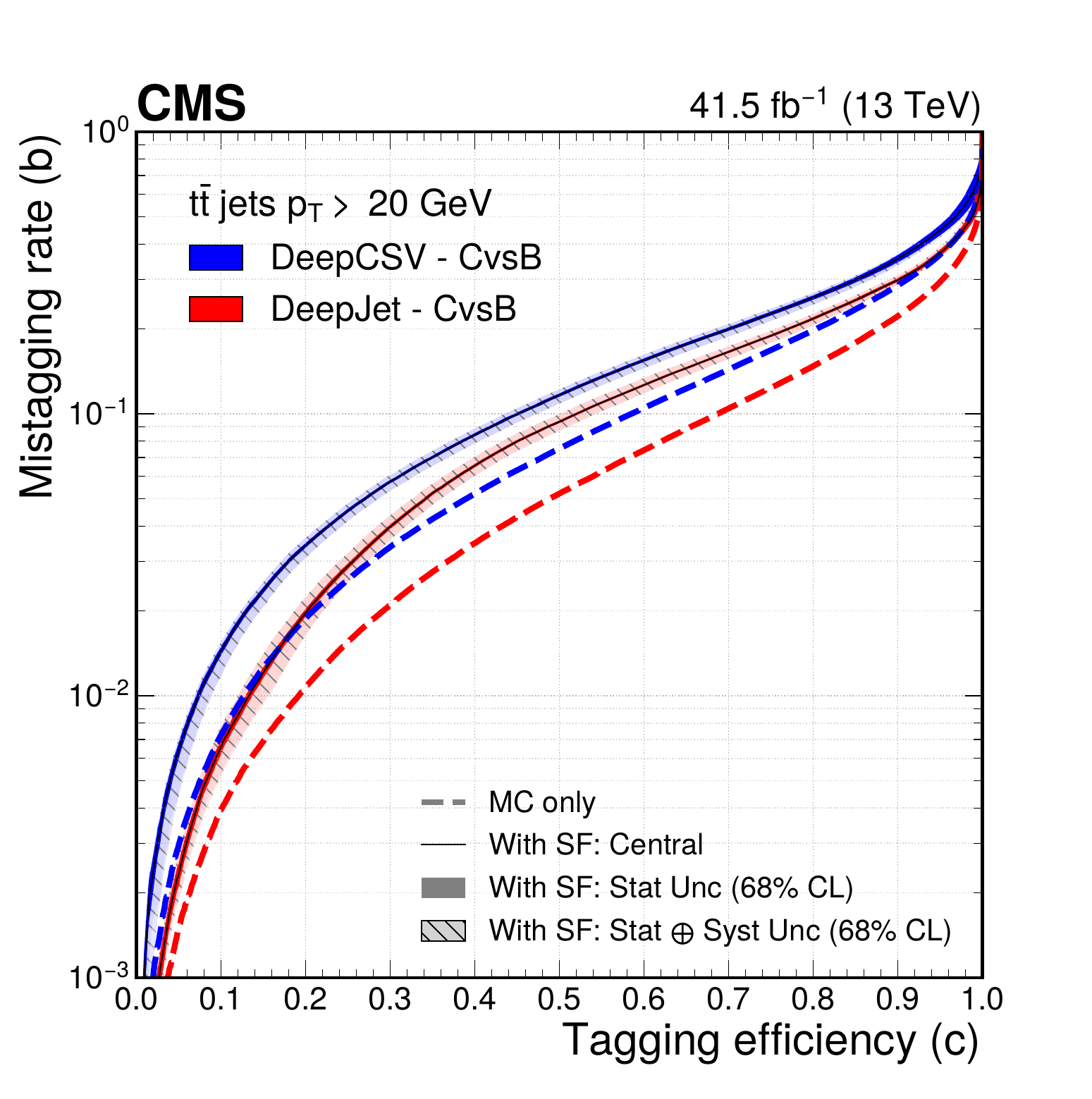} 
\caption{\label{fig:CvsXPerf}The ROC curves showing the individual performance of the CvsL (left) and CvsB (right) discriminators for the DeepCSV (blue) and DeepJet (red) algorithms
for simulated jets (the dashed lines) and the estimation of the
same for jets in data (the solid lines). The solid uncertainty bands around the solid lines represent
the statistical uncertainties in the SFs propagated to the ROCs, and the hatched semi-transparent bands represent statistical and
systematic uncertainties in the SFs propagated to the ROCs and added in quadrature.}
\end{figure}

\begin{figure}
	\centering
		\includegraphics[width=0.475\textwidth]{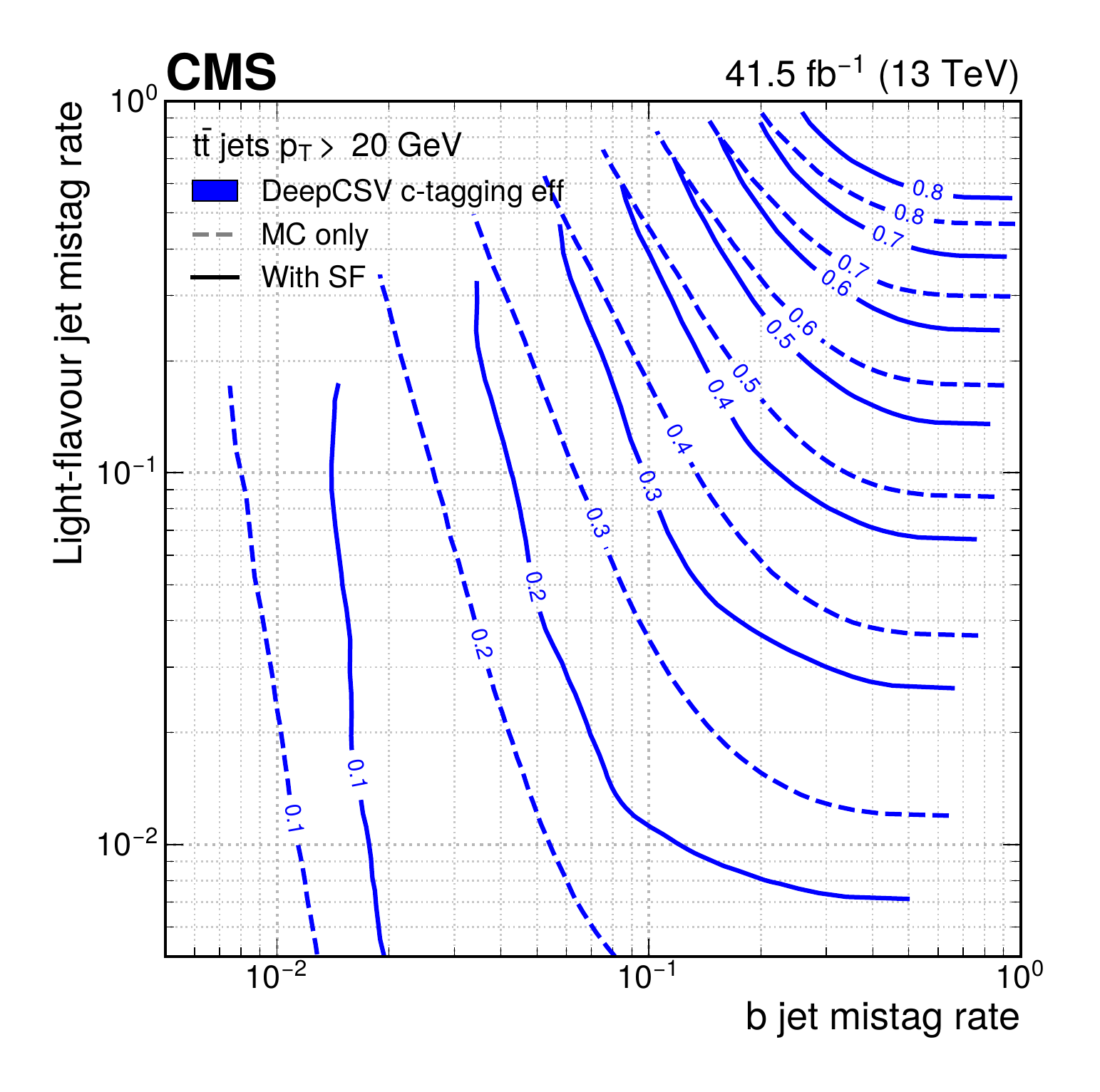} 
		\includegraphics[width=0.475\textwidth]{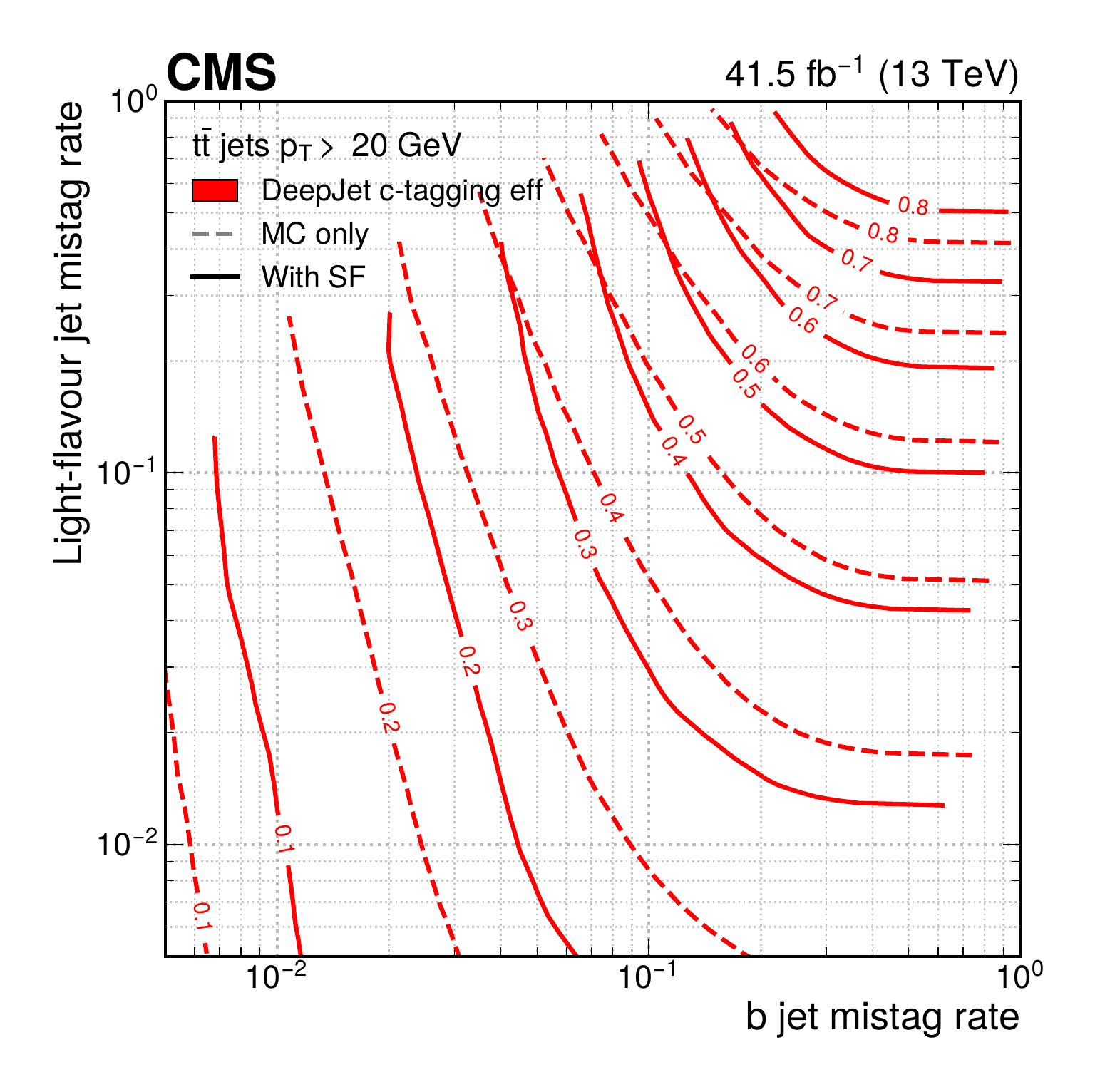} 
	\caption{\label{fig:CvsXPerf2D}The ROC contours showing {\PQc} tagging efficiencies as functions
		of {\PQb} and light-flavour jet misidentification rates, for the DeepCSV (left) and DeepJet (right) algorithms
		for simulated jets (the dashed lines) and the estimation of the
		same for jets in data (the solid lines). Each line represents points in the plane that correspond to a fixed value of the {\PQc} tagging efficiency, which is shown as a number at the centre of each line.}
\end{figure}

The {\PQc} tagging shape calibration SFs are used to adjust the
tagger discriminator distributions of simulated jets of each flavour
to obtain an estimate of the corresponding distributions
of jets in data. This is performed using an independent sample of jets
taken from simulated hadronically decaying \ttbar events and
the adjusted distributions are then used to plot the ROC curves to estimate {\PQc} tagging performance
in data. Furthermore, the statistical and systematic uncertainties
in the SFs are suitably propagated to the ROC curves. The
adjusted ROC curves for CvsL and CvsB discrimination
along with their variations because of statistical and systematic
uncertainties are shown in Fig.~\ref{fig:CvsXPerf}. In a similar fashion,
the adjusted {\PQc} tagging efficiencies, as functions of the
adjusted {\PQb} and light-flavour
jet misidentification rates are shown in Fig.~\ref{fig:CvsXPerf2D}.

\begin{figure}
\centering
\includegraphics[width=0.70\textwidth]{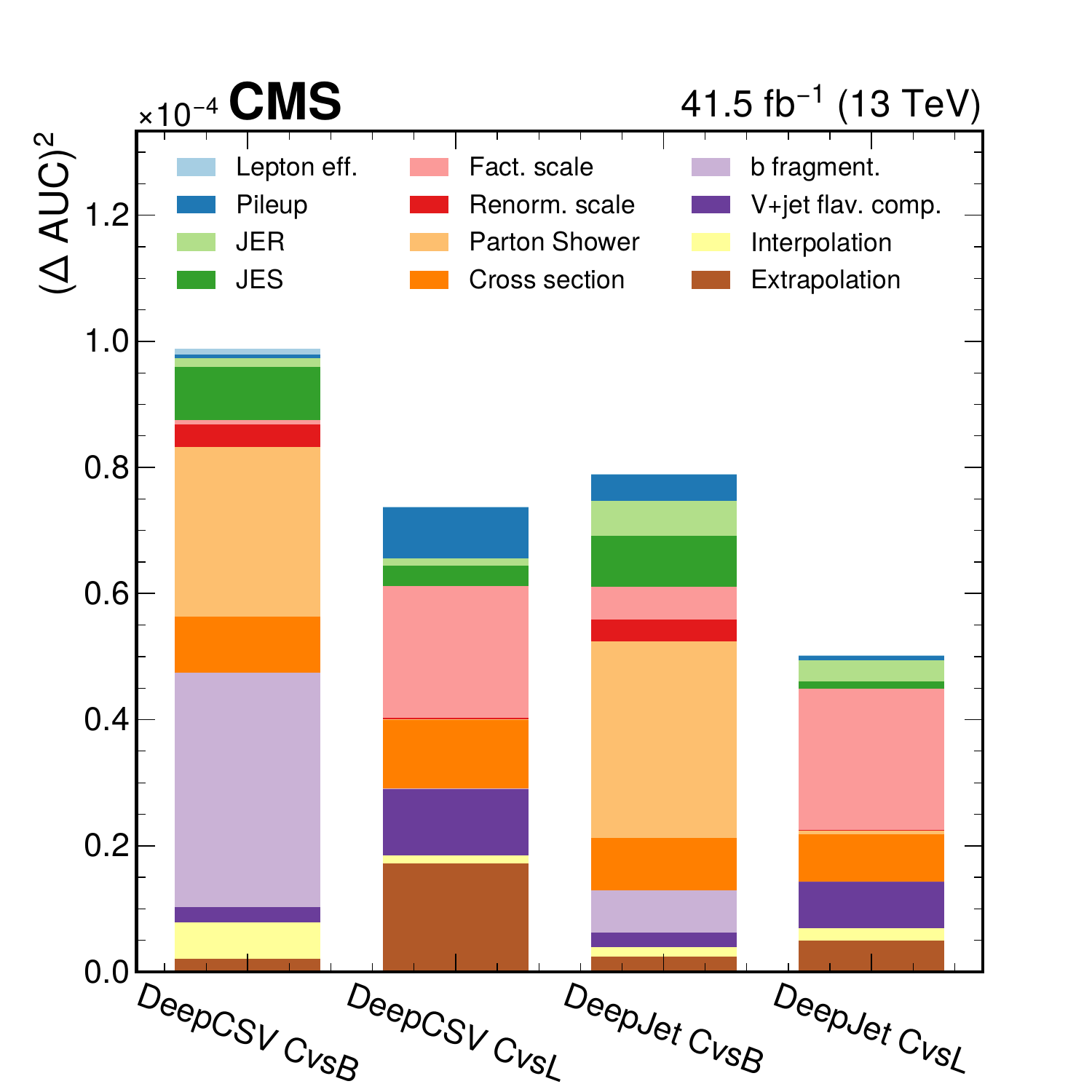} 
\caption{\label{fig:UncCont}Relative contributions of each source of uncertainty
to the total uncertainty (statistical + systematic) for both CvsL and CvsB discrimination and
for both DeepCSV and DeepJet taggers, quantified by the square of the change in area
under ROC curves.}
\end{figure}

Individual contributions from each source of uncertainty are also quantified
by the change in the area under the ROC curve from that
of the central ROC curve. The relative contributions, including that of statistical
uncertainties, evaluated using the ROC curve variations
as described here are presented graphically in Fig.~\ref{fig:UncCont}.
The approach of presenting uncertainties as a function of discriminator values in 
Figs.~\ref{fig:UncCont11} and \ref{fig:UncCont12} is essentially different
from this interpretation, because the former disregards the relative abundance of jets
at different discriminator values, whereas the latter includes the absolute
change in the jet yields in different parts of the phase space and could therefore be more representative of
the corresponding effect in a physics analysis.
In this approach, uncertainties in the factorisation scale are the dominant
contributions to the overall uncertainties in CvsL discrimination, whereas
uncertainties in {\PQb} fragmentation and in the ISR and FSR in the PS are among
the highest contributors to the uncertainties in the CvsB discriminator, owing to
their large contributions to the {\PQb} jet SFs.

\section{Validation\label{sec:Validation}}

\subsection{Closure test\label{subsec:closure}}

\begin{figure}[p!]
	\centering
		\includegraphics[width=0.435\textwidth]{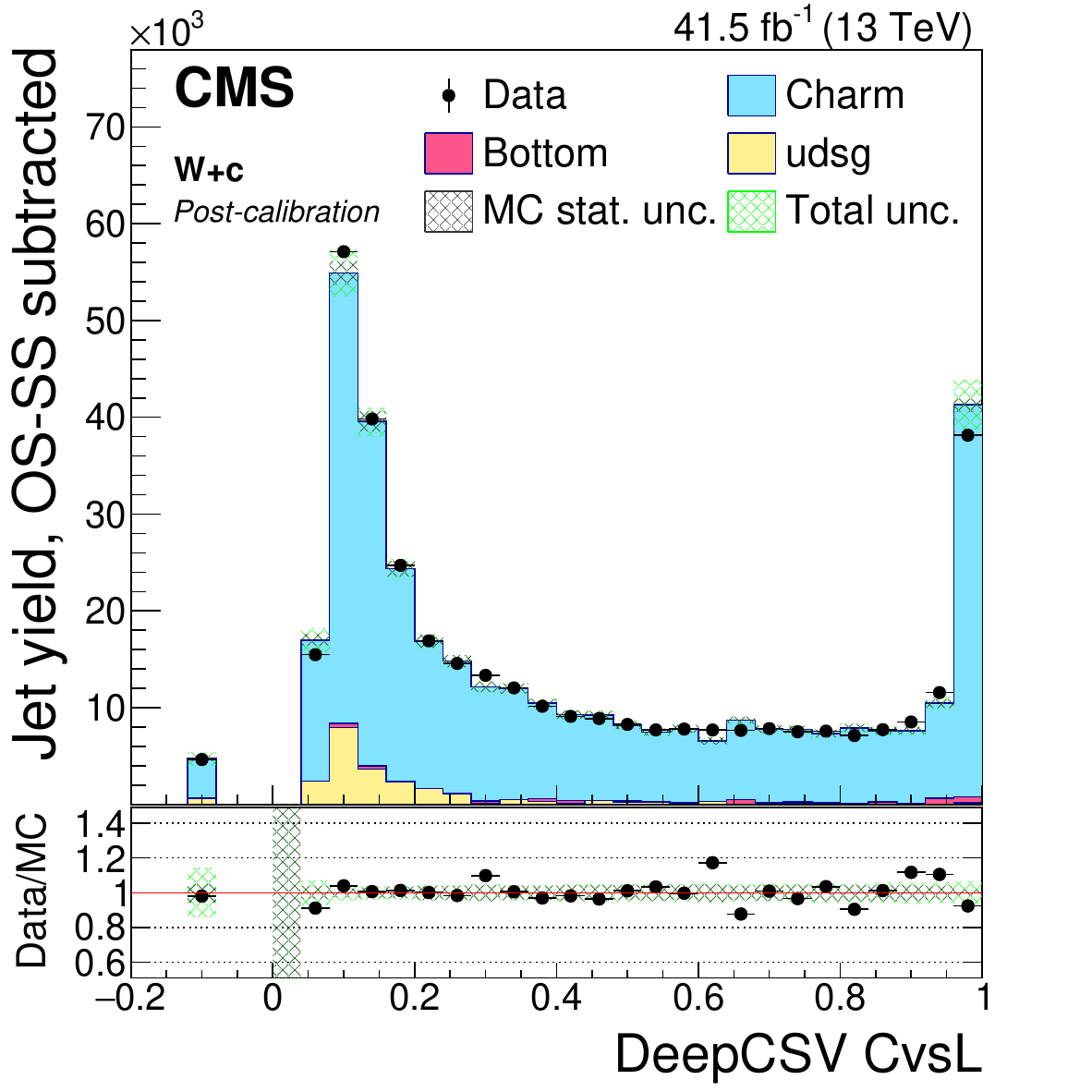}
    	        \includegraphics[width=0.435\textwidth]{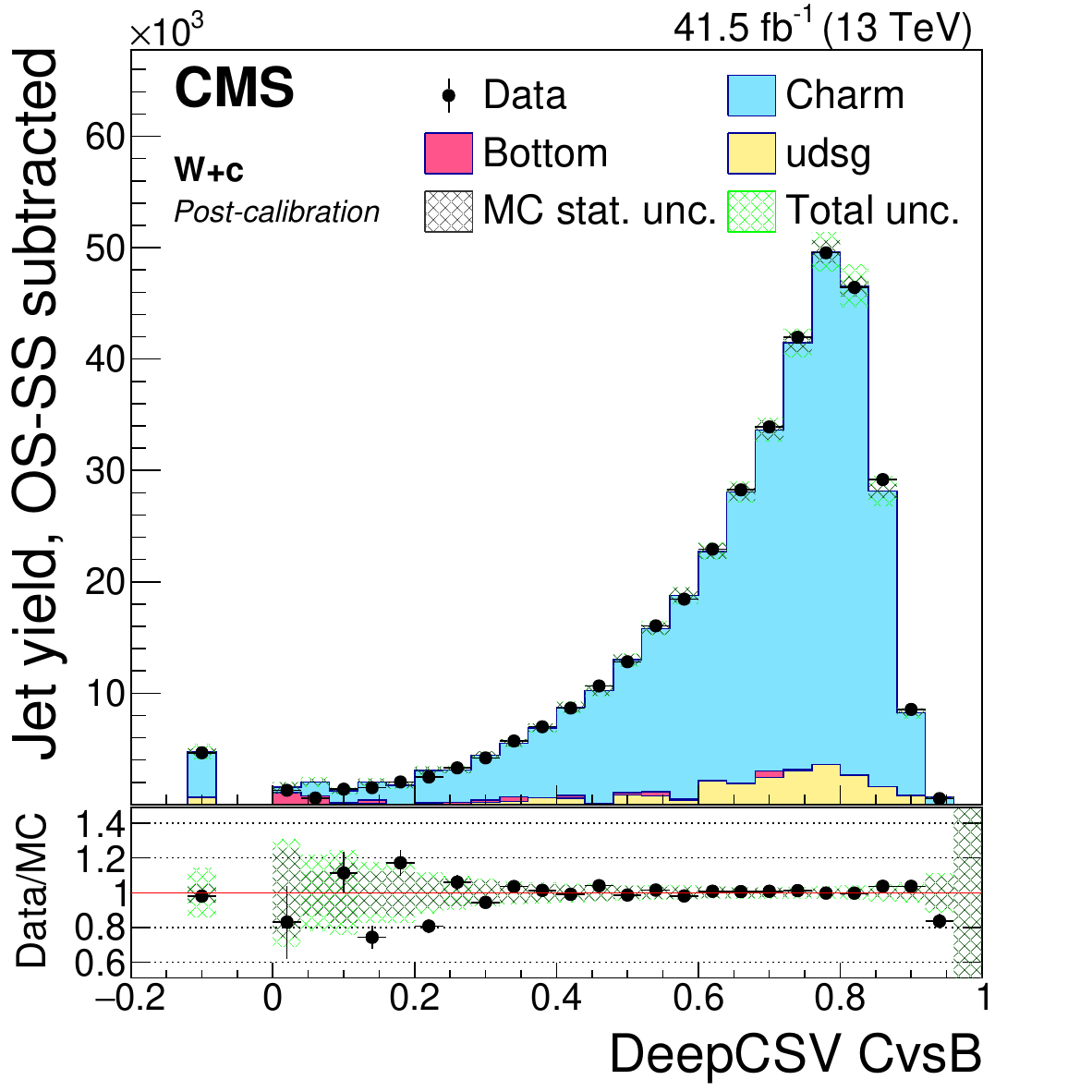}\\
		\includegraphics[width=0.435\textwidth]{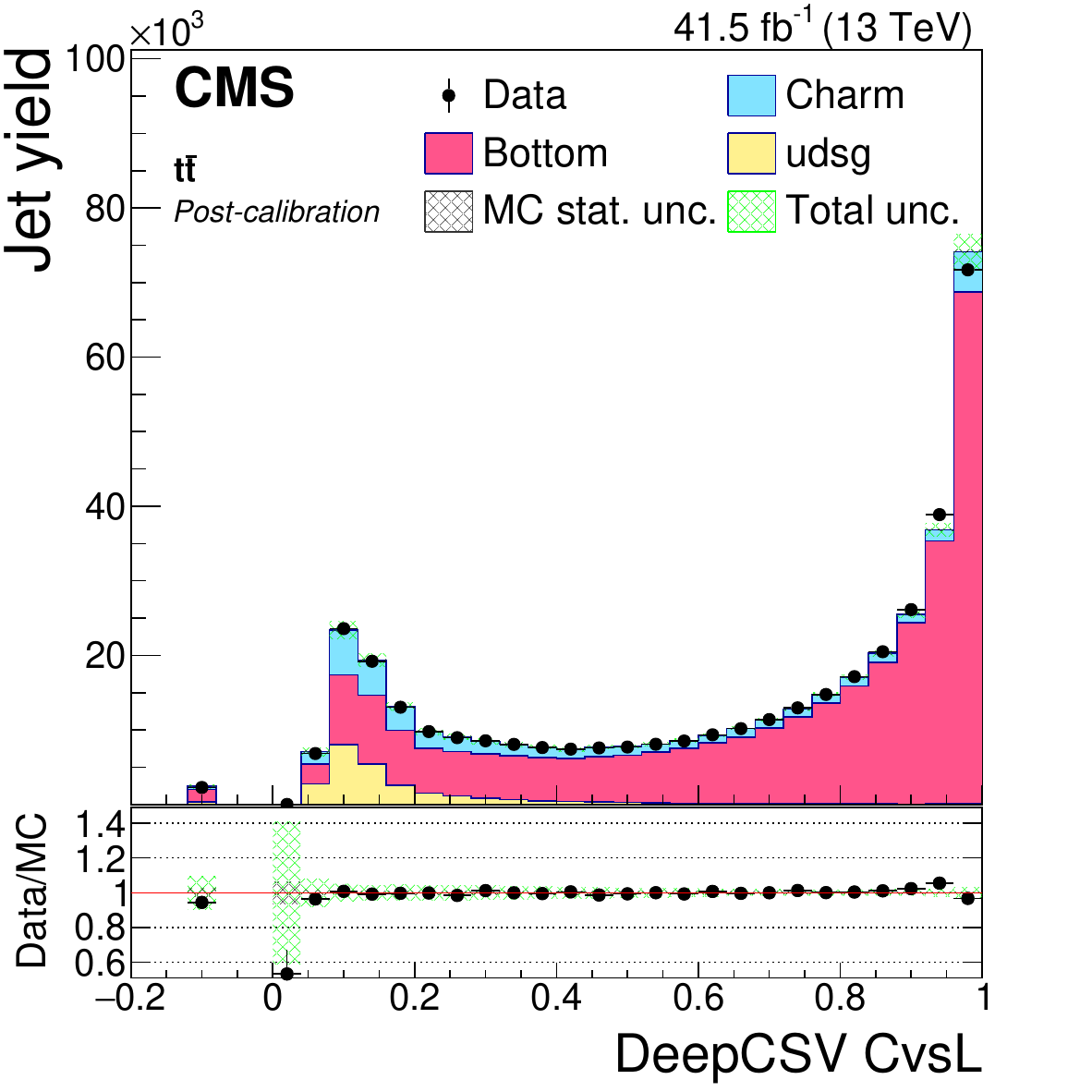}
		\includegraphics[width=0.435\textwidth]{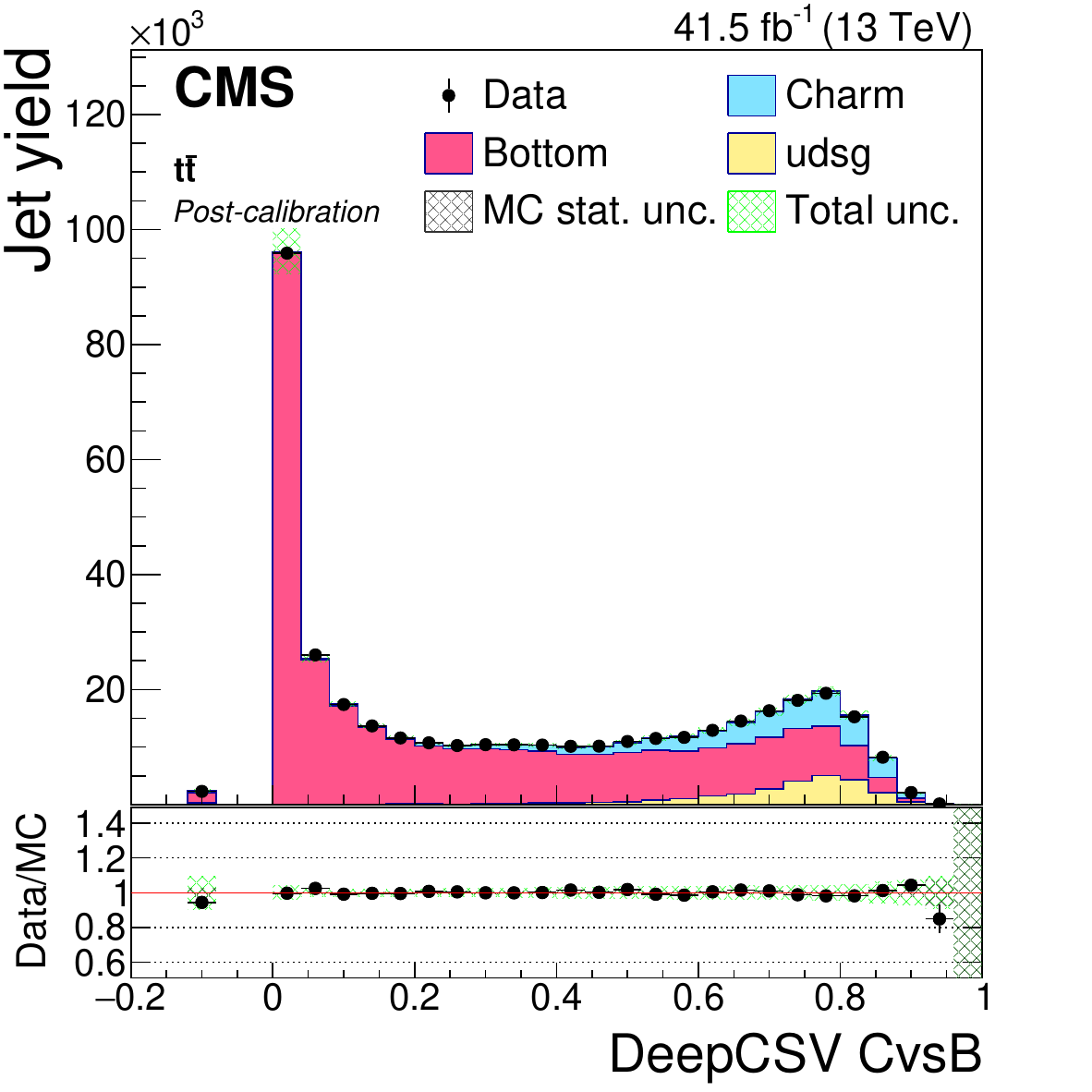}\\
		\includegraphics[width=0.435\textwidth]{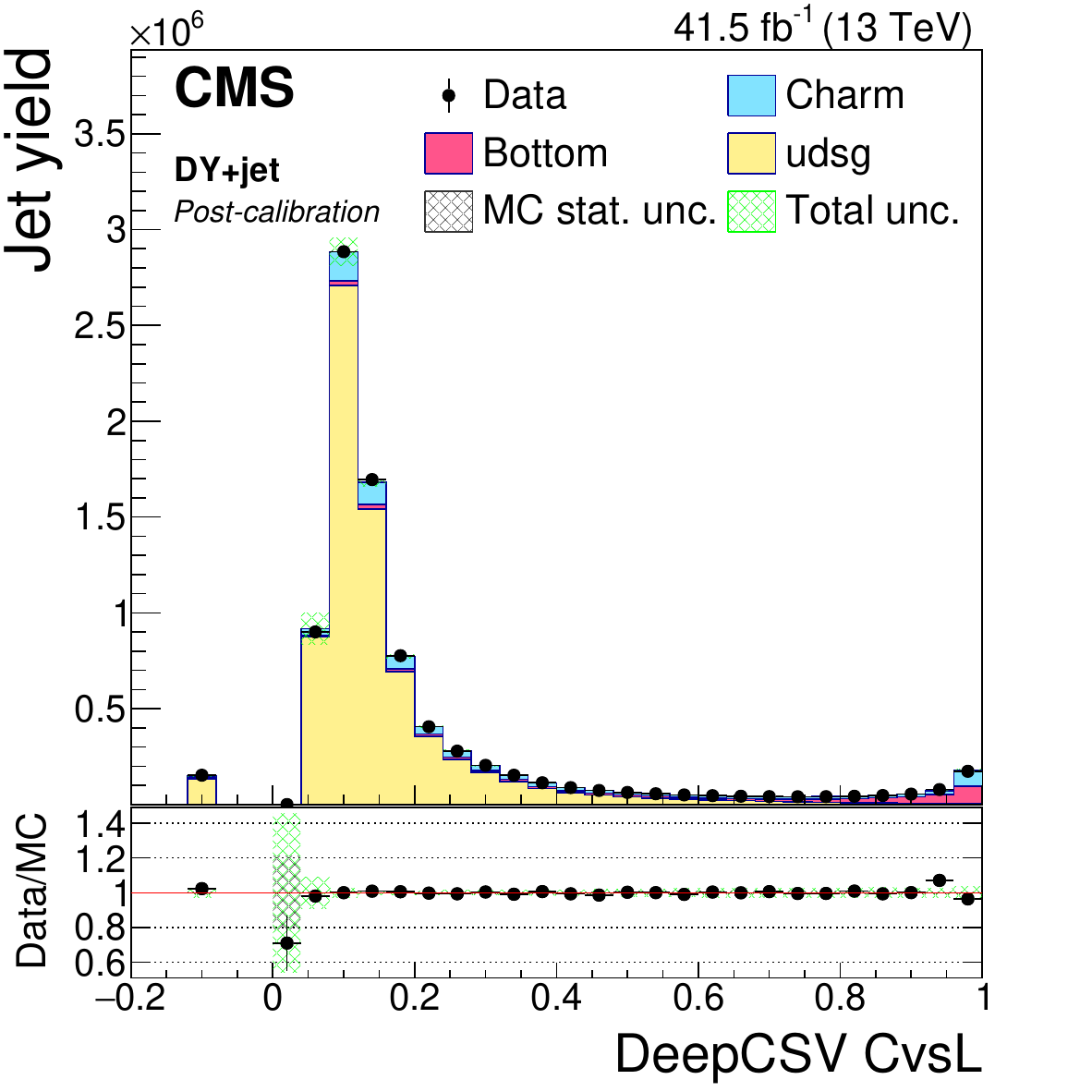}
		\includegraphics[width=0.435\textwidth]{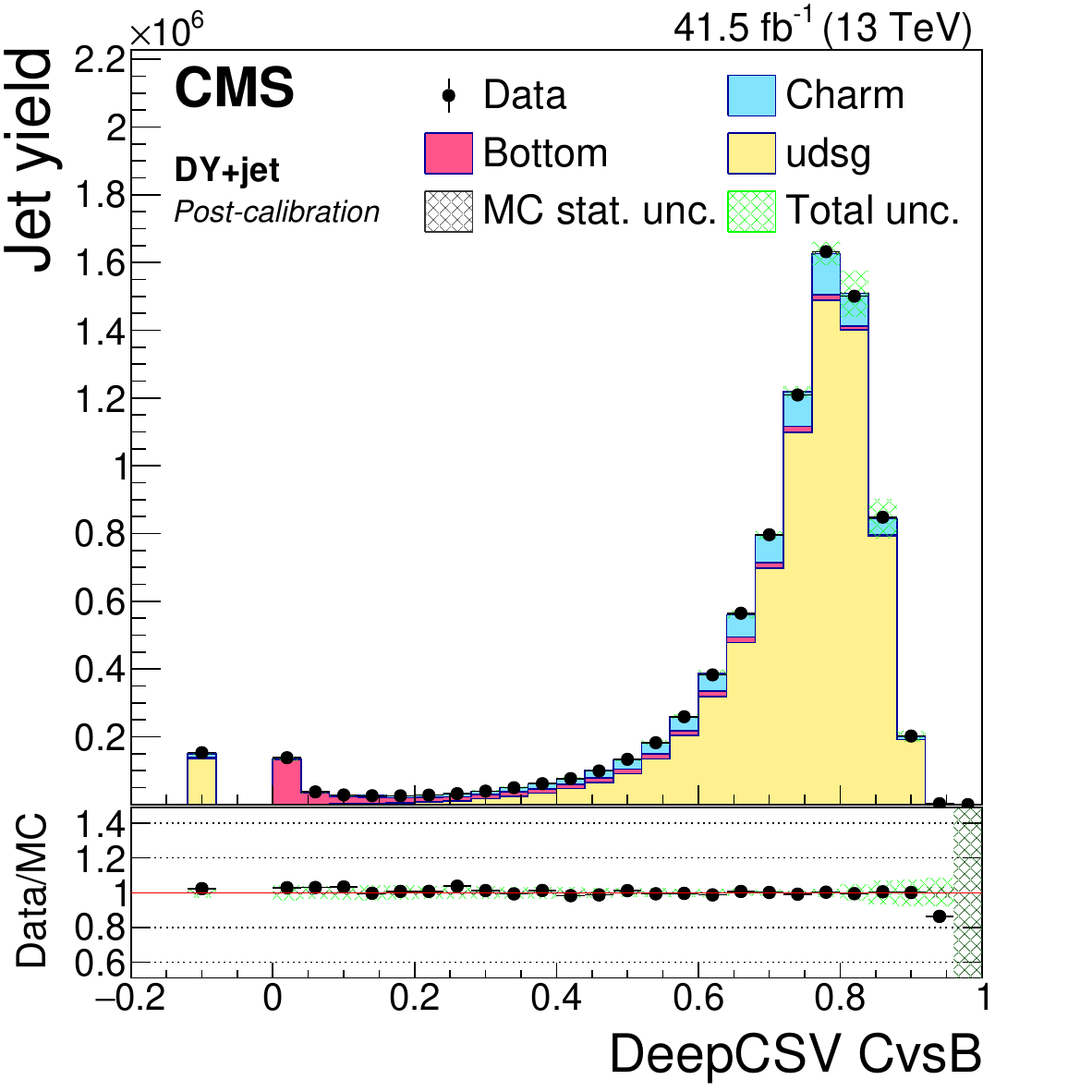}\\
	\centering{}\caption{\label{fig:AfterSF1}Post-calibration DeepCSV CvsL (left)
		and CvsB (right) distributions of jet samples
		selected from {\PW}+{\PQc} (upper), \ttbar semi- and dileptonic (middle), and $\text{DY}+\text{jet}$
		(lower) events after application of DeepCSV {\PQc} tagger shape calibration SFs.
		The bin corresponding to a tagger value of $-1$ is plotted at $-0.1$.
		Vertical error bars in data represent statistical uncertainties in data. The simulations are shown as stacked histograms.}
\end{figure}

\begin{figure}[p!]
	\centering
		\includegraphics[width=0.435\textwidth]{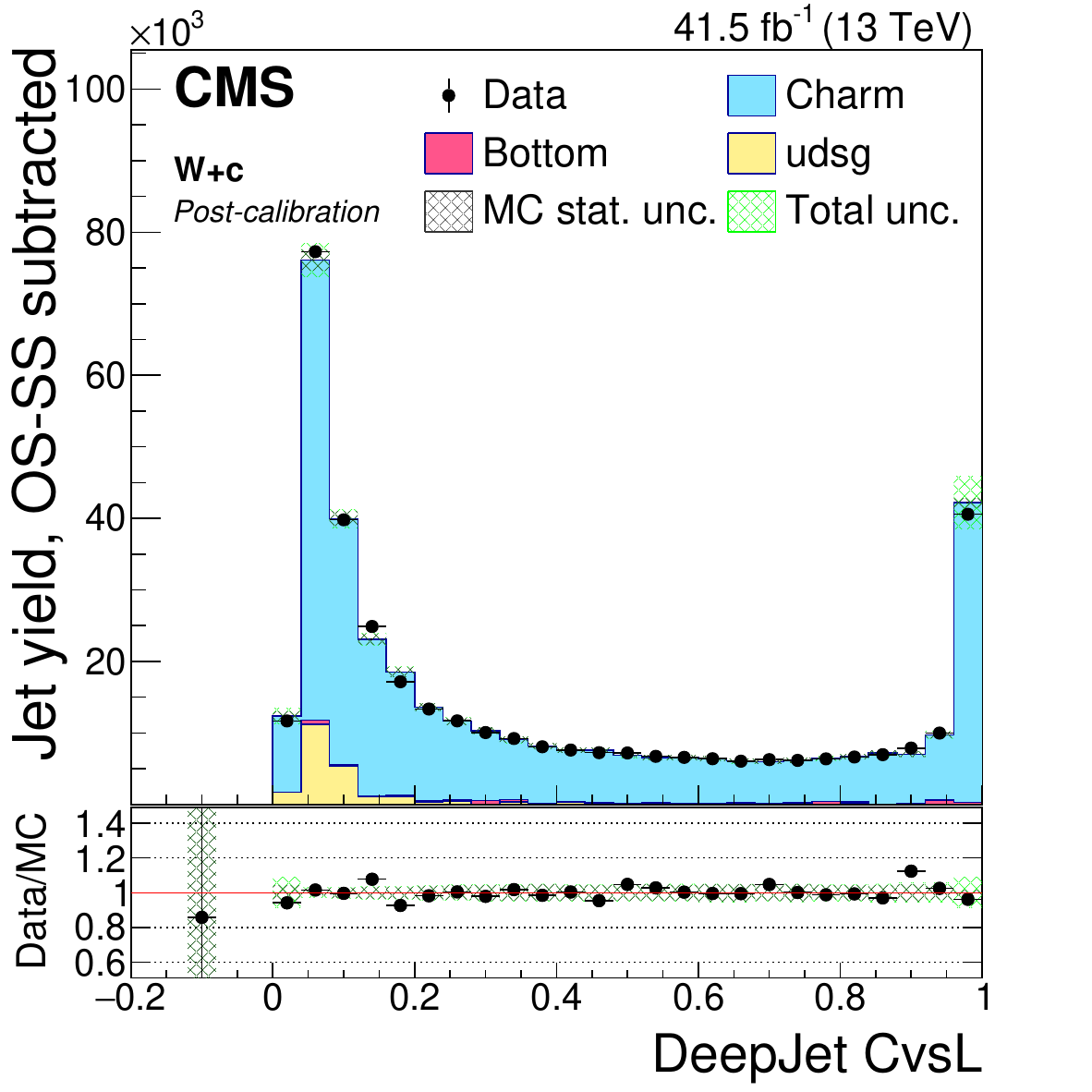}
		\includegraphics[width=0.435\textwidth]{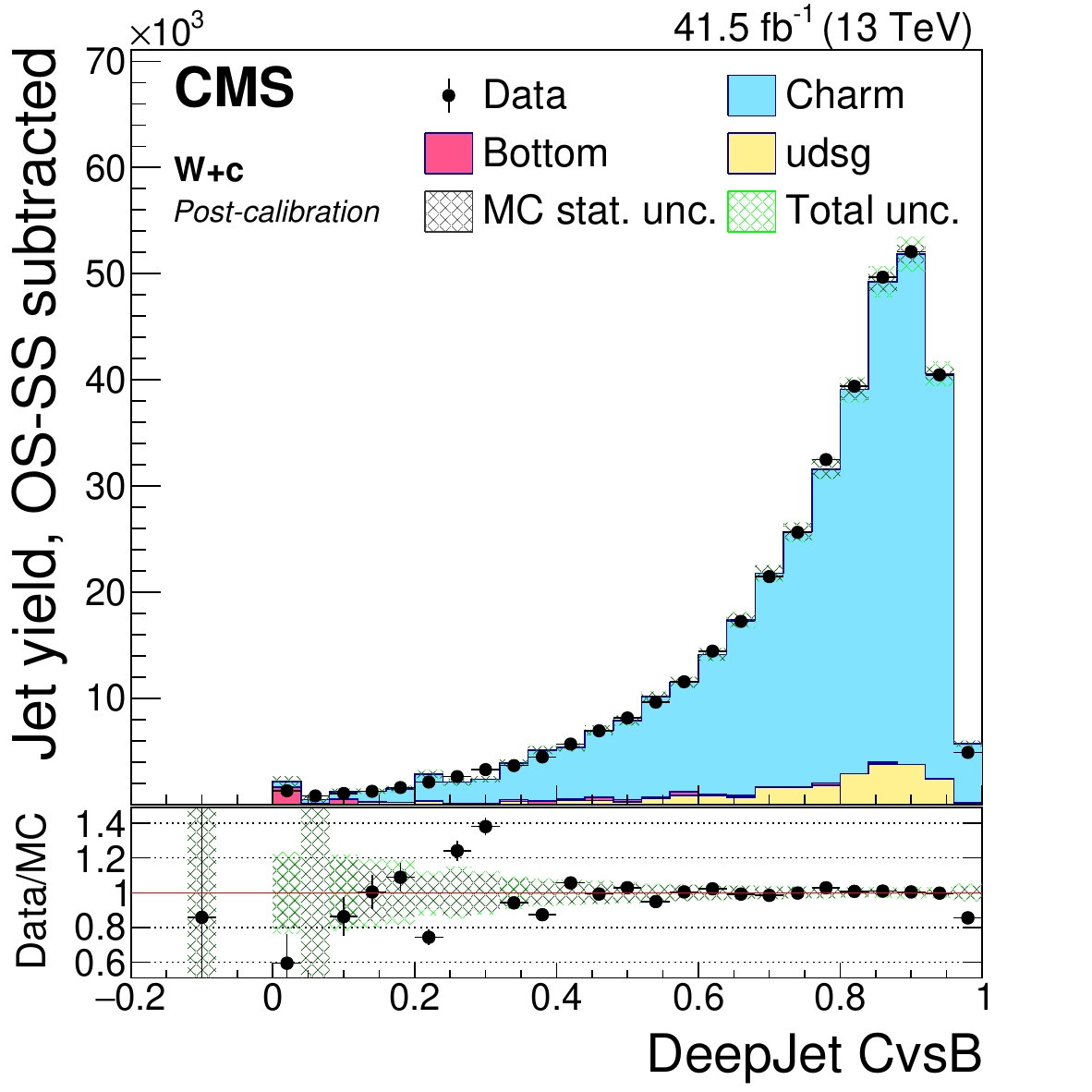}\\
		\includegraphics[width=0.435\textwidth]{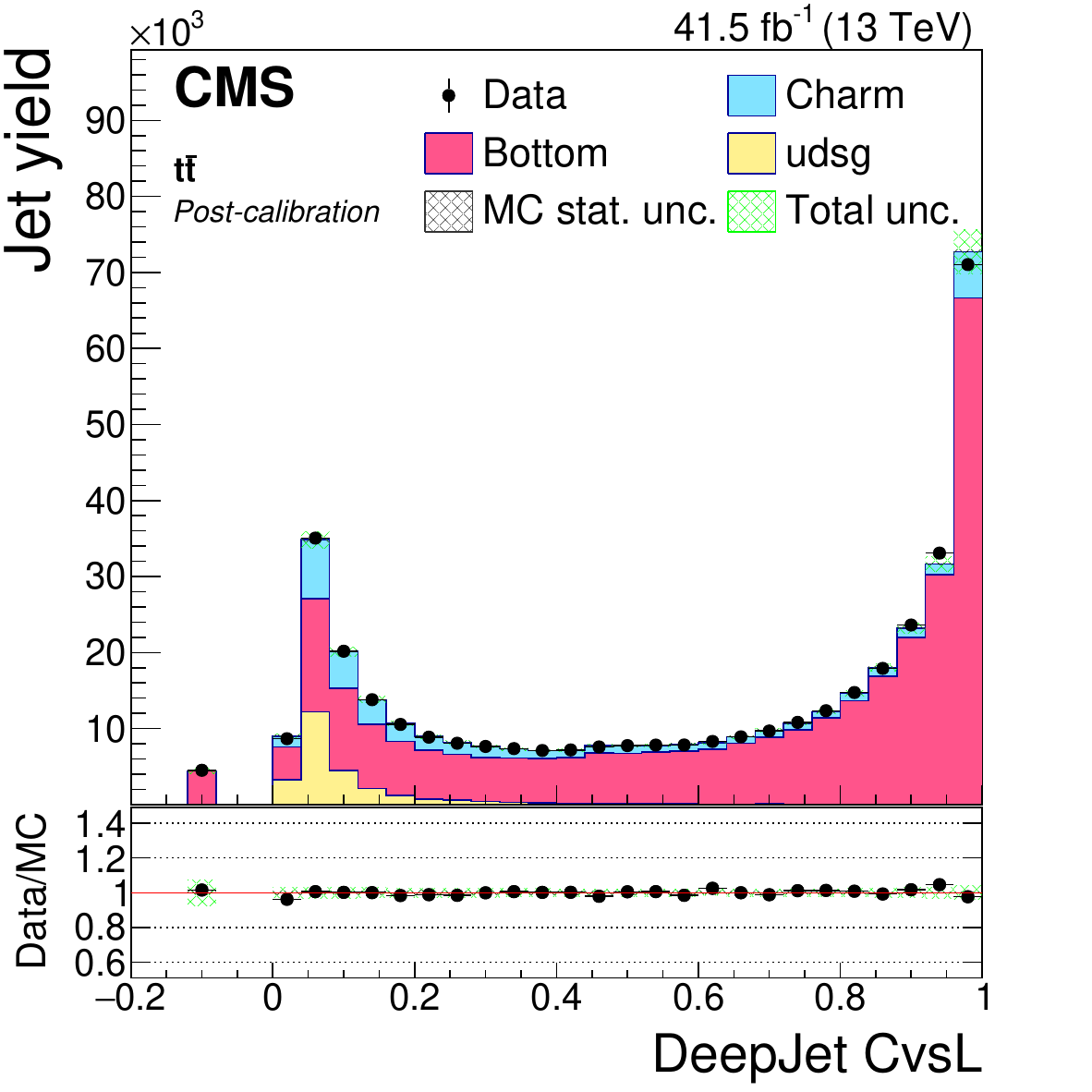}
		\includegraphics[width=0.435\textwidth]{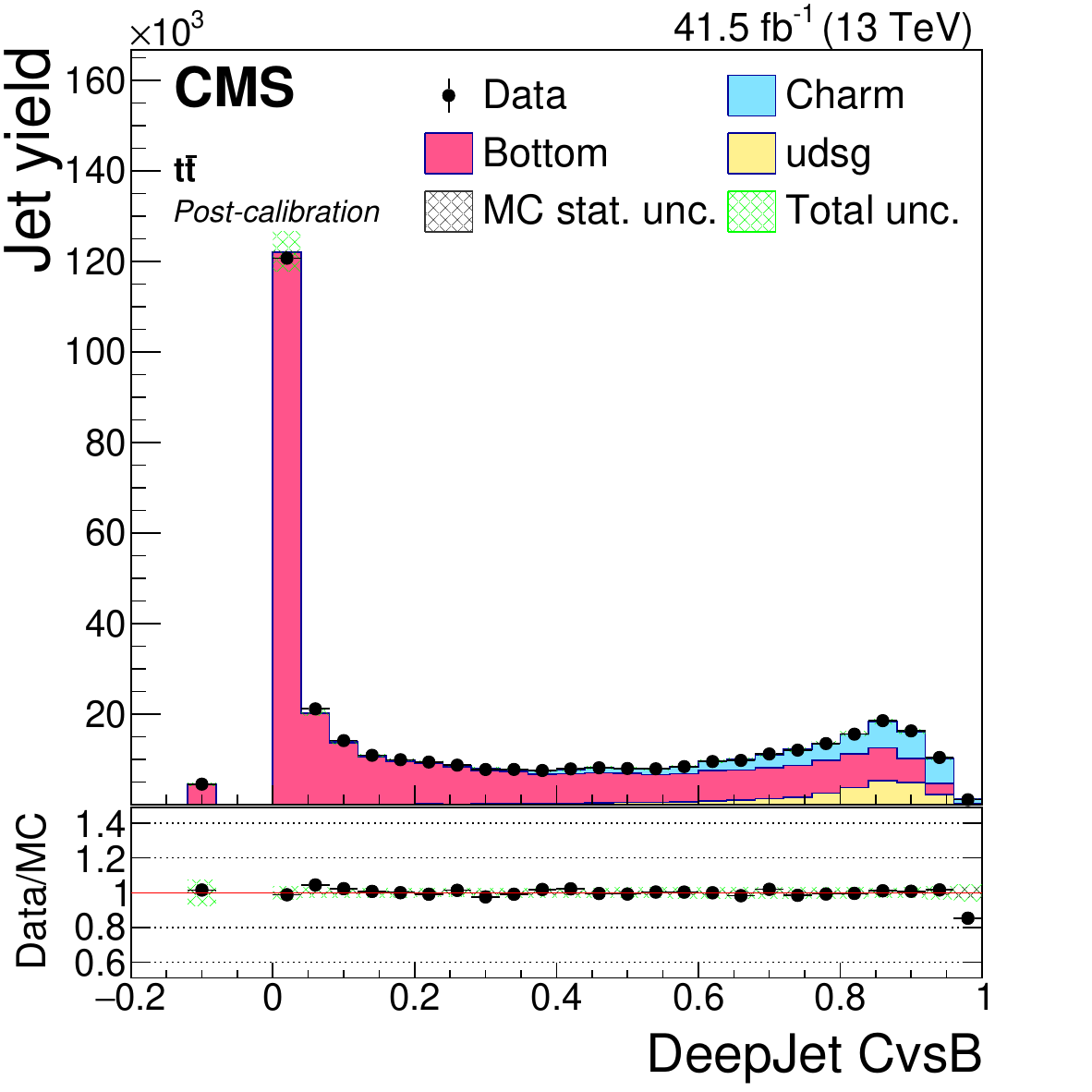}\\
		\includegraphics[width=0.435\textwidth]{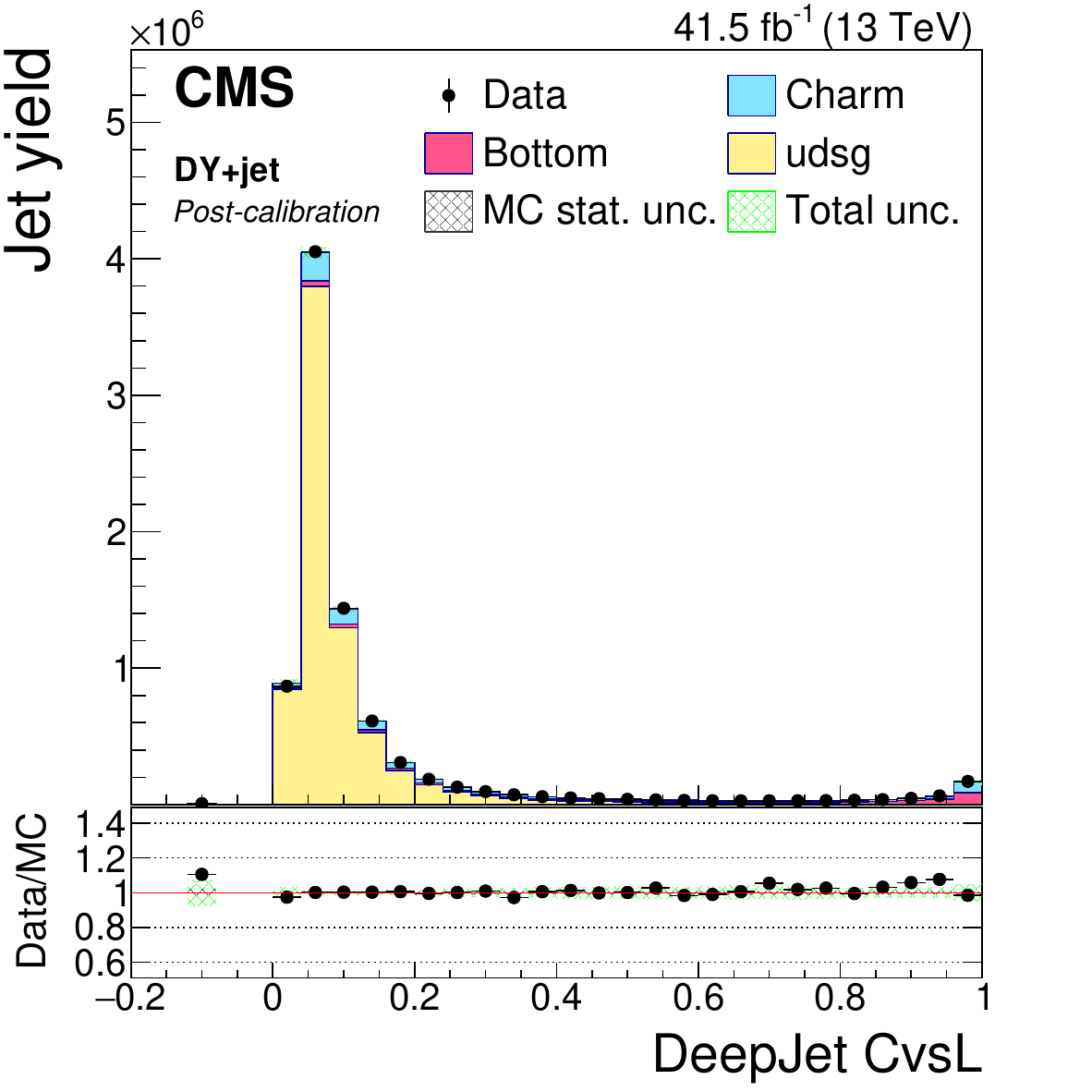}
		\includegraphics[width=0.435\textwidth]{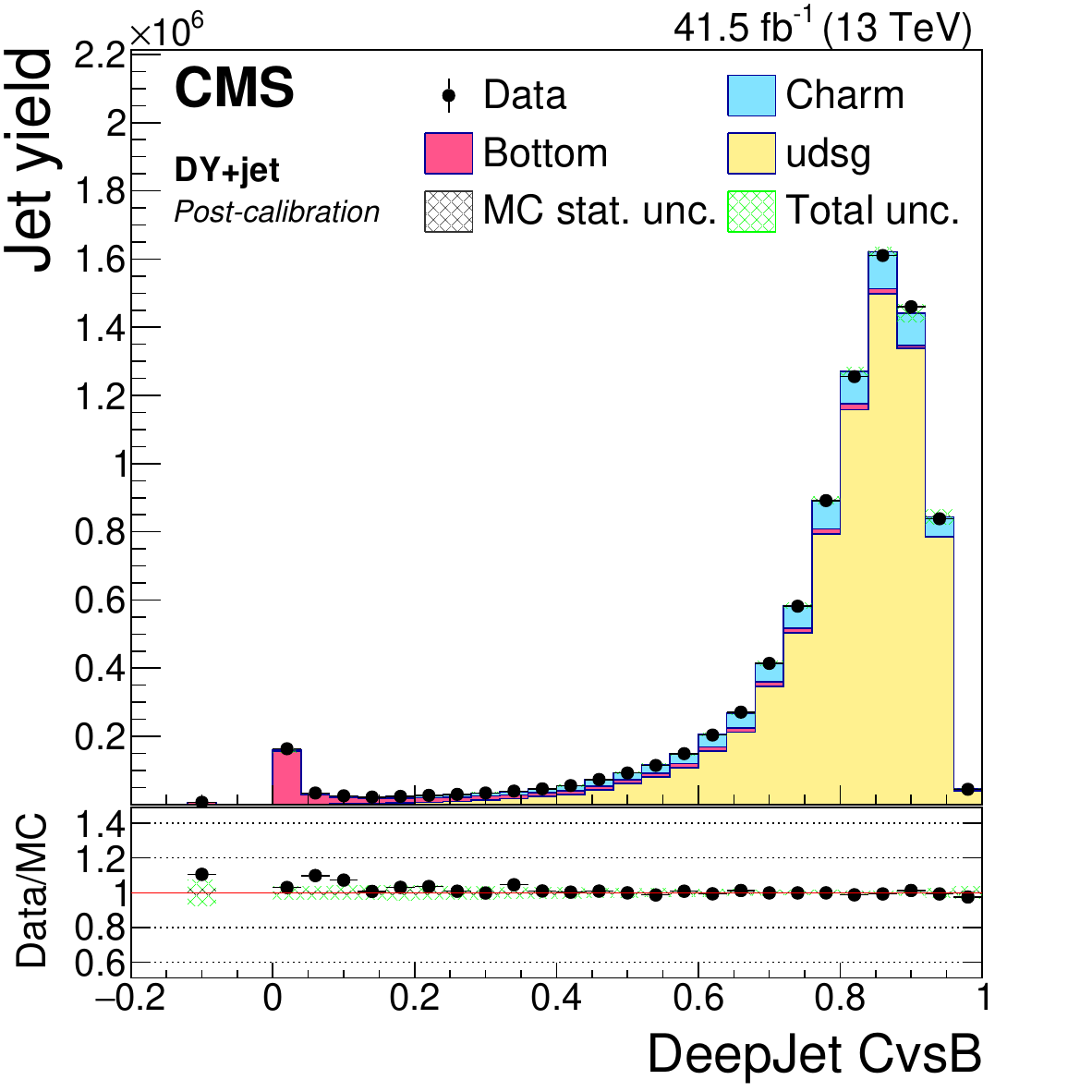}\\
    \caption{\label{fig:AfterSF2}Post-calibration DeepJet CvsL (left)
		and CvsB (right) distributions of jet samples
		selected from {\PW}+{\PQc} (upper), \ttbar semi- and dileptonic (middle), and $\text{DY}+\text{jet}$
		(lower) events after application of DeepJet {\PQc} tagger shape calibration SFs.
	The bin corresponding to a tagger value of $-1$ is plotted at $-0.1$.
	Vertical error bars in data represent statistical uncertainties in data. The simulations are shown as stacked histograms.}
\end{figure}

A closure test is performed by applying the derived shape calibration 
SFs to the CvsL and CvsB distributions of the same jet selections
for which the SFs were derived. The effect of applying the
DeepCSV (DeepJet) SFs on the DeepCSV (DeepJet)
{\PQc} tagger distributions is shown in Fig.~\ref{fig:AfterSF1}
(\ref{fig:AfterSF2}). A good agreement between the {\PQc} tagging 
discriminator distributions of simulated jets and those of jets
in data establishes the closure of this method.

While the derivation of correction factors as well as this closure test were performed inclusively in jet \pt, we have additionally verified that the same inclusively-derived corrections result in a good agreement between the discriminator distributions in data and simulation in each case when applied to jets in exclusive ranges of jet \pt in between 20--30\GeV, 30--50\GeV, 50--80\GeV, and greater than 80\GeV. This indicates that there is no strong kinematical dependence of these corrections.

\subsection{Validation on jets not biased with muons\label{subsec:MuBias}}

Two additional tests are performed to demonstrate applicability of the SFs,
which have been derived using {\PQb} and {\PQc} jets containing a soft
muon inside them, when applied to jet samples that do not necessarily
contain a soft muon. To construct a sample of jets enriched in
{\PQb} jets without using soft muons to identify them, the same
\ttbar dileptonic selection is used, where the 
jet with the highest \pt that was not tagged with a soft muon in it is treated
as the ``second'' {\PQb} jet candidate.

Similarly, a second sample relatively enriched in {\PQc} jets is constructed
from the \ttbar semileptonic selection already described in Section~\ref{SemittSel}.
All possible triplets of jets in the events are considered to reconstruct the hadronically decaying {\PW} boson candidate,
and hence, the top quark candidate. For each combination, the first two jets are considered as the {\PQc} and {\PQs} candidates to reconstruct the {\PW} boson, whereas the third jet is considered as the {\PQb} jet candidate. All combinations where the jet tagged with a soft muon is selected as the {\PQc} jet or {\PQs} jet candidate are rejected. For all other combinations, a mass-based $\chi^2$ is defined as

\begin{equation*}
\chi^2 = \left(\frac{m_\PW-80.3}{20}\right)^2 + \left(\frac{m_\PQt-172.5}{30}\right)^2,
\end{equation*}

with $m_\PW$ and $m_\PQt$ as the invariant masses of the reconstructed {\PW} boson and top quark
candidates, respectively, in GeV. The combination of jets that yields the lowest value of $\chi^2$
for a given event is considered the best combination, and the highest-\pt jet assigned to the
hadronically decaying {\PW} boson candidate is considered the {\PQc} jet candidate.

Each jet in both these samples may or may not contain a soft muon,
and the proportion of jets with and without a soft muon should be 
distributed as it would be in any inclusive selection of jets with
similar event selections and flavour composition. The first sample is expected to contain mainly {\PQb} jets
and light-flavour jets. Since light-flavour jet SFs are already known to be free from any bias from soft muons,
a validation for this sample would demonstrate applicability of the
derived SFs on {\PQb} jets that are selected without any requirement related to the presence of a soft muon.
The second sample, on the other hand, is expected to contain {\PQc} jets with significant
contamination from {\PQb} and light-flavour jets. A validation using this sample, in combination with the conclusion from the aforementioned validation on {\PQb} jets, would demonstrate the applicability of the SFs for all {\PQc} jets as well.

\begin{figure}[p!]
	\centering
		\includegraphics[width=0.32\textwidth]{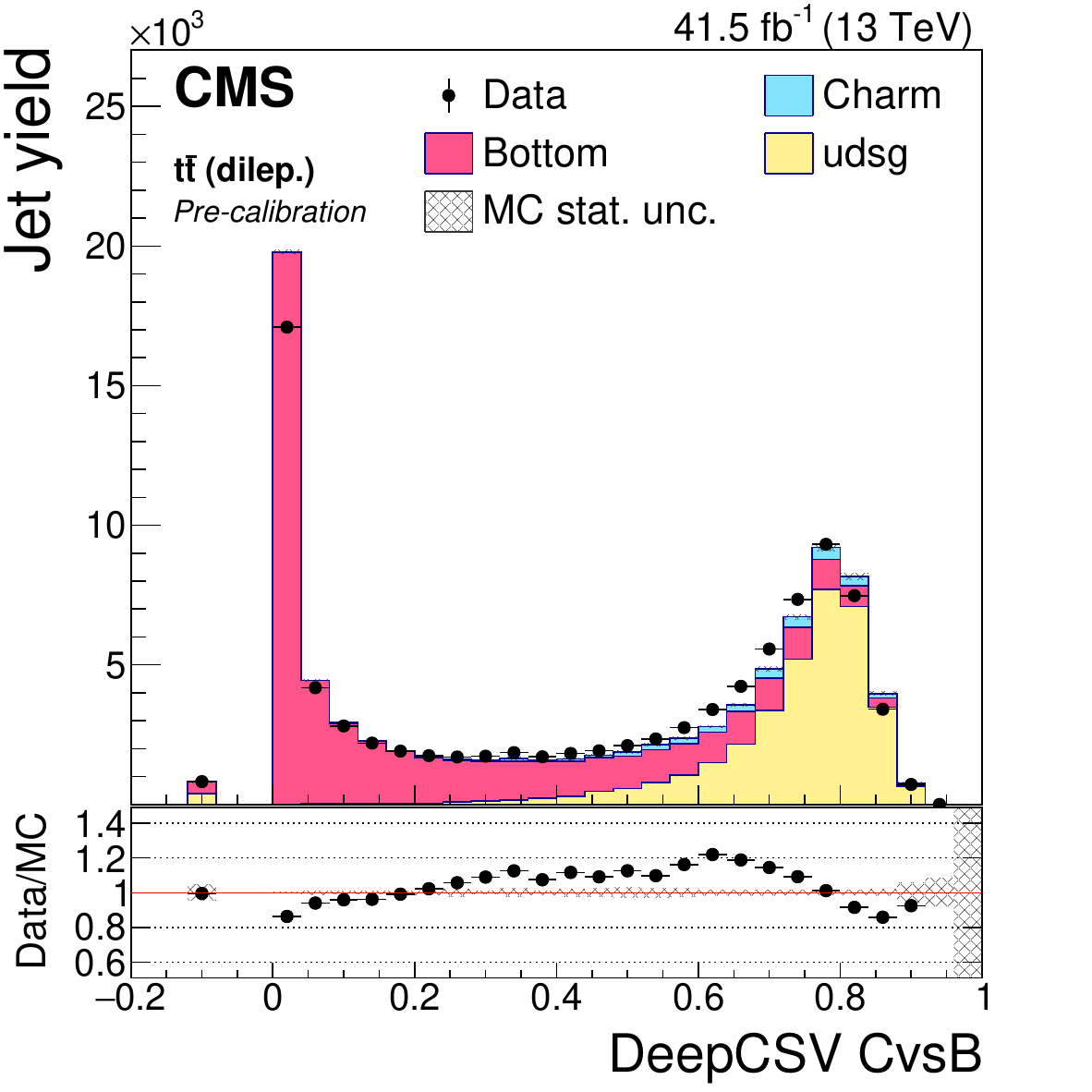}
                \includegraphics[width=0.32\textwidth]{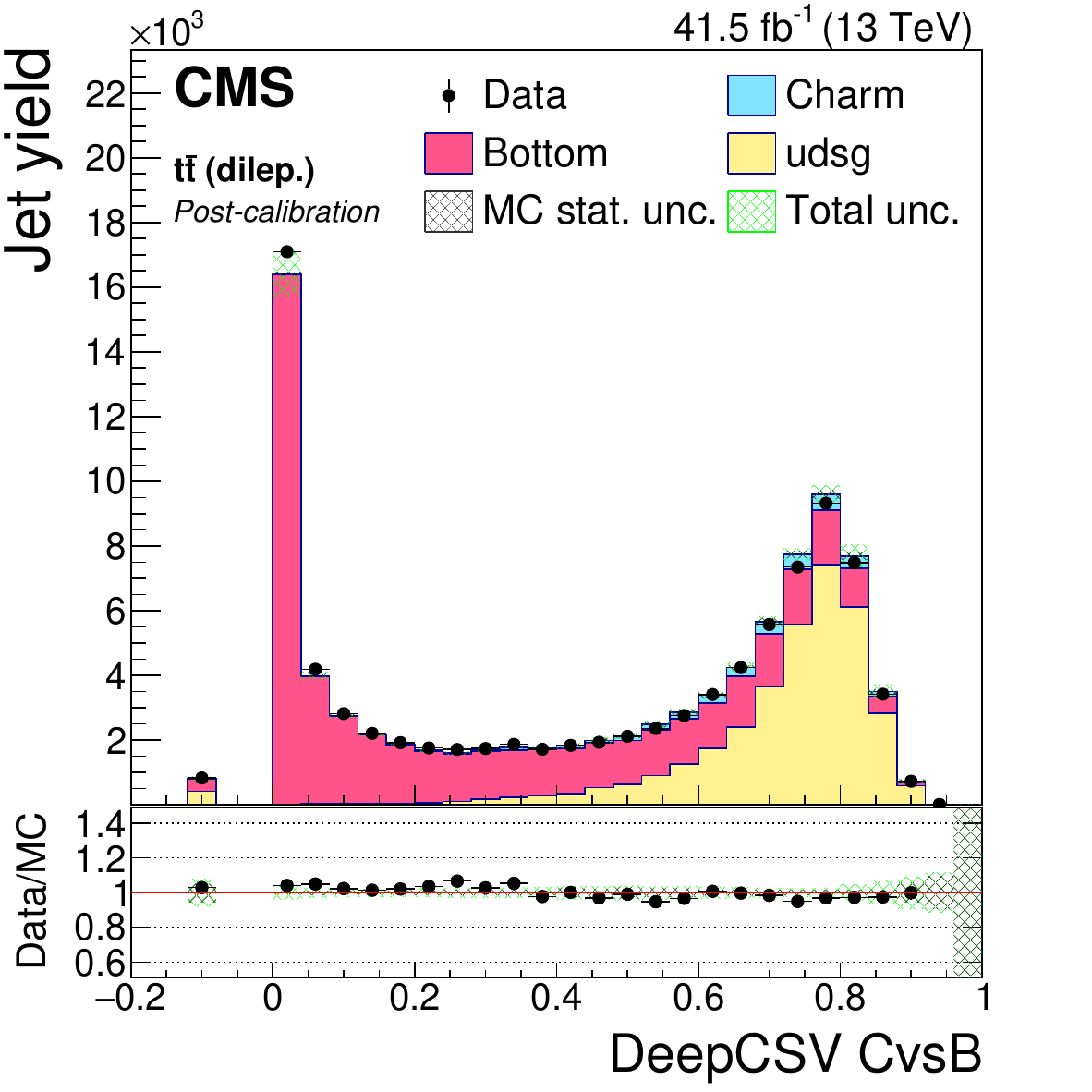}\\
		\includegraphics[width=0.32\textwidth]{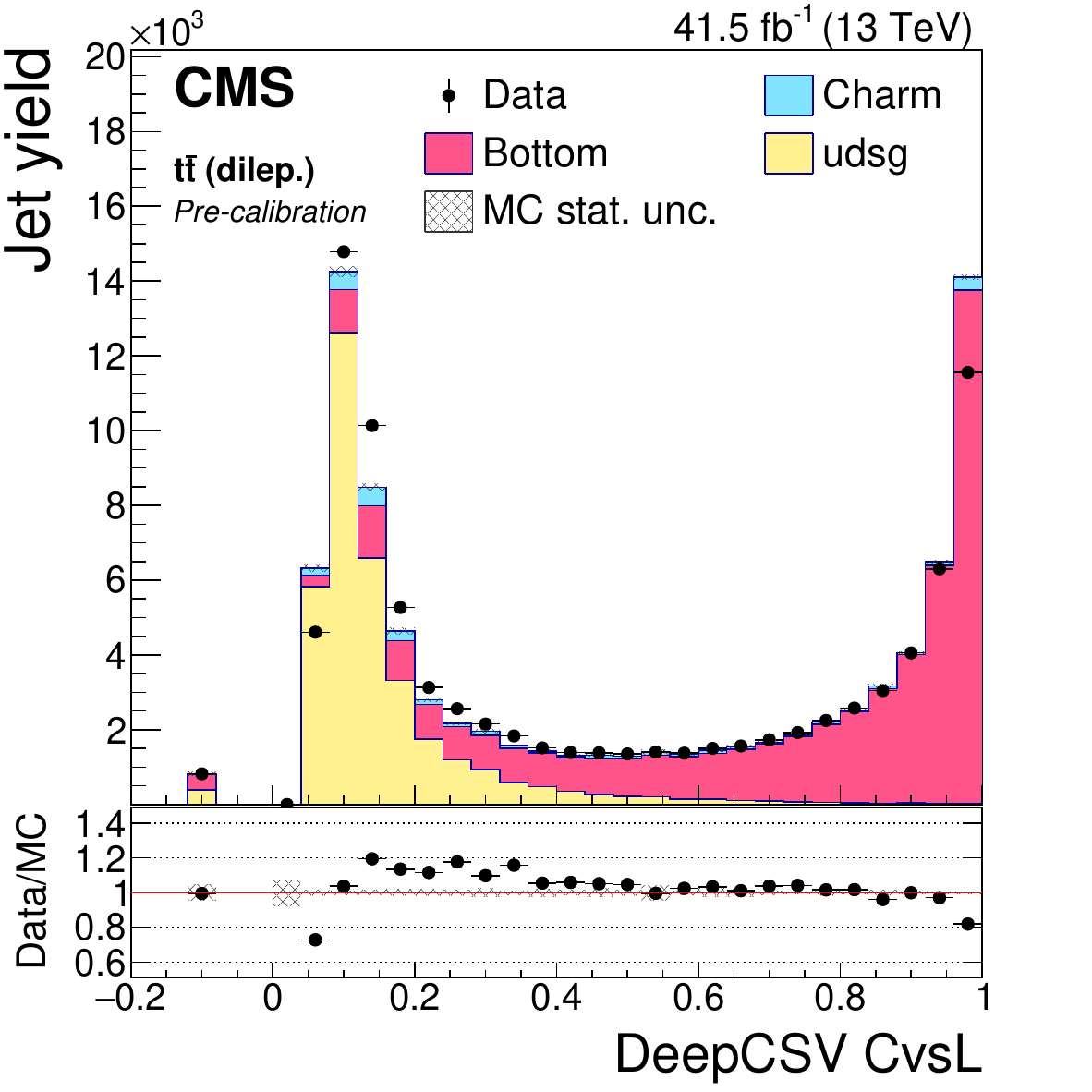}
                \includegraphics[width=0.32\textwidth]{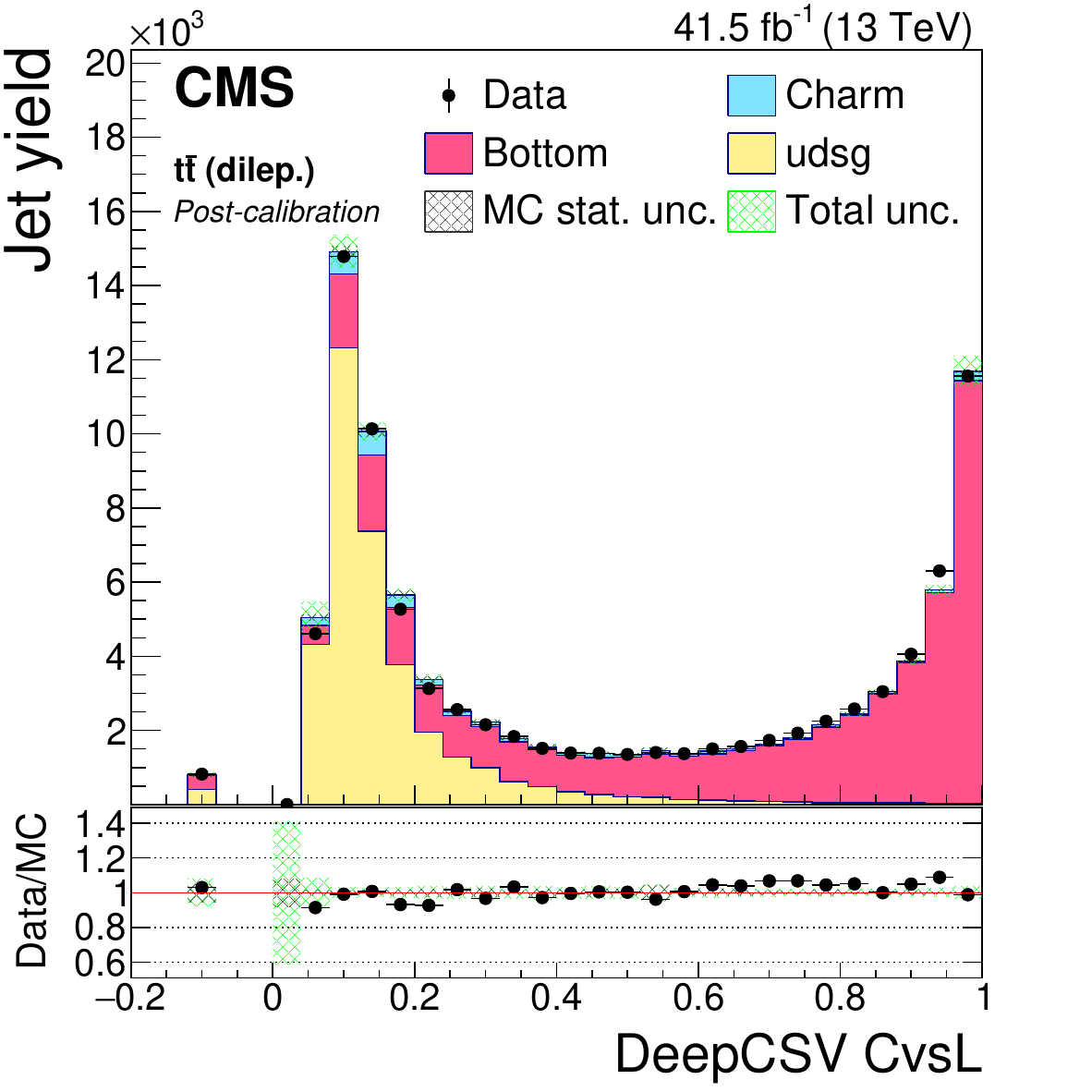}\\
		\includegraphics[width=0.32\textwidth]{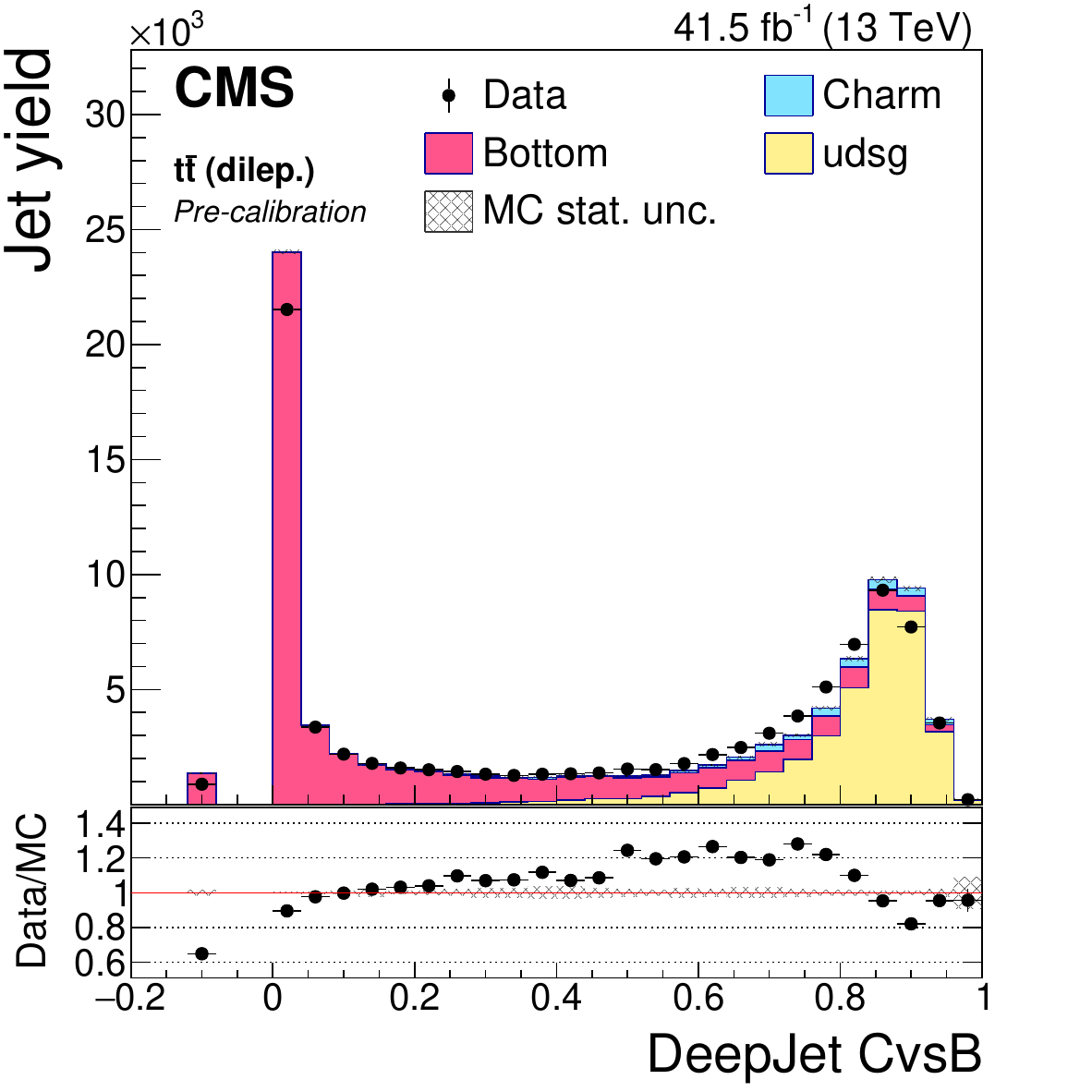}
                \includegraphics[width=0.32\textwidth]{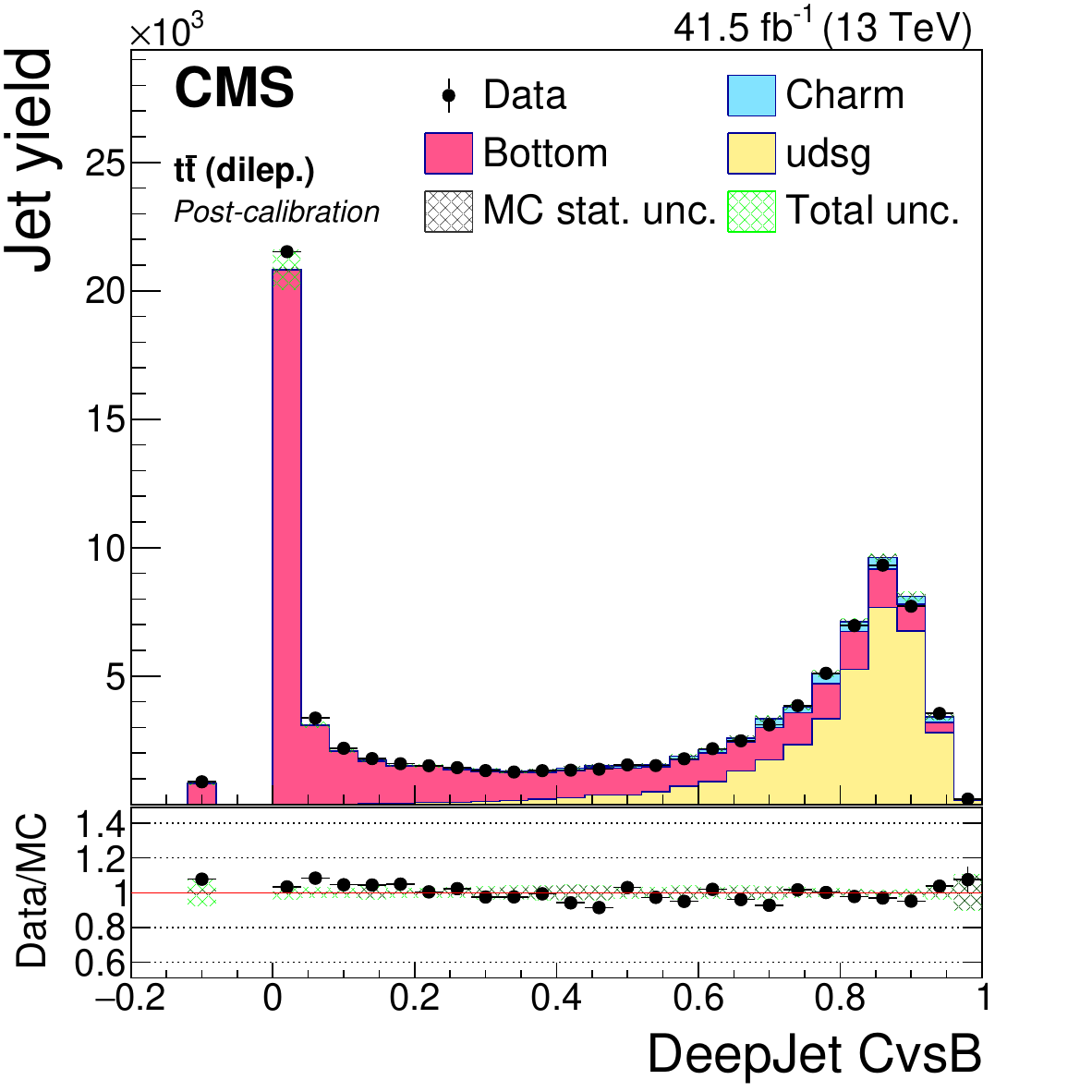}\\
		\includegraphics[width=0.32\textwidth]{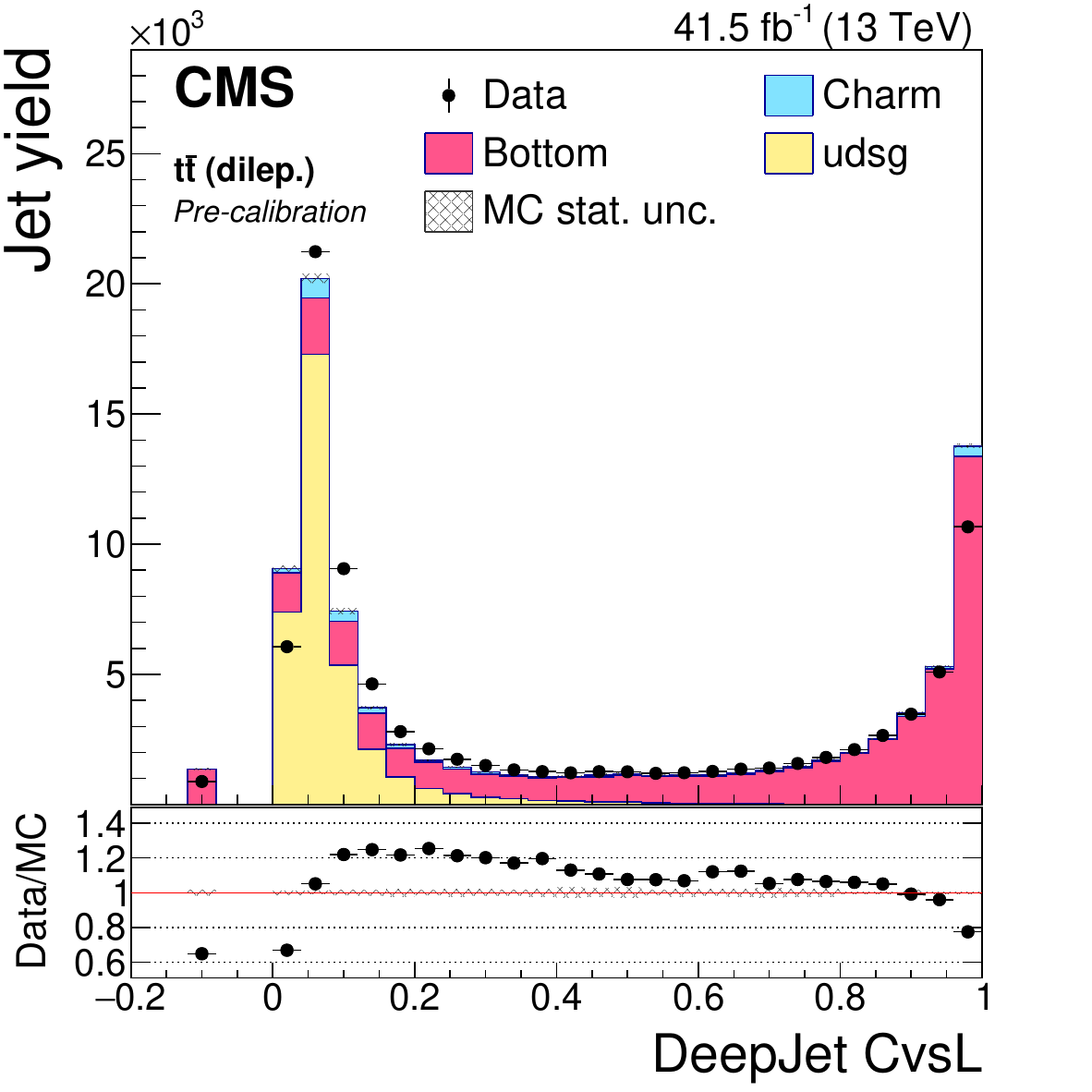}
                \includegraphics[width=0.32\textwidth]{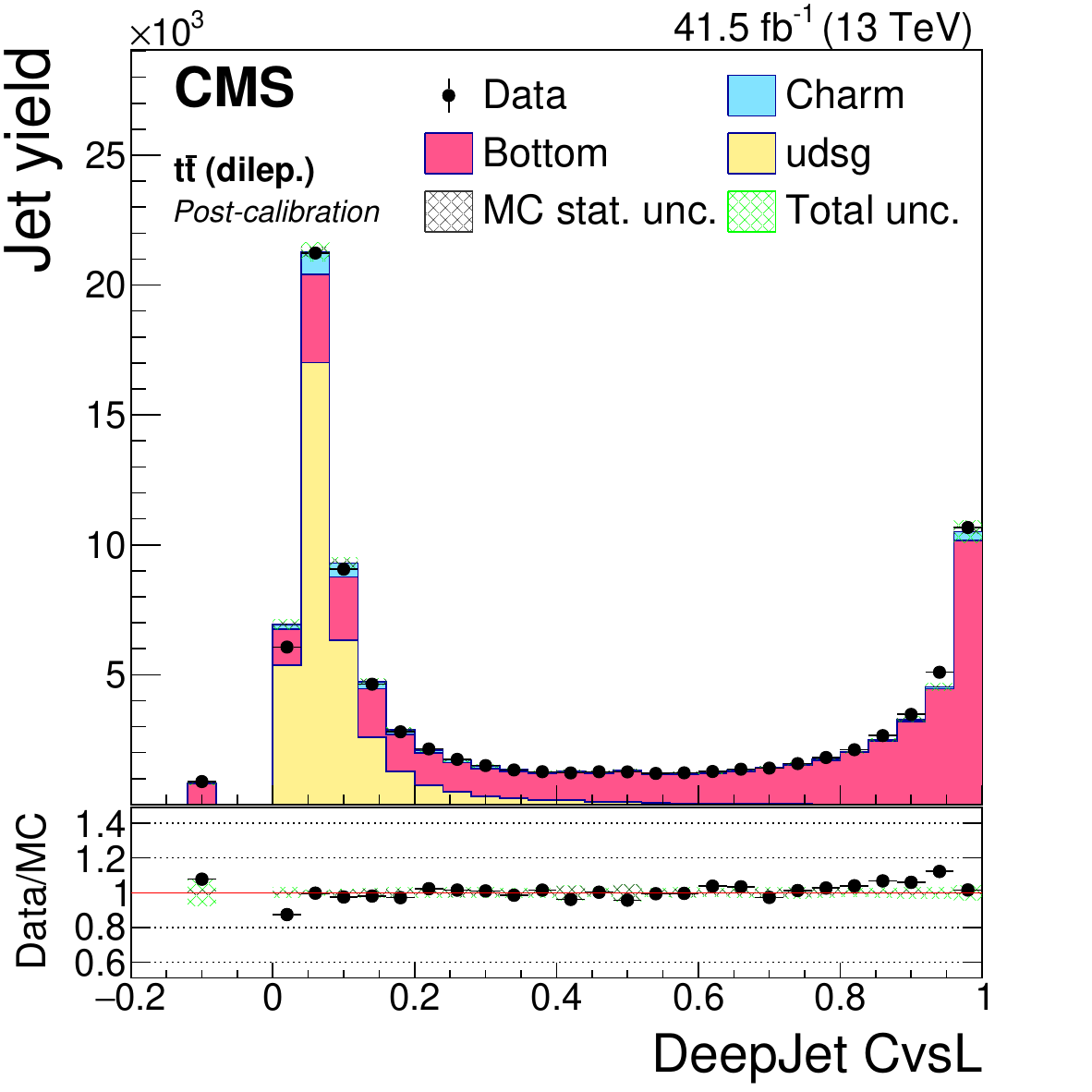}\\
			\caption{\label{fig:muBias1}DeepCSV CvsB (first row), DeepCSV CvsL (second row),
		DeepJet CvsB (third row) and DeepJet CvsL (fourth row)
		discriminators of dileptonic \ttbar jets not biased with soft muons, before (left) and after
		(right) application of SFs. The bin corresponding to a tagger value of $-1$ is plotted at $-0.1$.
		Vertical error bars in data represent statistical uncertainties in data. The simulations are shown as stacked histograms.}
\end{figure}

\begin{figure}[p!]
	\centering
		\includegraphics[width=0.32\textwidth]{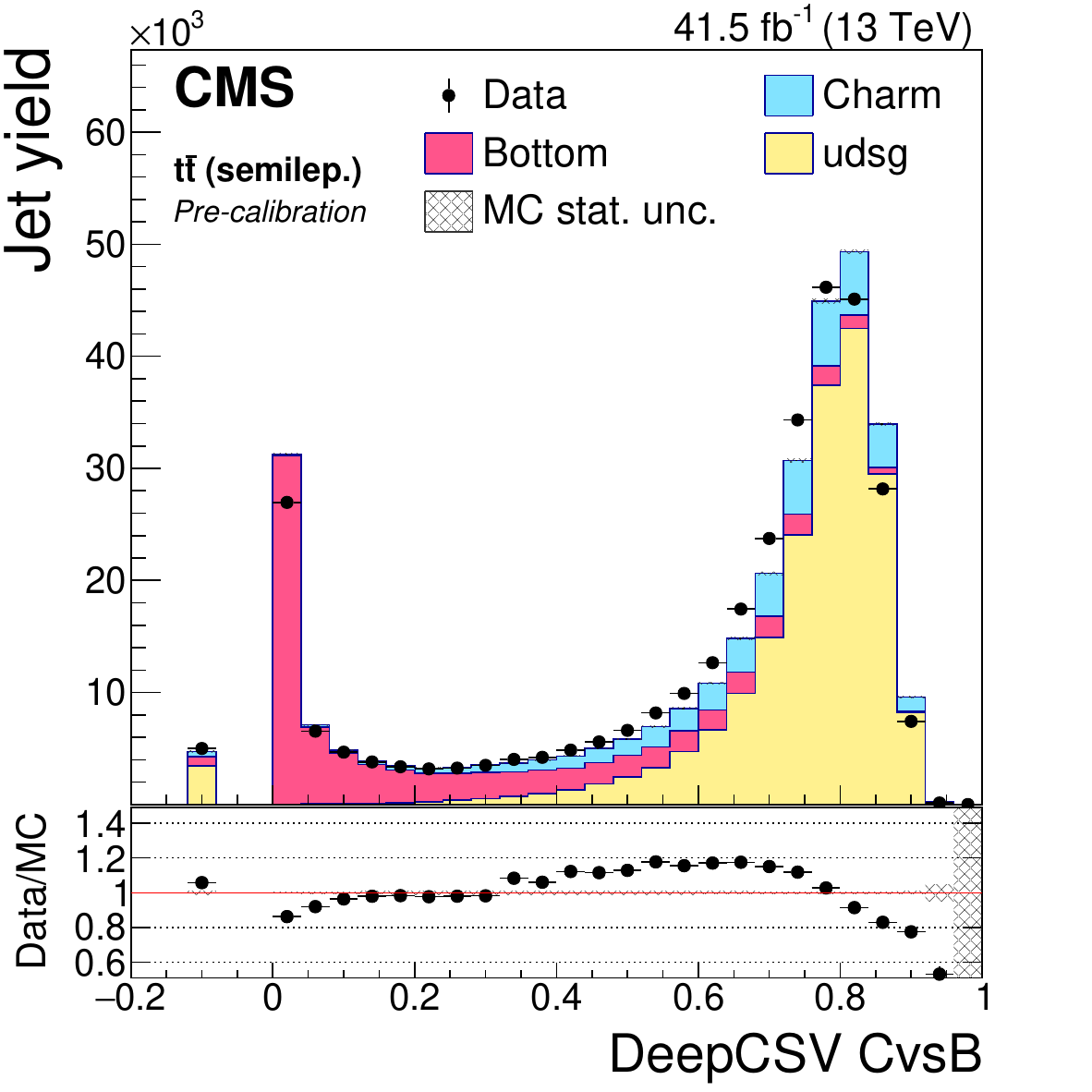}
                \includegraphics[width=0.32\textwidth]{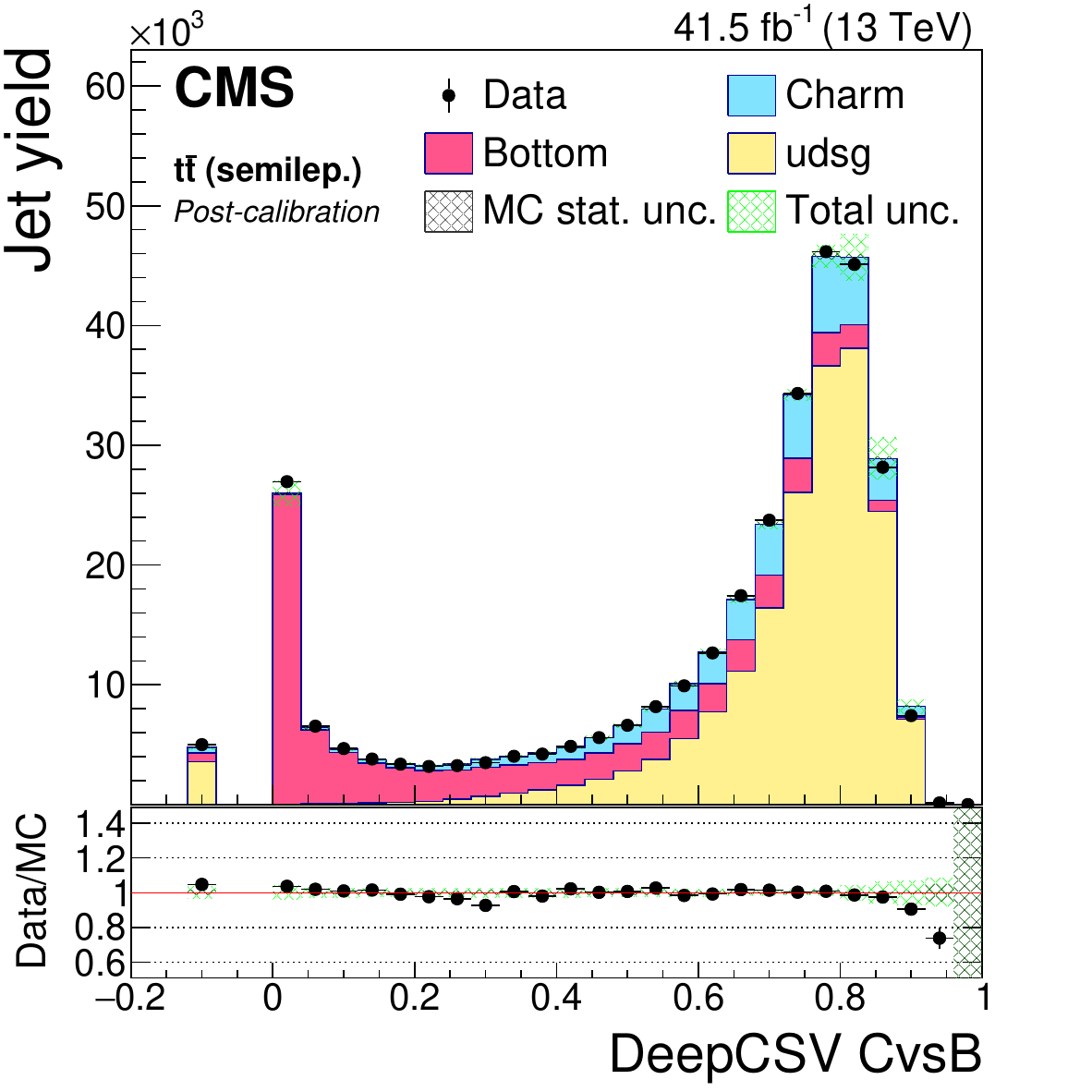}\\
		\includegraphics[width=0.32\textwidth]{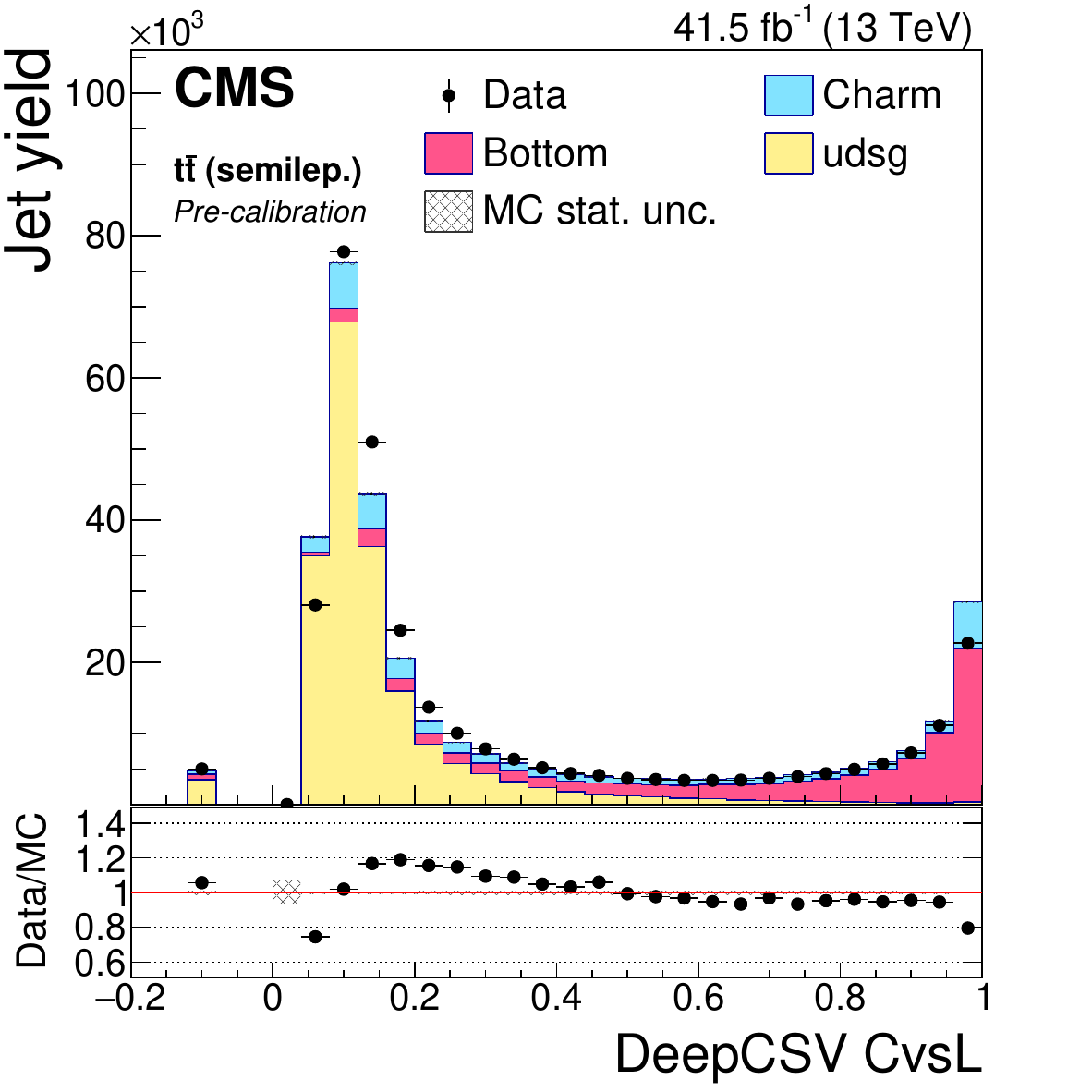}
                \includegraphics[width=0.32\textwidth]{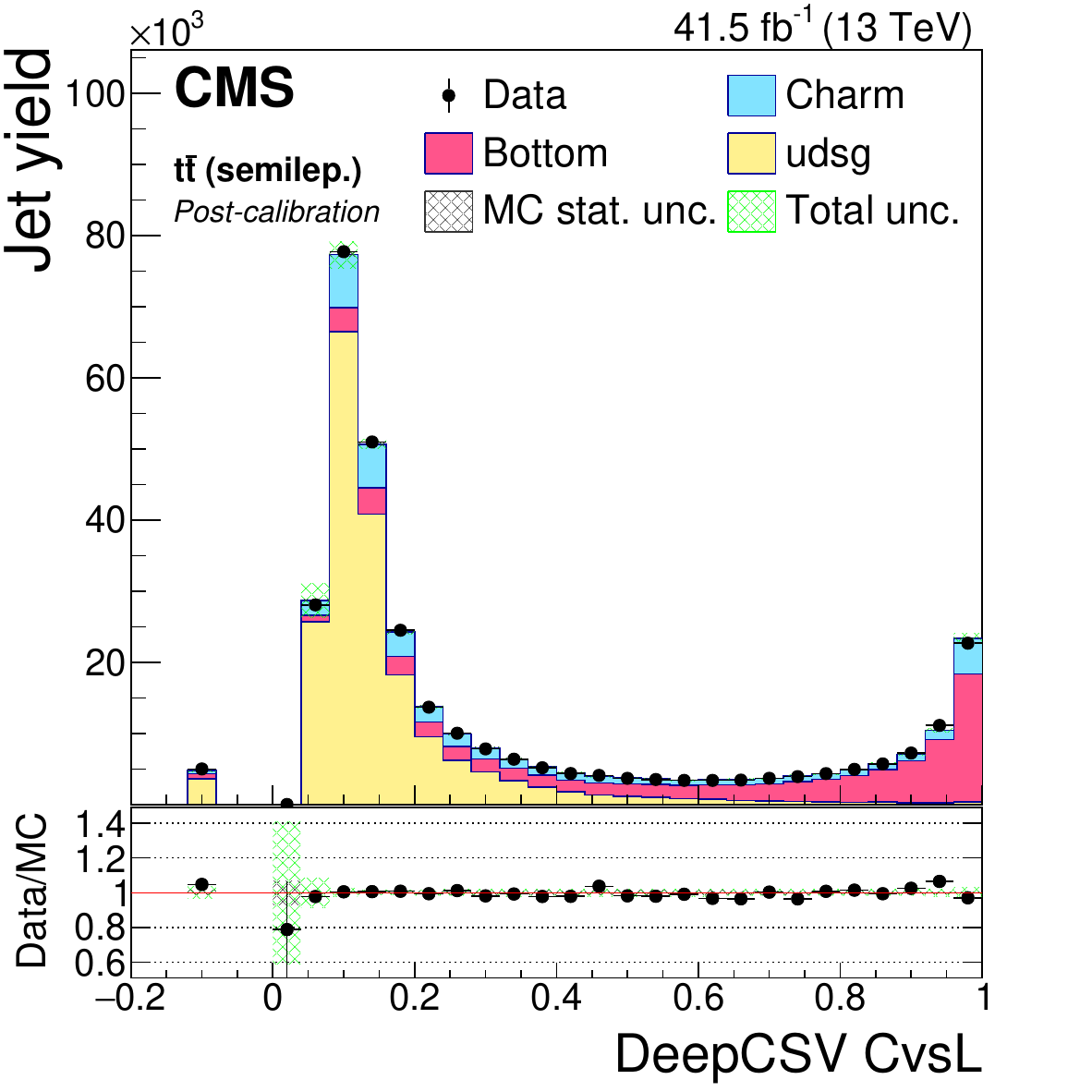}\\
		\includegraphics[width=0.32\textwidth]{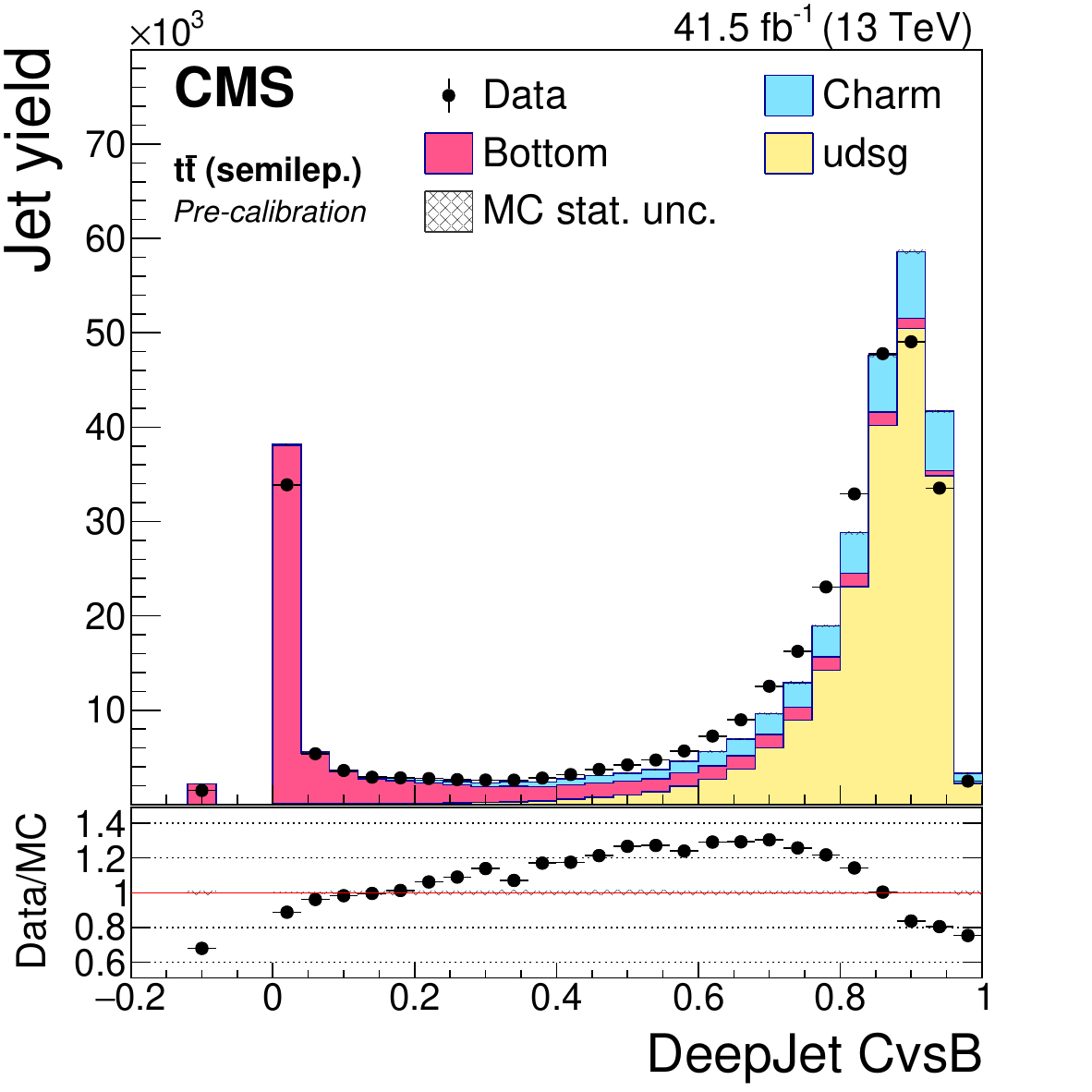} 
                \includegraphics[width=0.32\textwidth]{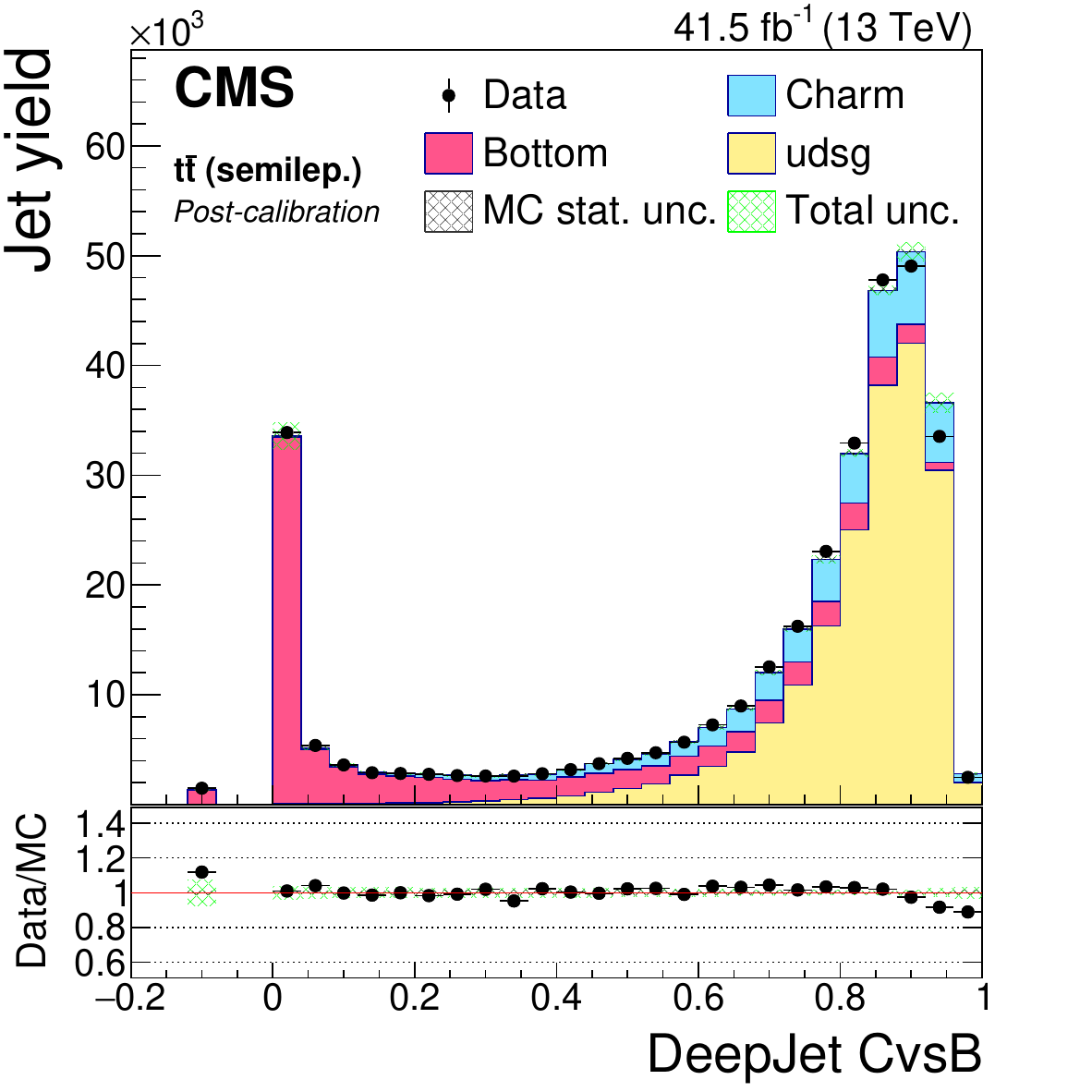}\\
		\includegraphics[width=0.32\textwidth]{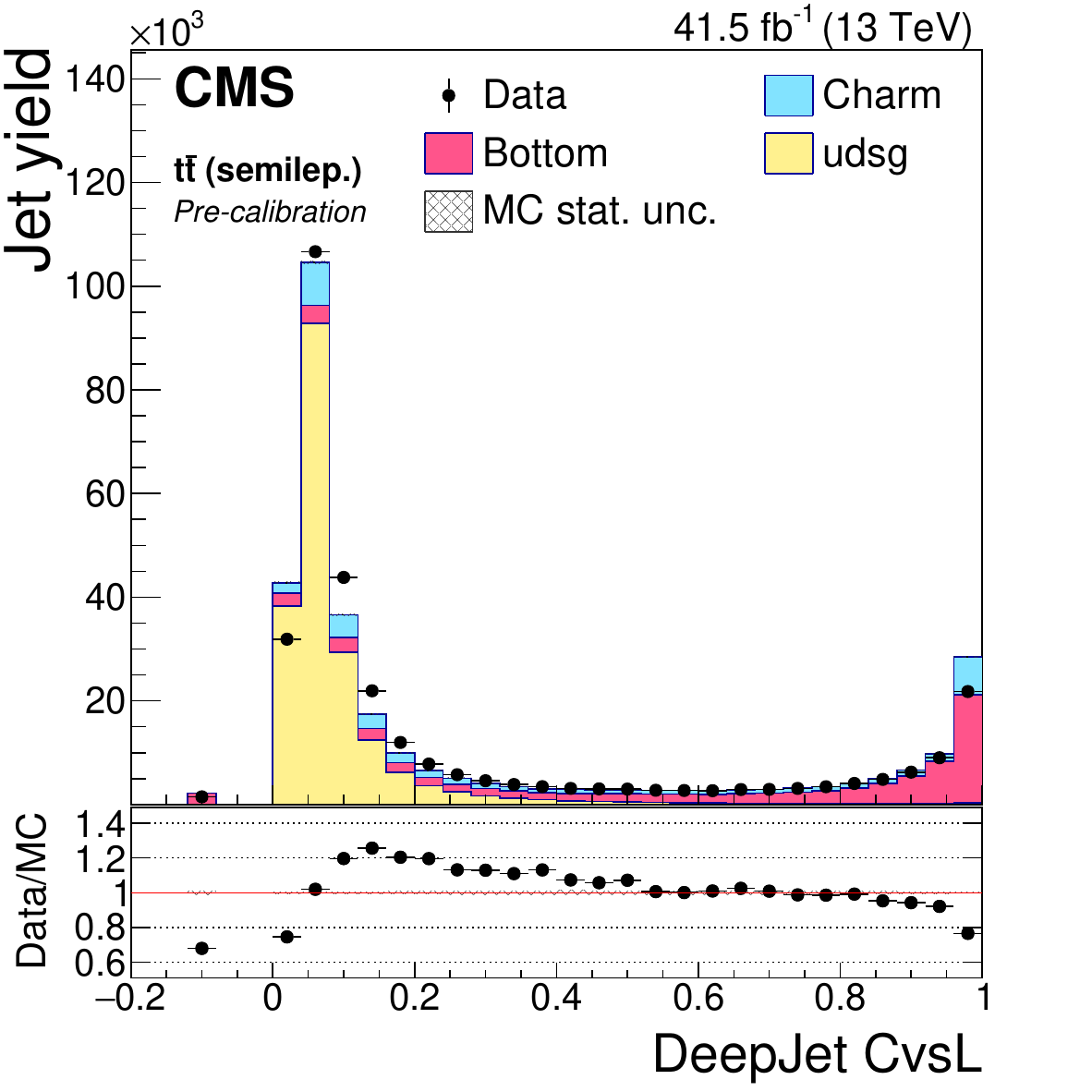}
                \includegraphics[width=0.32\textwidth]{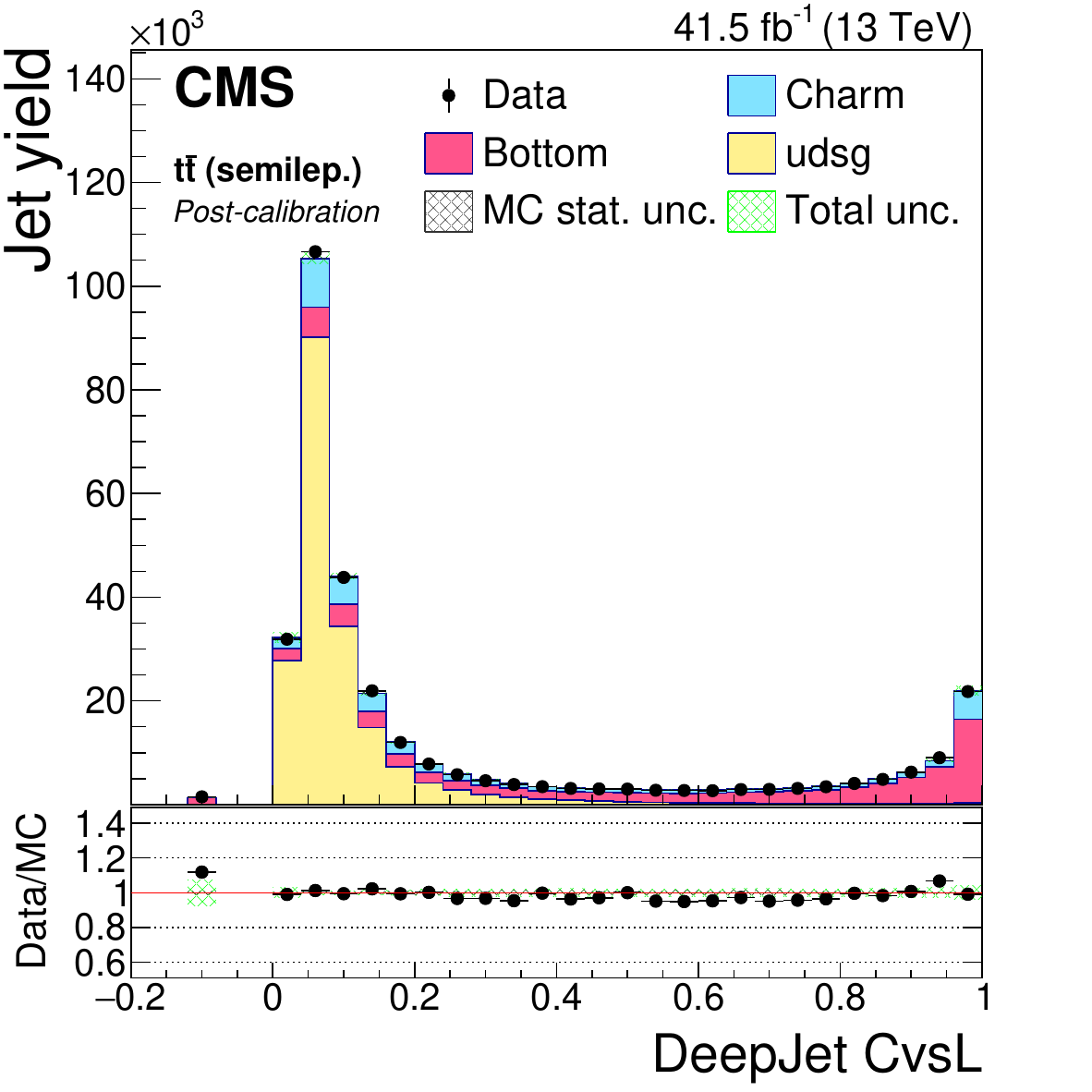}\\
			\caption{\label{fig:muBias2}DeepCSV CvsB (first row), DeepCSV CvsL (second row),
		DeepJet CvsB (third row) and DeepJet CvsL (last row)
		discriminators of semileptonic \ttbar jets not biased with soft muons, before (left) and after
		(right) application of SFs. The bin corresponding to a tagger value of $-1$ is plotted at $-0.1$.
		Vertical error bars in data represent statistical uncertainties in data. The simulations are shown as stacked histograms.}
\end{figure}

The pre- and post-calibration distributions of the {\PQc} tagging discriminator
values of the two samples of jets thus selected are shown in Figs.~\ref{fig:muBias1}
(dileptonic \ttbar) and \ref{fig:muBias2} (semileptonic \ttbar). A much flatter
data-to-simulation ratio is achieved after the corrections are applied, compatible with
unity within a few percent for most of the discriminator values.

This extension to a more general selection of jets, however, does not take into account
potential differences in mismodelling across different selections. Such differences need
to be accounted for in physics analyses by constructing separate control regions and
studying the data-to-simulation agreement in them after application of these SFs, and hence
fitting the discriminator shapes in the control regions simultaneously with
those in the signal regions, during the extraction of the signal.

\subsection{Fit validation with pseudodata}

A test is performed to validate the accuracy of the SF
values from the iterative convergence. A set of distributions
of ``pseudodata'' is created by applying artificial, handcrafted
SF maps to simulated jet distributions. To mimic the statistical independence of data from simulation, 
the central value of every bin of the pseudodata is set by sampling a random
number from a normal distribution, whose mean is equal to the central value of the 
corresponding bin in the rescaled simulation and standard deviation is equal to the statistical uncertainty
in the same bin. The relative uncertainty of every bin in the pseudodata, in turn,
is set to the relative uncertainty in the corresponding bin of data.
The iterative fit algorithm is now allowed to run on the same
selections replacing data with pseudodata.

The artificial SFs are generated as an arbitrary, continuous 2D function
of DeepCSV CvsL and CvsB values, individually for each jet flavour.
A single arbitrary number is assigned as the SF for jets
with the default DeepCSV discriminator value for each flavour. Subsequently,
the SF map of a given flavour is normalised in such a way
that the total jet yield of an independent selection of simulated
jets of that flavour taken from a hadronically decaying \ttbar sample remains
unchanged upon application of the SF map under consideration.

\begin{figure}
	\centering
		\includegraphics[width=0.48\textwidth]{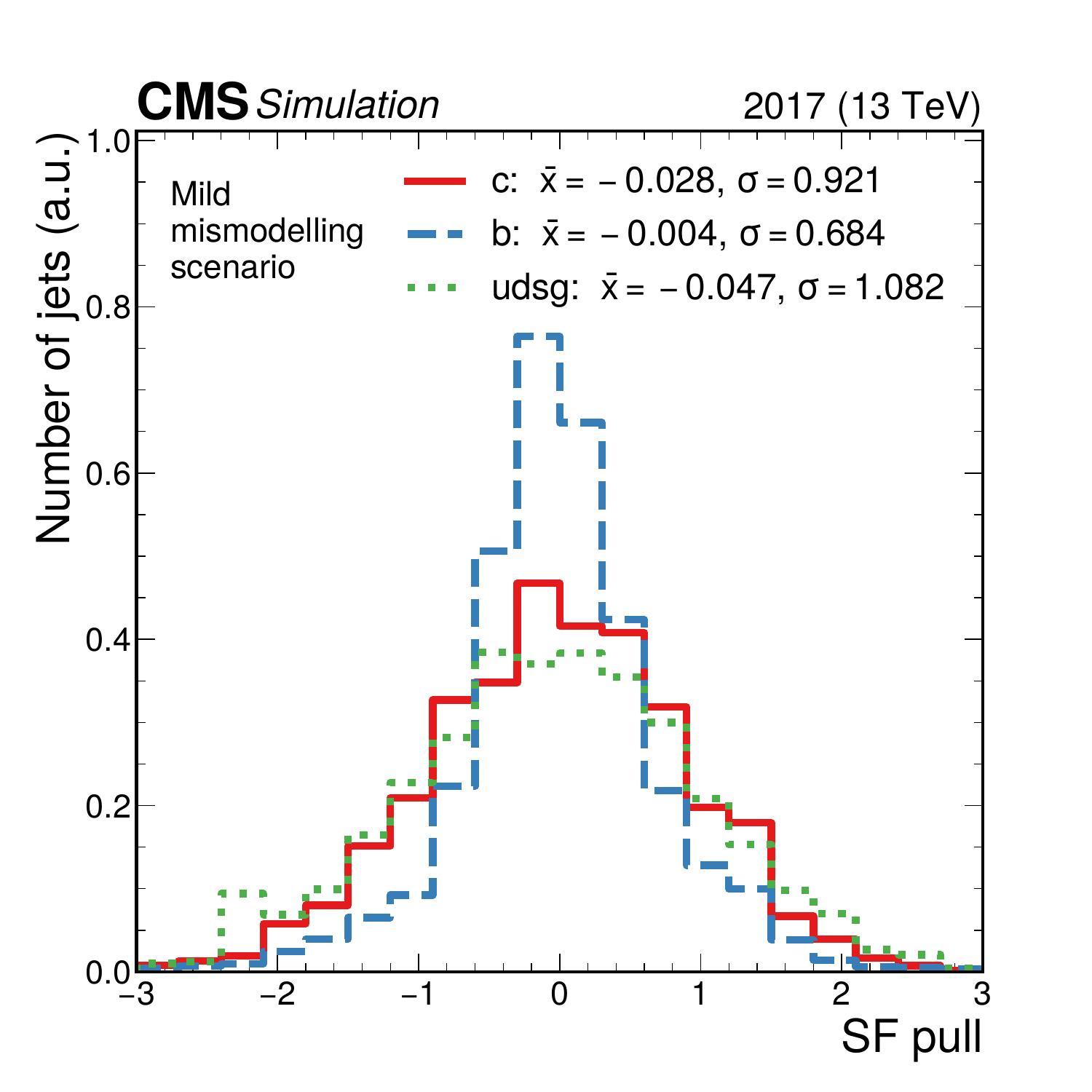}
		\includegraphics[width=0.48\textwidth]{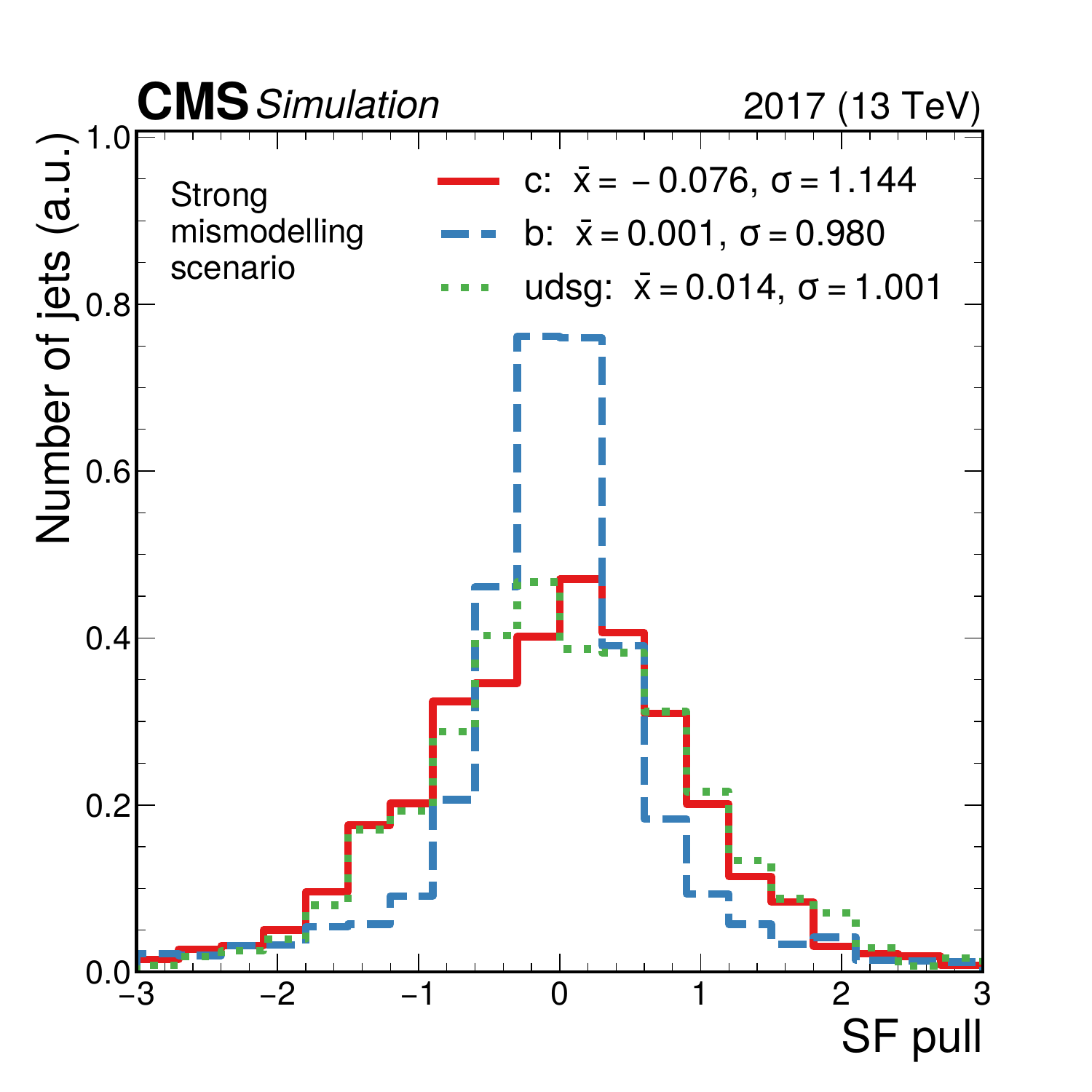}
	\caption{\label{fig:Pseudodata-diff}Distribution of the SF pulls, quantified as the differences between the injected SFs and
		the SFs retrieved by the fits in units of the statistical uncertainties
		in the latter ($[\mathrm{SF}_\text{extracted}-\mathrm{SF}_\text{injected}]/\sigma_{\text{extracted}}$), across all bins in the CvsL-CvsB plane, for the SF map with ``mild'' (left) and ``strong'' (right) SFs. The $\bar{x}$ and $\sigma$ denote the mean and standard deviation of the distributions, respectively.}
\end{figure}

Two such maps of artificial SFs are designed, the first
(second) of which consists of ``mild'' (``strong'') SFs, that is,
SFs with values in the range 0.5--1.4 (0.1--3.0).
For each map, a total of 50 pseudodata models are created injecting the same
set of artificial SFs but using a different random number seed in each
model. For each map, the difference between the injected SFs and the extracted SFs (``SF pull'') is plotted in Fig.~\ref{fig:Pseudodata-diff}. The differences are measured in units of the statistical uncertainty of the extracted SFs across all bins, weighted by the relative jet abundance in each bin in the CvsL-CvsB plane. The figure shows the cumulative results from all 50 models for {\PQc}, {\PQb}, and light favours separately.
A mean of around 0 of the SF pulls
illustrates a correct convergence of the method. A standard deviation of around
1 illustrates an optimal estimation of the SF statistical uncertainties in most cases. A standard
deviation smaller than 1 for {\PQb} jets in case of the ``mild'' map illustrates a
conservative estimation of SF statistical uncertainties in case of ``mild'' deviations from data in simulation.

Whereas these studies show that the statistical uncertainties might be somewhat conservative in
regions of the phase space where SFs for {\PQb} jets are close to 1, this might not be the case for
other regions where SFs show larger deviations.
Since the SFs that map simulated {\PQb} jets to {\PQb} jet candidates in collision data (Fig.~\ref{fig:SFb})
are measured to have high central values close to 2 at some parts of the phase space,
we have decided to not apply any a posteriori reduction of the uncertainties in the
case of collision data.

\section{Conclusion \label{sec:conclusion}}
This paper presents a novel method to calibrate the full differential shape of the discriminator distributions used for charm ({\PQc}) jet identification at CMS. The method uses three different sets of event selection criteria, targeting topologies enriched in {\PW}+{\PQc}, top quark pairs, and Drell--Yan+jet events. These topologies are highly enriched in {\PQc}, bottom ({\PQb}) and light-flavour jets, respectively, resulting in purities of a given jet-flavour that range between 81 and 93\%. By employing an iterative fitting approach in each of these three regions, scale factors (SFs) are derived to match the simulated discriminator distributions to those observed in data. Since the {\PQc} tagging algorithm is composed of two discriminators, one to discriminate {\PQc} from {\PQb} jets (CvsB) and another to discriminate {\PQc} from light-flavour and gluon jets (CvsL), the SFs are derived as functions of CvsL and CvsB discriminator values. An adaptive binning is used to optimise the granularity of the provided calibration with respect to the statistical uncertainty in each bin. Finally, an interpolation is performed to obtain more representative corrections over the entire two-dimensional plane.

Validation and closure tests confirm the robustness of the method. Although this paper reports calibration results with only 2017 data, similar calibrations are obtained with 2016 and 2018 data separately that are used for the analysis of data collected in the respective years. The calibration of the full differential discriminator shape allows the use of the {\PQc} tagging discriminators as inputs to multivariate techniques (based on machine learning) or by fitting the discriminator shapes to data to extract observables that are sensitive to the jet flavour. The shape calibration extends the use of {\PQc} tagging algorithms beyond the application of discrete working points, and facilitates more advanced uses for {\PQc} jet identification in physics analyses.

\begin{acknowledgments}
\hyphenation{Bundes-ministerium Forschungs-gemeinschaft Forschungs-zentren Rachada-pisek} We congratulate our colleagues in the CERN accelerator departments for the excellent performance of the LHC and thank the technical and administrative staffs at CERN and at other CMS institutes for their contributions to the success of the CMS effort. In addition, we gratefully acknowledge the computing centres and personnel of the Worldwide LHC Computing Grid and other centres for delivering so effectively the computing infrastructure essential to our analyses. Finally, we acknowledge the enduring support for the construction and operation of the LHC, the CMS detector, and the supporting computing infrastructure provided by the following funding agencies: the Austrian Federal Ministry of Education, Science and Research and the Austrian Science Fund; the Belgian Fonds de la Recherche Scientifique, and Fonds voor Wetenschappelijk Onderzoek; the Brazilian Funding Agencies (CNPq, CAPES, FAPERJ, FAPERGS, and FAPESP); the Bulgarian Ministry of Education and Science, and the Bulgarian National Science Fund; CERN; the Chinese Academy of Sciences, Ministry of Science and Technology, and National Natural Science Foundation of China; the Ministerio de Ciencia Tecnolog\'ia e Innovaci\'on (MINCIENCIAS), Colombia; the Croatian Ministry of Science, Education and Sport, and the Croatian Science Foundation; the Research and Innovation Foundation, Cyprus; the Secretariat for Higher Education, Science, Technology and Innovation, Ecuador; the Ministry of Education and Research, Estonian Research Council via PRG780, PRG803 and PRG445 and European Regional Development Fund, Estonia; the Academy of Finland, Finnish Ministry of Education and Culture, and Helsinki Institute of Physics; the Institut National de Physique Nucl\'eaire et de Physique des Particules~/~CNRS, and Commissariat \`a l'\'Energie Atomique et aux \'Energies Alternatives~/~CEA, France; the Bundesministerium f\"ur Bildung und Forschung, the Deutsche Forschungsgemeinschaft (DFG), under Germany's Excellence Strategy -- EXC 2121 ``Quantum Universe" -- 390833306, and under project number 400140256 - GRK2497, and Helmholtz-Gemeinschaft Deutscher Forschungszentren, Germany; the General Secretariat for Research and Innovation, Greece; the National Research, Development and Innovation Fund, Hungary; the Department of Atomic Energy and the Department of Science and Technology, India; the Institute for Studies in Theoretical Physics and Mathematics, Iran; the Science Foundation, Ireland; the Istituto Nazionale di Fisica Nucleare, Italy; the Ministry of Science, ICT and Future Planning, and National Research Foundation (NRF), Republic of Korea; the Ministry of Education and Science of the Republic of Latvia; the Lithuanian Academy of Sciences; the Ministry of Education, and University of Malaya (Malaysia); the Ministry of Science of Montenegro; the Mexican Funding Agencies (BUAP, CINVESTAV, CONACYT, LNS, SEP, and UASLP-FAI); the Ministry of Business, Innovation and Employment, New Zealand; the Pakistan Atomic Energy Commission; the Ministry of Science and Higher Education and the National Science Centre, Poland; the Funda\c{c}\~ao para a Ci\^encia e a Tecnologia, grants CERN/FIS-PAR/0025/2019 and CERN/FIS-INS/0032/2019, Portugal; JINR, Dubna; the Ministry of Education and Science of the Russian Federation, the Federal Agency of Atomic Energy of the Russian Federation, Russian Academy of Sciences, the Russian Foundation for Basic Research, and the National Research Center ``Kurchatov Institute"; the Ministry of Education, Science and Technological Development of Serbia; the Secretar\'{\i}a de Estado de Investigaci\'on, Desarrollo e Innovaci\'on, Programa Consolider-Ingenio 2010, Plan Estatal de Investigaci\'on Cient\'{\i}fica y T\'ecnica y de Innovaci\'on 2017--2020, research project IDI-2018-000174 del Principado de Asturias, and Fondo Europeo de Desarrollo Regional, Spain; the Ministry of Science, Technology and Research, Sri Lanka; the Swiss Funding Agencies (ETH Board, ETH Zurich, PSI, SNF, UniZH, Canton Zurich, and SER); the Ministry of Science and Technology, Taipei; the Thailand Center of Excellence in Physics, the Institute for the Promotion of Teaching Science and Technology of Thailand, Special Task Force for Activating Research and the National Science and Technology Development Agency of Thailand; the Scientific and Technical Research Council of Turkey, and Turkish Atomic Energy Authority; the National Academy of Sciences of Ukraine; the Science and Technology Facilities Council, UK; the US Department of Energy, and the US National Science Foundation.

{\tolerance=1200
Individuals have received support from the Marie-Curie programme and the European Research Council and Horizon 2020 Grant, contract Nos.\ 675440, 724704, 752730, 758316, 765710, 824093, 884104, and COST Action CA16108 (European Union) the Leventis Foundation; the Alfred P.\ Sloan Foundation; the Alexander von Humboldt Foundation; the Belgian Federal Science Policy Office; the Fonds pour la Formation \`a la Recherche dans l'Industrie et dans l'Agriculture (FRIA-Belgium); the Agentschap voor Innovatie door Wetenschap en Technologie (IWT-Belgium); the F.R.S.-FNRS and FWO (Belgium) under the ``Excellence of Science -- EOS" -- be.h project n.\ 30820817; the Beijing Municipal Science \& Technology Commission, No. Z191100007219010; the Ministry of Education, Youth and Sports (MEYS) of the Czech Republic; the Lend\"ulet (``Momentum") Programme and the J\'anos Bolyai Research Scholarship of the Hungarian Academy of Sciences, the New National Excellence Program \'UNKP, the NKFIA research grants 123842, 123959, 124845, 124850, 125105, 128713, 128786, and 129058 (Hungary); the Council of Scientific and Industrial Research, India; the Latvian Council of Science; the National Science Center (Poland), contracts Opus 2014/15/B/ST2/03998 and 2015/19/B/ST2/02861; the Funda\c{c}\~ao para a Ci\^encia e a Tecnologia, grant FCT CEECIND/01334/2018; the National Priorities Research Program by Qatar National Research Fund; the Ministry of Science and Higher Education, projects no. 14.W03.31.0026 and no. FSWW-2020-0008, and the Russian Foundation for Basic Research, project No.19-42-703014 (Russia); the Programa de Excelencia Mar\'{i}a de Maeztu, and the Programa Severo Ochoa del Principado de Asturias; the Stavros Niarchos Foundation (Greece); the Rachadapisek Sompot Fund for Postdoctoral Fellowship, Chulalongkorn University, and the Chulalongkorn Academic into Its 2nd Century Project Advancement Project (Thailand); the Kavli Foundation; the Nvidia Corporation; the SuperMicro Corporation; the Welch Foundation, contract C-1845; and the Weston Havens Foundation (USA).
\par}
\end{acknowledgments}

\bibliography{auto_generated}

\providecommand{\href}[2]{#2}\begingroup\raggedright\begin{thebibliography}{10}%
\makeatletter
\providecommand{\hrefCMSnoop }[0]{\@secondoftwo}%
\makeatother
\providecommand{\doi}{\texttt{doi:}\begingroup \urlstyle{tt}\Url}

\bibitem{Chatrchyan:2012jua}
\hrefCMSnoop {}{{CMS Collaboration}, ``{Identification of b-quark jets with the
  CMS experiment}'',} \textit{ JINST} \textbf{ 8} (2013) P04013,
  \href{http://dx.doi.org/10.1088/1748-0221/8/04/P04013}{\doi{10.1088/1748-0221/8/04/P04013}},
\href{http://www.arXiv.org/abs/1211.4462}{\texttt{arXiv:1211.4462}}.

\bibitem{Sirunyan:2017ezt}
\hrefCMSnoop {}{{CMS Collaboration}, ``{Identification of heavy-flavour jets
  with the CMS detector in pp collisions at 13 \TeV}'',} \textit{ JINST}
  \textbf{ 13} (2018) P05011,
  \href{http://dx.doi.org/10.1088/1748-0221/13/05/P05011}{\doi{10.1088/1748-0221/13/05/P05011}},
\href{http://www.arXiv.org/abs/1712.07158}{\texttt{arXiv:1712.07158}}.

\bibitem{CMS-PAS-BTV-16-001}
\href {http://cds.cern.ch/record/2205149}{{CMS Collaboration},
  ``{Identification of c-quark jets at the CMS experiment}'',} {CMS Physics
  Analysis Summary} CMS-PAS-BTV-16-001, 2016.

\bibitem{bols2020jet}
E.~Bols\hrefCMSnoop {}{ {et~al.}, ``{Jet flavour classification using
  DeepJet}'',} \textit{ JINST} \textbf{ 15} (2020) P12012,
  \href{http://dx.doi.org/10.1088/1748-0221/15/12/P12012}{\doi{10.1088/1748-0221/15/12/P12012}},
  \href{http://www.arXiv.org/abs/2008.10519}{\texttt{arXiv:2008.10519}}.

\bibitem{CMS-DP-2018-058}
\href {http://cds.cern.ch/record/2646773}{{CMS Collaboration}, ``{Performance
  of the DeepJet b tagging algorithm using 41.9/fb of data from proton-proton
  collisions at 13\TeV with Phase 1 CMS detector}'',} {CMS Detector Performance
  Note} CMS-DP-2018-058, 2018.

\bibitem{Guest_2016}
D.~Guest\hrefCMSnoop {}{ {et~al.}, ``Jet flavor classification in high-energy
  physics with deep neural networks'',} \textit{ Phys. Rev. D} \textbf{ 94}
  (2016) 112002,
  \href{http://dx.doi.org/10.1103/physrevd.94.112002}{\doi{10.1103/physrevd.94.112002}},
  \href{http://www.arXiv.org/abs/1607.08633}{\texttt{arXiv:1607.08633}}.

\bibitem{Sirunyan:2018kst}
\hrefCMSnoop {}{{CMS Collaboration}, ``{Observation of Higgs boson decay to
  bottom quarks}'',} \textit{ Phys. Rev. Lett.} \textbf{ 121} (2018) 121801,
  \href{http://dx.doi.org/10.1103/PhysRevLett.121.121801}{\doi{10.1103/PhysRevLett.121.121801}},
\href{http://www.arXiv.org/abs/1808.08242}{\texttt{arXiv:1808.08242}}.

\bibitem{Moortgat:2676133}
\href {https://cds.cern.ch/record/2676133}{S.~Moortgat, ``{When charm and
  beauty adjoin the top. First measurement of the cross section of top quark
  pair production with additional charm jets with the CMS experiment.}''}.
\newblock PhD thesis, Vrije U., Brussels, May, 2019.
\newblock CERN-THESIS-2019-051.

\bibitem{Sirunyan:2019qia}
\hrefCMSnoop {}{{CMS Collaboration}, ``{A search for the standard model Higgs
  boson decaying to charm quarks}'',} \textit{ JHEP} \textbf{ 03} (2020) 131,
  \href{http://dx.doi.org/10.1007/JHEP03(2020)131}{\doi{10.1007/JHEP03(2020)131}},
  \href{http://www.arXiv.org/abs/1912.01662}{\texttt{arXiv:1912.01662}}.

\bibitem{CMS-TDR-011}
\href {https://cds.cern.ch/record/1481838}{{CMS Collaboration}, ``{CMS
  technical design report for the pixel detector upgrade}'',} CMS Technical
  Design Report CERN-LHCC-2012-016, CMS-TDR-011, 2012.

\bibitem{CMS-DP-2020-049}
\href {https://cds.cern.ch/record/2743740}{{CMS Collaboration}, ``{Track impact
  parameter resolution for the full pseudo rapidity coverage in the 2017
  dataset with the CMS Phase-1 pixel detector}'',} {CMS Detector Performance
  Note} CMS-DP-2020-049, 2020.

\bibitem{Chatrchyan:2008zzk}
\hrefCMSnoop {}{{CMS Collaboration}, ``The {CMS} experiment at the {CERN}
  {LHC}'',} \textit{ JINST} \textbf{ 3} (2008) S08004,
  \href{http://dx.doi.org/10.1088/1748-0221/3/08/S08004}{\doi{10.1088/1748-0221/3/08/S08004}}.

\bibitem{Khachatryan:2016bia}
\hrefCMSnoop {}{{CMS Collaboration}, ``{The CMS trigger system}'',} \textit{
  JINST} \textbf{ 12} (2017) P01020,
  \href{http://dx.doi.org/10.1088/1748-0221/12/01/P01020}{\doi{10.1088/1748-0221/12/01/P01020}},
\href{http://www.arXiv.org/abs/1609.02366}{\texttt{arXiv:1609.02366}}.

\bibitem{Sirunyan:2020zal}
\hrefCMSnoop {}{{CMS Collaboration}, ``{Performance of the CMS Level-1 trigger
  in proton-proton collisions at $\sqrt{s} =$ 13 TeV}'',} \textit{ JINST}
  \textbf{ 15} (2020) P10017,
  \href{http://dx.doi.org/10.1088/1748-0221/15/10/P10017}{\doi{10.1088/1748-0221/15/10/P10017}},
  \href{http://www.arXiv.org/abs/2006.10165}{\texttt{arXiv:2006.10165}}.

\bibitem{CMS:2018elu}
\href {https://cdsweb.cern.ch/record/2621960}{{CMS Collaboration}, ``{CMS
  luminosity measurement for the 2017 data-taking period at $\sqrt{s} =
  13~\mathrm{TeV}$}'',} CMS Physics Analysis Summary CMS-PAS-LUM-17-004, 2018.

\bibitem{SJOSTRAND2015159}
T.~Sj{\"o}strand\hrefCMSnoop {}{ {et~al.}, ``An introduction to {PYTHIA}
  8.2'',} \textit{ Comput. Phys. Commun.} \textbf{ 191} (2015) 159,
  \href{http://dx.doi.org/10.1016/j.cpc.2015.01.024}{\doi{10.1016/j.cpc.2015.01.024}},
\href{http://www.arXiv.org/abs/1410.3012}{\texttt{arXiv:1410.3012}}.

\bibitem{Skands:2014pea}
\hrefCMSnoop {}{P.~Skands, S.~Carrazza, and J.~Rojo, ``{Tuning PYTHIA 8.1: the
  Monash 2013 tune}'',} \textit{ Eur. Phys. J. C} \textbf{ 74} (2014) 3024,
  \href{http://dx.doi.org/10.1140/epjc/s10052-014-3024-y}{\doi{10.1140/epjc/s10052-014-3024-y}},
\href{http://www.arXiv.org/abs/1404.5630}{\texttt{arXiv:1404.5630}}.

\bibitem{Ball:2014uwa}
\hrefCMSnoop {}{{NNPDF} Collaboration, ``{Parton distributions for the LHC Run
  II}'',} \textit{ JHEP} \textbf{ 04} (2015) 040,
  \href{http://dx.doi.org/10.1007/JHEP04(2015)040}{\doi{10.1007/JHEP04(2015)040}},
\href{http://www.arXiv.org/abs/1410.8849}{\texttt{arXiv:1410.8849}}.

\bibitem{Nason:2004rx}
\hrefCMSnoop {}{P.~Nason, ``{A new method for combining NLO QCD with shower
  Monte Carlo algorithms}'',} \textit{ JHEP} \textbf{ 11} (2004) 040,
  \href{http://dx.doi.org/10.1088/1126-6708/2004/11/040}{\doi{10.1088/1126-6708/2004/11/040}},
\href{http://www.arXiv.org/abs/hep-ph/0409146}{\texttt{arXiv:hep-ph/0409146}}.

\bibitem{Frixione:2007vw}
\hrefCMSnoop {}{S.~Frixione, P.~Nason, and C.~Oleari, ``{Matching NLO QCD
  computations with parton shower simulations: the POWHEG method}'',} \textit{
  JHEP} \textbf{ 11} (2007) 070,
  \href{http://dx.doi.org/10.1088/1126-6708/2007/11/070}{\doi{10.1088/1126-6708/2007/11/070}},
\href{http://www.arXiv.org/abs/0709.2092}{\texttt{arXiv:0709.2092}}.

\bibitem{Alioli:2010xd}
\hrefCMSnoop {}{S.~Alioli, P.~Nason, C.~Oleari, and E.~Re, ``{A general
  framework for implementing NLO calculations in shower Monte Carlo programs:
  the POWHEG BOX}'',} \textit{ JHEP} \textbf{ 06} (2010) 043,
  \href{http://dx.doi.org/10.1007/JHEP06(2010)043}{\doi{10.1007/JHEP06(2010)043}},
\href{http://www.arXiv.org/abs/1002.2581}{\texttt{arXiv:1002.2581}}.

\bibitem{Campbell:2014kua}
\hrefCMSnoop {}{J.~M. Campbell, R.~K. Ellis, P.~Nason, and E.~Re, ``{Top-pair
  production and decay at NLO matched with parton showers}'',} \textit{ JHEP}
  \textbf{ 04} (2015) 114,
  \href{http://dx.doi.org/10.1007/JHEP04(2015)114}{\doi{10.1007/JHEP04(2015)114}},
\href{http://www.arXiv.org/abs/1412.1828}{\texttt{arXiv:1412.1828}}.

\bibitem{Alwall:2014hca}
J.~Alwall\hrefCMSnoop {}{ {et~al.}, ``{The automated computation of tree-level
  and next-to-leading order differential cross sections, and their matching to
  parton shower simulations}'',} \textit{ JHEP} \textbf{ 07} (2014) 079,
  \href{http://dx.doi.org/10.1007/JHEP07(2014)079}{\doi{10.1007/JHEP07(2014)079}},
\href{http://www.arXiv.org/abs/1405.0301}{\texttt{arXiv:1405.0301}}.

\bibitem{Alwall:2007fs}
J.~Alwall\hrefCMSnoop {}{ {et~al.}, ``Comparative study of various algorithms
  for the merging of parton showers and matrix elements in hadronic
  collisions'',} \textit{ Eur. Phys. J. C} \textbf{ 53} (2008) 473,
  \href{http://dx.doi.org/10.1140/epjc/s10052-007-0490-5}{\doi{10.1140/epjc/s10052-007-0490-5}},
\href{http://www.arXiv.org/abs/0706.2569}{\texttt{arXiv:0706.2569}}.

\bibitem{Li:2012wna}
\hrefCMSnoop {}{Y.~Li and F.~Petriello, ``{Combining QCD and electroweak
  corrections to dilepton production in FEWZ}'',} \textit{ Phys. Rev. D}
  \textbf{ 86} (2012) 094034,
  \href{http://dx.doi.org/10.1103/PhysRevD.86.094034}{\doi{10.1103/PhysRevD.86.094034}},
\href{http://www.arXiv.org/abs/1208.5967}{\texttt{arXiv:1208.5967}}.

\bibitem{Frixione:2008yi}
S.~Frixione\hrefCMSnoop {}{ {et~al.}, ``{Single-top hadroproduction in
  association with a W boson}'',} \textit{ JHEP} \textbf{ 07} (2008) 029,
  \href{http://dx.doi.org/10.1088/1126-6708/2008/07/029}{\doi{10.1088/1126-6708/2008/07/029}},
\href{http://www.arXiv.org/abs/0805.3067}{\texttt{arXiv:0805.3067}}.

\bibitem{Re:2010bp}
\hrefCMSnoop {}{E.~Re, ``{Single-top Wt-channel production matched with parton
  showers using the POWHEG method}'',} \textit{ Eur. Phys. J. C} \textbf{ 71}
  (2011) 1547,
  \href{http://dx.doi.org/10.1140/epjc/s10052-011-1547-z}{\doi{10.1140/epjc/s10052-011-1547-z}},
\href{http://www.arXiv.org/abs/1009.2450}{\texttt{arXiv:1009.2450}}.

\bibitem{Kidonakis:2012rm}
\hrefCMSnoop {}{N.~Kidonakis, ``{NNLL threshold resummation for top-pair and
  single-top production}'',} \textit{ Phys. Part. Nucl.} \textbf{ 45} (2014)
  714,
  \href{http://dx.doi.org/10.1134/S1063779614040091}{\doi{10.1134/S1063779614040091}},
\href{http://www.arXiv.org/abs/1210.7813}{\texttt{arXiv:1210.7813}}.

\bibitem{Campbell:2010ff}
\hrefCMSnoop {}{J.~M. Campbell and R.~K. Ellis, ``{MCFM for the Tevatron and
  the LHC}'',} \textit{ Nucl. Phys. Proc. Suppl.} \textbf{ 205-206} (2010) 10,
  \href{http://dx.doi.org/10.1016/j.nuclphysbps.2010.08.011}{\doi{10.1016/j.nuclphysbps.2010.08.011}},
\href{http://www.arXiv.org/abs/1007.3492}{\texttt{arXiv:1007.3492}}.

\bibitem{Gehrmann:2014fva}
T.~Gehrmann\hrefCMSnoop {}{ {et~al.}, ``{W$^+$W$^-$ production at hadron
  colliders in next to next to leading order QCD}'',} \textit{ Phys. Rev.
  Lett.} \textbf{ 113} (2014) 212001,
  \href{http://dx.doi.org/10.1103/PhysRevLett.113.212001}{\doi{10.1103/PhysRevLett.113.212001}},
\href{http://www.arXiv.org/abs/1408.5243}{\texttt{arXiv:1408.5243}}.

\bibitem{AGOSTINELLI2003250}
\hrefCMSnoop {}{{GEANT4} Collaboration, ``{GEANT4}---a simulation toolkit'',}
  \textit{ Nucl. Instrum. Meth. A} \textbf{ 506} (2003) 250,
\href{http://dx.doi.org/10.1016/S0168-9002(03)01368-8}{\doi{10.1016/S0168-9002(03)01368-8}}.

\bibitem{CMS-PRF-14-001}
\hrefCMSnoop {}{{CMS Collaboration}, ``{Particle-flow reconstruction and global
  event description with the CMS detector}'',} \textit{ JINST} \textbf{ 12}
  (2017) P10003,
  \href{http://dx.doi.org/10.1088/1748-0221/12/10/P10003}{\doi{10.1088/1748-0221/12/10/P10003}},
\href{http://www.arXiv.org/abs/1706.04965}{\texttt{arXiv:1706.04965}}.

\bibitem{Sirunyan:2019kia}
\hrefCMSnoop {}{{CMS Collaboration}, ``Performance of missing transverse
  momentum reconstruction in proton-proton collisions at $\sqrt{s} = 13$\,{TeV}
  using the {CMS} detector'',} \textit{ JINST} \textbf{ 14} (2019) P07004,
  \href{http://dx.doi.org/10.1088/1748-0221/14/07/P07004}{\doi{10.1088/1748-0221/14/07/P07004}},
\href{http://www.arXiv.org/abs/1903.06078}{\texttt{arXiv:1903.06078}}.

\bibitem{Cacciari:2008gp}
\hrefCMSnoop {}{M.~Cacciari, G.~P. Salam, and G.~Soyez, ``{The anti-$\kt$ jet
  clustering algorithm}'',} \textit{ JHEP} \textbf{ 04} (2008) 063,
  \href{http://dx.doi.org/10.1088/1126-6708/2008/04/063}{\doi{10.1088/1126-6708/2008/04/063}},
  \href{http://www.arXiv.org/abs/0802.1189}{\texttt{arXiv:0802.1189}}.

\bibitem{Cacciari:2011ma}
\hrefCMSnoop {}{M.~Cacciari, G.~P. Salam, and G.~Soyez, ``{FastJet user
  manual}'',} \textit{ Eur. Phys. J. C} \textbf{ 72} (2012) 1896,
  \href{http://dx.doi.org/10.1140/epjc/s10052-012-1896-2}{\doi{10.1140/epjc/s10052-012-1896-2}},
\href{http://www.arXiv.org/abs/1111.6097}{\texttt{arXiv:1111.6097}}.

\bibitem{Khachatryan:2016kdb}
\hrefCMSnoop {}{{CMS Collaboration}, ``{Jet energy scale and resolution in the
  CMS experiment in pp collisions at 8\TeV}'',} \textit{ JINST} \textbf{ 12}
  (2017) P02014,
  \href{http://dx.doi.org/10.1088/1748-0221/12/02/P02014}{\doi{10.1088/1748-0221/12/02/P02014}},
\href{http://www.arXiv.org/abs/1607.03663}{\texttt{arXiv:1607.03663}}.

\bibitem{Sirunyan:2020foa}
\hrefCMSnoop {}{{CMS Collaboration}, ``{Pileup mitigation at CMS in 13 TeV
  data}'',} \textit{ JINST} \textbf{ 15} (2020) P09018,
  \href{http://dx.doi.org/10.1088/1748-0221/15/09/P09018}{\doi{10.1088/1748-0221/15/09/P09018}},
  \href{http://www.arXiv.org/abs/2003.00503}{\texttt{arXiv:2003.00503}}.

\bibitem{Cacciari:2007fd}
\hrefCMSnoop {}{M.~Cacciari and G.~P. Salam, ``{Pileup subtraction using jet
  areas}'',} \textit{ Phys. Lett. B} \textbf{ 659} (2008) 119,
  \href{http://dx.doi.org/10.1016/j.physletb.2007.09.077}{\doi{10.1016/j.physletb.2007.09.077}},
\href{http://www.arXiv.org/abs/0707.1378}{\texttt{arXiv:0707.1378}}.

\bibitem{Khachatryan:2011wq}
\hrefCMSnoop {}{{CMS Collaboration}, ``{Measurement of
  ${\text{B}}\overline{\text{B}}$ angular correlations based on secondary
  vertex reconstruction at $\sqrt{s}=7$ \TeV}'',} \textit{ JHEP} \textbf{ 03}
  (2011) 136,
  \href{http://dx.doi.org/10.1007/JHEP03(2011)136}{\doi{10.1007/JHEP03(2011)136}},
\href{http://www.arXiv.org/abs/1102.3194}{\texttt{arXiv:1102.3194}}.

\bibitem{CERN-EP-2018-282}
\hrefCMSnoop {}{{CMS Collaboration}, ``{Measurement of associated production of
  a W boson and a charm quark in proton-proton collisions at $\sqrt{s} =
  13\,\text {Te}\text {V} $}'',} \textit{ Eur. Phys. J. C} \textbf{ 79} (2019)
  269,
  \href{http://dx.doi.org/10.1140/epjc/s10052-019-6752-1}{\doi{10.1140/epjc/s10052-019-6752-1}},
  \href{http://www.arXiv.org/abs/1811.10021}{\texttt{arXiv:1811.10021}}.

\bibitem{CMS:2021vhb}
\hrefCMSnoop {}{{CMS Collaboration}, ``{Measurement of differential \ttbar
  production cross sections in the full kinematic range using lepton+jets
  events from proton-proton collisions at $\sqrt{s}=13$ TeV}'',} \textit{ Phys.
  Rev. D} \textbf{ 104} (2021) 092013,
  \href{http://dx.doi.org/10.1103/PhysRevD.104.092013}{\doi{10.1103/PhysRevD.104.092013}},
  \href{http://www.arXiv.org/abs/2108.02803}{\texttt{arXiv:2108.02803}}.

\bibitem{CMS:2019zct}
\hrefCMSnoop {}{{CMS Collaboration}, ``{Search for new physics in top quark
  production in dilepton final states in proton-proton collisions at
  $\sqrt{s}=13$ TeV}'',} \textit{ Eur. Phys. J. C} \textbf{ 79} (2019) 886,
  \href{http://dx.doi.org/10.1140/epjc/s10052-019-7387-y}{\doi{10.1140/epjc/s10052-019-7387-y}},
  \href{http://www.arXiv.org/abs/1903.11144}{\texttt{arXiv:1903.11144}}.

\bibitem{Sirunyan_2020}
\hrefCMSnoop {}{{CMS Collaboration}, ``{Measurement of the associated
  production of a Z boson with charm or bottom quark jets in proton-proton
  collisions at $\sqrt{s} = $ 13 TeV}'',} \textit{ Phys. Rev. D} \textbf{ 102}
  (2020) 032007,
  \href{http://dx.doi.org/10.1103/physrevd.102.032007}{\doi{10.1103/physrevd.102.032007}},
  \href{http://www.arXiv.org/abs/2001.06899}{\texttt{arXiv:2001.06899}}.

\bibitem{CMS-DP-2018-017}
\hrefCMSnoop {}{{CMS Collaboration}, ``{Electron and photon reconstruction and
  identification with the CMS experiment at the CERN LHC}'',} \textit{ JINST}
  \textbf{ 16} (2021) P05014,
  \href{http://dx.doi.org/10.1088/1748-0221/16/05/p05014}{\doi{10.1088/1748-0221/16/05/p05014}},
  \href{http://www.arXiv.org/abs/2012.06888}{\texttt{arXiv:2012.06888}}.

\bibitem{2020SciPy-NMeth}
\hrefCMSnoop {}{P.~Virtanen {et~al.}, ``{{SciPy} 1.0: Fundamental algorithms
  for scientific computing in Python}'',} \textit{ Nature Methods} \textbf{ 17}
  (2020) 261,
  \href{http://dx.doi.org/10.1038/s41592-019-0686-2}{\doi{10.1038/s41592-019-0686-2}}.

\bibitem{Barber96thequickhull}
\hrefCMSnoop {}{C.~B. Barber, D.~P. Dobkin, and H.~Huhdanpaa, ``The quickhull
  algorithm for convex hulls'',} \textit{ ACM Trans. Math. Softw} \textbf{ 22}
  (1996) 469,
  \href{http://dx.doi.org/10.1145/235815.235821}{\doi{10.1145/235815.235821}}.

\bibitem{bezier1970numerical}
P.~B{\'e}zier, ``Numerical control: mathematics and applications''.
\newblock Wiley, London, 1970.

\bibitem{clough1965finite}
\hrefCMSnoop {}{R.~Clough and J.~Tocher, ``Finite element stiffness matricess
  for analysis of plate bending'',} in \textit{ Proc. of the First Conf. on
  Matrix Methods in Struct. Mech.}, p.~515.
\newblock 1965.

\bibitem{10.2307/2007373}
\hrefCMSnoop {}{G.~M. Nielson, ``A method for interpolating scattered data
  based upon a minimum norm network'',} \textit{ Math. Comput.} \textbf{ 40}
  (1983) 253,
  \href{http://dx.doi.org/10.1090/S0025-5718-1983-0679444-7}{\doi{10.1090/S0025-5718-1983-0679444-7}}.

\bibitem{10.2307/44236796}
\hrefCMSnoop {}{R.~J. Renka and A.~K. Cline, ``{A triangle-based $C^1$
  interpolation method}'',} \textit{ Rocky Mountain J. Math.} \textbf{ 14}
  (1984) 223,
  \href{http://dx.doi.org/10.1216/RMJ-1984-14-1-223}{\doi{10.1216/RMJ-1984-14-1-223}}.

\bibitem{Sirunyan:2018fpa}
\hrefCMSnoop {}{{CMS Collaboration}, ``Performance of the {CMS} muon detector
  and muon reconstruction with proton-proton collisions at $\sqrt{s}=13$
  {TeV}'',} \textit{ JINST} \textbf{ 13} (2018) P06015,
  \href{http://dx.doi.org/10.1088/1748-0221/13/06/P06015}{\doi{10.1088/1748-0221/13/06/P06015}},
  \href{http://www.arXiv.org/abs/1804.04528}{\texttt{arXiv:1804.04528}}.

\bibitem{Sirunyan:2019yvv}
\hrefCMSnoop {}{{CMS Collaboration}, ``{Performance of the reconstruction and
  identification of high-momentum muons in proton-proton collisions at
  $\sqrt{s} =$ 13 TeV}'',} \textit{ JINST} \textbf{ 15} (2020) P02027,
  \href{http://dx.doi.org/10.1088/1748-0221/15/02/P02027}{\doi{10.1088/1748-0221/15/02/P02027}},
  \href{http://www.arXiv.org/abs/1912.03516}{\texttt{arXiv:1912.03516}}.

\bibitem{Sirunyan:2018nqx}
\hrefCMSnoop {}{{CMS Collaboration}, ``{Measurement of the inelastic
  proton-proton cross section at $\sqrt{\text{s}}\text{ = 13} $\TeV}'',}
  \textit{ JHEP} \textbf{ 07} (2018) 161,
  \href{http://dx.doi.org/10.1007/JHEP07(2018)161}{\doi{10.1007/JHEP07(2018)161}},
\href{http://www.arXiv.org/abs/1802.02613}{\texttt{arXiv:1802.02613}}.

\bibitem{Bowler:1981sb}
\hrefCMSnoop {}{M.~G. Bowler, ``{e+ e- production of heavy quarks in the string
  model}'',} \textit{ Z. Phys. C} \textbf{ 11} (1981) 169,
  \href{http://dx.doi.org/10.1007/BF01574001}{\doi{10.1007/BF01574001}}.

\bibitem{Andersson:1983ia}
\hrefCMSnoop {}{B.~Andersson, G.~Gustafson, G.~Ingelman, and T.~Sj{\"o}strand,
  ``{Parton fragmentation and string dynamics}'',} \textit{ Phys. Rept.}
  \textbf{ 97} (1983) 31,
  \href{http://dx.doi.org/10.1016/0370-1573(83)90080-7}{\doi{10.1016/0370-1573(83)90080-7}}.

\bibitem{Sjostrand:1984ic}
\hrefCMSnoop {}{T.~Sj{\"o}strand, ``{Jet fragmentation of nearby partons}'',}
  \textit{ Nucl. Phys. B} \textbf{ 248} (1984) 469,
  \href{http://dx.doi.org/10.1016/0550-3213(84)90607-2}{\doi{10.1016/0550-3213(84)90607-2}}.

\bibitem{Heister:2001jg}
\hrefCMSnoop {}{{ALEPH} Collaboration, ``{Study of the fragmentation of \PQb
  quarks into B mesons at the \PZ peak}'',} \textit{ Phys. Lett. B} \textbf{
  512} (2001) 30,
  \href{http://dx.doi.org/10.1016/S0370-2693(01)00690-6}{\doi{10.1016/S0370-2693(01)00690-6}},
  \href{http://www.arXiv.org/abs/hep-ex/0106051}{\texttt{arXiv:hep-ex/0106051}}.

\bibitem{DELPHI:2011aa}
\hrefCMSnoop {}{{DELPHI} Collaboration, ``{A study of the \PQb quark
  fragmentation function with the DELPHI detector at LEP I and an averaged
  distribution obtained at the \PZ pole}'',} \textit{ Eur. Phys. J. C} \textbf{
  71} (2011) 1557,
  \href{http://dx.doi.org/10.1140/epjc/s10052-011-1557-x}{\doi{10.1140/epjc/s10052-011-1557-x}},
  \href{http://www.arXiv.org/abs/1102.4748}{\texttt{arXiv:1102.4748}}.

\bibitem{Abbiendi:2002vt}
\hrefCMSnoop {}{{OPAL} Collaboration, ``{Inclusive analysis of the \PQb quark
  fragmentation function in \PZ decays at LEP}'',} \textit{ Eur. Phys. J. C}
  \textbf{ 29} (2003) 463,
  \href{http://dx.doi.org/10.1140/epjc/s2003-01229-x}{\doi{10.1140/epjc/s2003-01229-x}},
  \href{http://www.arXiv.org/abs/hep-ex/0210031}{\texttt{arXiv:hep-ex/0210031}}.

\bibitem{Abe:2002iq}
\hrefCMSnoop {}{{SLD} Collaboration, ``{Measurement of the \PQb quark
  fragmentation function in \PZz decays}'',} \textit{ Phys. Rev. D} \textbf{
  65} (2002) 092006,
  \href{http://dx.doi.org/10.1103/PhysRevD.66.079905}{\doi{10.1103/PhysRevD.66.079905}},
  \href{http://www.arXiv.org/abs/hep-ex/0202031}{\texttt{arXiv:hep-ex/0202031}}.

\bibitem{hepdata}
\hrefCMSnoop {}{}{HEPD}ata record for this measurement, 2021.
\newblock
  \href{http://dx.doi.org/10.17182/hepdata.114864}{\doi{10.17182/hepdata.114864}}.

\end{thebibliography}\endgroup
\cleardoublepage \appendix\section{The CMS Collaboration \label{app:collab}}\begin{sloppypar}\hyphenpenalty=5000\widowpenalty=500\clubpenalty=5000\cmsinstitute{Yerevan~Physics~Institute, Yerevan, Armenia}
A.~Tumasyan
\cmsinstitute{Institut~f\"{u}r~Hochenergiephysik, Vienna, Austria}
W.~Adam\cmsorcid{0000-0001-9099-4341}, J.W.~Andrejkovic, T.~Bergauer\cmsorcid{0000-0002-5786-0293}, S.~Chatterjee\cmsorcid{0000-0003-2660-0349}, M.~Dragicevic\cmsorcid{0000-0003-1967-6783}, A.~Escalante~Del~Valle\cmsorcid{0000-0002-9702-6359}, R.~Fr\"{u}hwirth\cmsAuthorMark{1}, M.~Jeitler\cmsAuthorMark{1}\cmsorcid{0000-0002-5141-9560}, N.~Krammer, L.~Lechner\cmsorcid{0000-0002-3065-1141}, D.~Liko, I.~Mikulec, P.~Paulitsch, F.M.~Pitters, J.~Schieck\cmsAuthorMark{1}\cmsorcid{0000-0002-1058-8093}, R.~Sch\"{o}fbeck\cmsorcid{0000-0002-2332-8784}, M.~Spanring\cmsorcid{0000-0001-6328-7887}, S.~Templ\cmsorcid{0000-0003-3137-5692}, W.~Waltenberger\cmsorcid{0000-0002-6215-7228}, C.-E.~Wulz\cmsAuthorMark{1}\cmsorcid{0000-0001-9226-5812}
% ...
\vskip\cmsinstskip
% ...
1:  Also at TU~Wien, Wien, Austria\\
% ...
\end{sloppypar}
\end{document}